\documentclass[%
reprint,
superscriptaddress,
nofootinbib,
amsmath,amssymb,
aps,
floatfix,
10pt
]{revtex4-2} 
\usepackage[latin1]{inputenc}
\usepackage{amsmath,amssymb}
\usepackage{mathrsfs} 
\usepackage[capitalise]{cleveref}
\usepackage{siunitx}
\usepackage{braket}
\usepackage{calc}
\usepackage{tabularx}
\usepackage{dsfont}
\usepackage{color}
\usepackage{seqsplit}
\usepackage{ifthen}
\usepackage[normalem]{ulem}
\usepackage{graphicx}
\usepackage{mathtools} 

\allowdisplaybreaks
\newcommand{\ii}{\mathrm{i}}%

\newcommand{\dif}{\mathrm{d}}%
\newcommand{\Laplace}{\boldsymbol{\triangle}}%
\newcommand{\abs}[1]{\lvert#1\rvert}%
\newcommand{\Tr}{\operatorname{Tr}}%
\newcommand{\ZT}[1]{\textquotedblleft#1\textquotedblright}%

\newcolumntype{Y}{>{\centering\arraybackslash}X}%
\newcolumntype{Z}{>{\raggedright\arraybackslash}X}%

\newlength{\myl}%
\newcommand{\SUM}[2]{{\setlength{\myl}{\widthof{$\displaystyle\sum_{#1}^{#2}$}*\real{0.5}-\widthof{$\displaystyle\sum$}*\real{0.5}}\sum_{#1}^{#2}\;\hspace{-\the\myl}}}
\newcommand{\INT}[3]{\settowidth{\myl}{$\displaystyle\int_{#1}^{#2}$}{\int_{#1}^{#2}\;\;\;\hspace{-\the\myl}\dif #3}\,}
\newcommand{\TINT}[3]{\settowidth{\myl}{$\int_{#1}^{#2}$}{\int_{#1}^{#2}\!\ifthenelse{\equal{#1#2}{}}{}{\;\;\;\;\hspace{-\the\myl}}\dif #3}\,}%
\newcommand{\EINT}[3]{\settowidth{\myl}{$\int_{#1}^{#2}$}{\int_{#1}^{#2}\;\;\;\,\hspace{-\the\myl}\dif #3}\,}

\begin{document}
\title{Dynamical field theories for biaxial liquid crystals}

\author{Anouar El Moumane}
\affiliation{Physical and Theoretical Chemistry Laboratory, Department of Chemistry, University of Oxford, South Parks Road, OX1 3QZ, United Kingdom}

\author{Ren\'e Wittmann}
\affiliation{Institut f\"ur Sicherheit und Qualit\"at bei Fleisch, Max Rubner-Institut, 95326 Kulmbach, Germany}
\affiliation{Institut f\"{u}r Theoretische Physik II: Weiche Materie, Heinrich-Heine-Universit\"{a}t D\"{u}sseldorf, 40225 D\"{u}sseldorf, Germany}

\author{Hartmut L\"owen}
\affiliation{Institut f\"{u}r Theoretische Physik II: Weiche Materie, Heinrich-Heine-Universit\"{a}t D\"{u}sseldorf, 40225 D\"{u}sseldorf, Germany}

\author{Michael te Vrugt}
\email[]{Corresponding author: tevrugtm@uni-mainz.de}
\affiliation{Institut f\"ur Physik, Johannes Gutenberg-Universit\"at Mainz, 55128 Mainz, Germany}

\begin{abstract}
Phase field crystal (PFC) models constitute central tools for a microscopic understanding of the dynamics of complex systems in soft matter physics. They have found widespread application in the modeling of the uniaxial orientational ordering of liquid crystals. 
However, only very limited progress has been made in applying them to the more complex cases of biaxial phases and biaxial particles. Here, we discuss the microscopic derivation of PFC models for biaxial liquid crystals. We illustrate it by presenting two models, one involving four scalar orientational order parameters relevant for the dynamics of biaxial particles, and one involving two scalar order parameters and a director field to describe biaxial phases in a three-dimensional uniaxial nematic liquid crystal. These models allow for an efficient simulation of spatially inhomogeneous  biaxial orientational ordering dynamics. 
 We also combine a microscopic and macroscopic approach to extract  
model coefficients for a full biaxial model from the 
microscopic derivation for a simple special case. This universal method also enables to perform derivations for other low-symmetry particles where, due to the complexity of the general case, this has not been previously attempted. 
\end{abstract}
\maketitle

\section{Introduction}

The study of liquid crystalline ordering phenomena \cite{DeGennesP1993} is one of the central areas of soft matter physics 
\cite{dussi2016entropy, tran2016lassoing, chiappini2019biaxial,de2022skyrmion, lavrentovich2020ferroelectric, caimi2023fluid, zappone2020analogy, xia2021structural, paget2023complex, jull2024curvature, doostmohammadi2018active, huang2022defect, julicher2022broken}.
Interest in liquid crystals is motivated both by their rich phase behavior \cite{vroege1992phase,singh2000phase} and by the possibility of 
exploiting their intriguing optical properties 
for a broad range of technological applications \cite{lagerwall2012new,uchida2022advanced}.  
Even in the absence of the positional order that is characteristic for an ordinary crystal,
liquid crystals can exhibit order-disorder phase transitions due to their orientational degrees of freedom.
The most prominent example is the nematic phase where particles preferably align along a common axis, which can be easily manipulated by external fields. The properties of liquid crystals motivating research on them gain additional significance in liquid crystals where the phase or the particle interactions are not axially symmetric.
Biaxial liquid crystals \cite{LuckhurstS2015} can display considerably more orientational ordered phases \cite{freiser1970ordered,alben1973phase,taylor1991nematic,vanakaras2003theory,madsen2004thermotropic,van2009experimental, martinez2011biaxial,belli2011polydispersity, peroukidis2013phase, op2014tuning, cuetos2017phase,el2024biaxial}.
  Due to experimental advances in synthesizing colloidal particles of nearly arbitrary shape \cite{kuijk2011synthesis, yu2017synthesis, chen2019recent}, 
  liquid crystal phases exhibiting biaxial order
 have the potential to be translated into further applications. 

The dynamics of phase transitions in liquid crystals can be modeled via field theories, for which dynamical density functional theory (DDFT) is a prime example. DDFT, developed in Refs.\ \cite{Evans1979,MarconiT1999,ArcherE2004} and reviewed in Refs.\ \cite{teVrugtLW2020,te2022perspective}, is a microscopic field-theoretical method that allows to model the dynamics of complex fluids by extending results from equilibrium density functional theory (DFT) to the nonequilibrium case. The application of DDFT to orientational dynamics has a long tradition \cite{ChandraB1989b,VijayadamodarB1989,ChandraB1990,BagchiC1993,ChandraB1988}. Usually, these studies assume the state of an individual particle to be determined by a position vector $\boldsymbol{R}$ and a single orientation vector $\boldsymbol{m}_3$, 
which in the three-dimensional case implies that the particles have an axis of continuous rotational symmetry (\textit{uniaxial particles}). The archetypal example for this are rod-like particles. In general, specifying the orientation of a hard particle without such a symmetry (\textit{biaxial particle}) in three dimensions requires three angles (\textit{Euler angles}) \cite{GrayG1984}. A DDFT for this case, describing particles with arbitrary shapes, was derived in Refs.\ \cite{WittkowskiL2011,DuranYGK2017}. DDFT is now a widely used method to model  particles with orientational degrees of freedom, with applications that include nematic and smectic liquid crystals \cite{BiervRDvdS2008}, deposition of hard spherocylinders \cite{KlopotekHGDSSO2017}, protein solvation \cite{MondalMB2017}, and active matter \cite{Wensink2008aggregation,te2022derivation}.

Phase field crystal (PFC) models \cite{ElderKHG2002,EmmerichEtAl2012}  offer a simpler, more phenomenological, description than DDFT, and can be 
be derived from it via a series of approximations 
\cite{ElderPBSG2007,vanTeeffelenBVL2009,ArcherRRS2019,te2022derivation}. 
PFC models for liquid crystals, which can be derived from a DDFT for particles with orientational degrees of freedom, usually feature one or several orientational order parameters. Work on this topic started in Ref.\ \cite{Loewen2010} with the development of a PFC model for nematic liquid crystals in two dimensions. Extensions considering three-dimensional \cite{WittkowskiLB2010} and then polar liquid crystals \cite{WittkowskiLB2011,WittkowskiLB2011b} 
paved the way to an active PFC model \cite{MenzelL2013}, the most widely used theory of this type,
which  has evolved into one of the central modeling frameworks in active matter physics \cite{MenzelOL2014,OphausGT2018,te2021jerky,te2022derivation,frohoff2023stationary,te2023passive,huang2020dynamical,arold2020active,HollAGKOT2021}.

Most of these PFC studies, however, were limited to particles with uniaxial symmetry and did not consider orientational order parameters other than polarization and nematic order. An exception for the two-dimensional case is the recent work by \citet{WeigelS2022}, who derived a PFC model for particles that have an $n$-fold rotational symmetry. This derivation exploited the fact that one can, in two dimensions, specify the orientation of a particle of an arbitrary shape using just one angle. 
The very few existing field-theoretical models for the three-dimensional case \cite{WittkowskiL2011}, while being very general, are also quite complicated and did therefore not allow for an efficient numerical treatment. Therefore, the study of biaxial liquid crystals in three spatial dimensions would benefit significantly from the development of field-theoretical models that allow to study them in a PFC-type framework.

In this work, we discuss how such models can be derived by combining the derivation strategy employed in Refs.\ \cite{Loewen2010,WittkowskiLB2010,WittkowskiLB2011,WittkowskiLB2011b,te2022derivation} with the orientational order parameter approach employed in Refs.\ \cite{rosso2007orientational,el2024biaxial}. As specific examples, we then present two models covering important scenarios. First, we obtain a PFC model for liquid crystals consisting of biaxial particles 
that describes the dynamics of the orientation-averaged density $\psi_1$
together with the four scalar orientational order parameters $S$, $U$, $P$, and $F$,
as used in earlier treatments of biaxial particles 
\cite{rosso2007orientational,el2024biaxial}. Second, also accounting for the director field dynamics, we obtain a PFC model for phase biaxiality, which generalizes the result of Ref.\ \cite{WittkowskiLB2010}. This model is obtained with an implicit derivation, as explained in Sec.\ref{model2section}.
Our results can be used for efficient analytical and numerical investigations of orientational ordering dynamics in biaxial liquid crystals.  We also develop an efficient 
method that allows us to arrive at the general full biaxial free energy by combining the restricted microscopic derivation of the first model with a macroscopic approach. This method is also applicable for other symmetry groups  
and might allow to greatly simplify derivations of the free energy of particles with even lower symmetry (such as
chiral particles or bent-core molecules in chiral phases). Thereby, the technique presented here 
allows to perform further derivations that have previously not been attempted.

\subsection{Density Functional Theory (DFT)} 
The Hohenberg-Kohn theorem \cite{Evans1979,hohenberg1964inhomogeneous} states, in the classical case, that there is a unique functional mapping between the one-particle-density $\rho$ and the many-body phase-space distribution. This implies the existence of a grand canonical potential $\Omega$, which is a functional of $\rho$ and which is minimized by the equilibrium density: 
\begin{equation}\label{HohenbergKohn1}
    \frac{\delta \Omega}{\delta \rho(\boldsymbol{R}, \boldsymbol{O})} = 0.
\end{equation}
Here, $\boldsymbol{O}$ denotes the orientation and $\boldsymbol{R}$ the center-of-mass position of the particles. Via the transformation
\begin{equation}
    \Omega = \mathcal{F} + \INT{}{}{\boldsymbol{R}} \INT{}{}{\boldsymbol{O}} \rho(\boldsymbol{R},\boldsymbol{O})\left(V_{\textbf{ext}} -\mu \right),
\end{equation}
\cref{HohenbergKohn1} can be translated into the equation
\begin{equation}\label{HohenbergKohn2}
    \frac{\delta \mathcal{F}}{\delta \rho(\boldsymbol{R}, \boldsymbol{O})} = \mu-V_{\textbf{ext}},
\end{equation}
for the free energy functional $\mathcal{F}$ with external potential $V_{\textbf{ext}}$ and chemical potential $\mu$.
Since there exists an analytic representation of the free energy functional for an ideal gas, given by
\begin{align} 
 \mathcal{F}_{\mathbf{id}}= k_{\mathrm{B}}T\INT{}{}{\boldsymbol{R}} \INT{}{}{\boldsymbol{O}} \left(\rho(\boldsymbol{R},\boldsymbol{O})(\ln(\lambda^3 \rho(\boldsymbol{R},\boldsymbol{O}))-1)\right),
                \label{idealgasfreeenergy}
\end{align} 
we can split the free energy functional 
\begin{align} 
    \mathcal{F} &= \mathcal{F}_{\mathbf{id}}+\mathcal{F}_{\mathbf{exc}}
\end{align} 
into an ideal gas contribution $\mathcal{F}_{\mathrm{id}}$ and an unknown excess functional $\mathcal{F}_{\mathrm{exc}}$ accounting for further interactions not present in an ideal gas.

External potentials are ignored in this study, but can be easily added using the methods presented later on.
In \cref{idealgasfreeenergy}, $\lambda^3$ denotes the thermal de Broglie wavelength (which is required here only for dimensional reasons), $k_{\mathrm{B}}$ is the Boltzmann constant and $T$ is the temperature.
For the excess functional, there are multiple commonly used approximations such as Fundamental Measure Theory (FMT) \cite{FMTReview}, the Ramakrishnan-Yussouff functional \cite{RY1979} or the Onsager functional \cite{Onsager1949} for liquid crystals.

\subsection{Dynamical Density Functional Theory (DDFT)}

A widely used generalization of DFT is called dynamical density functional theory (DDFT) and is used to describe the dynamics of many-particle-systems as a field theory of the one-particle-density. An extensive review can be found in \cite{teVrugtLW2020}.
DDFT is, unlike DFT, not even in principle an exact theory, since it relies on the so-called "adiabatic approximation", which assumes that a certain equilibrium relation, the Yvon-Born-Green-relation (YBG-relation) \cite{HansenMD2009}, is valid in nonequilibrium \cite{TschoppB2022}, which is not the case. However, such a relation is approximately true for many systems not far from equilibrium.
Again, more details on the derivation are given in \cite{teVrugtLW2020}.

The DDFT equation for biaxial particles has, as shown in \cite{WittkowskiL2011}, the general form 
\begin{align}\label{generalddft} 
   \partial_{t} \rho(\boldsymbol{\mathcal{R}},t)=\beta \nabla_{{\boldsymbol{\mathcal{R}}}} \cdot \Big( D(\boldsymbol{\mathcal{R}})\cdot \Big(\rho(\boldsymbol{\mathcal{R}},t)\nabla_{\boldsymbol{\mathcal{R}}}\frac{\delta \mathcal{F}[\rho(\boldsymbol{\mathcal{R}},t)]}{\delta \rho(\boldsymbol{\mathcal{R}}, t)}\Big)\Big).
\end{align}
Here, $\boldsymbol{\mathcal{R}}$ denotes the vector $(\boldsymbol{R},\boldsymbol{O})^\mathrm{T}$ consisting of the spatial coordinate $\boldsymbol{R}$ and the angular orientation given by the three Euler angles $\boldsymbol{O} = (\phi, \theta, \chi)^\mathrm{T}$. The diffusion tensor $\boldsymbol{D}$ is given by
\begin{equation}
    \boldsymbol{D}(\boldsymbol{O})=\left( \begin{array}{cc}
        \boldsymbol{D}_{\textbf{TT}}(\boldsymbol{O}) & \boldsymbol{D}_{\textbf{TR}}(\boldsymbol{O}) \\
        \boldsymbol{D}_{\textbf{RT}}(\boldsymbol{O}) & \boldsymbol{D}_{\textbf{RR}}(\boldsymbol{O})
    \end{array}  \right),
\end{equation}
where $\boldsymbol{D}_{\textbf{TT}}$ is the translational diffusion tensor, $\boldsymbol{D}_{\textbf{RR}}$ the rotational diffusion tensor, and $\boldsymbol{D}_{\textbf{RT}}$ and $\boldsymbol{D}_{\textbf{TR}}$ are tensors coupling translational and rotational diffusion. The spatial dependency the diffusion coefficients might have in more complex cases is here ignored. Each of those matrices has three rows and columns (for three spatial coordinates and three Euler angles). 
The differential operator $\nabla_{\boldsymbol{\mathcal{R}}} = (\nabla,\nabla_{\boldsymbol{O}})^\mathrm{T}$ is a six-dimensional operator, which contains in addition to the spatial derivative $\nabla$ the angular derivative operator $\nabla_{\boldsymbol{O}}$ with components 
\begin{align}\label{angularmomentum}
    (\nabla_{\boldsymbol{O}})_{1}&=-\cos(\phi)\cot(\theta)\partial_{\phi}
    -\sin(\phi)\partial_{\theta}-\cos(\phi)\csc(\theta)\partial_{\chi},\cr
    (\nabla_{\boldsymbol{O}})_{2}&=-\sin(\phi)\cot(\theta)\partial_{\phi}+
     \cos(\phi)\partial_{\theta}+
     \sin(\phi)\csc(\theta)\partial_{\chi},\cr
    (\nabla_{\boldsymbol{O}})_{3}&=\partial_{\phi}.
\end{align} 
\subsection{Phase field crystal (PFC) models}\label{pfcmodels}
Equation \eqref{generalddft} is, while being very general, also very complicated -- it describes the dynamics of a density $\rho$ defined on a six-dimensional configuration space, involves a 6x6 diffusion tensor leading to couplings between spatial and orientational derivatives, and is (due to the sometimes complicated form of DFT free energy functionals) in general nonlocal. Therefore, practical applications require the development of a simpler model. Such a simplification is commonly achieved using phase field crystal models (PFC models) \cite{ElderKHG2002,EmmerichEtAl2012}, which can be derived via a series of approximations from (D)DFT \cite{ArcherRRS2019,teVrugtLW2020} (and which alternatively can be obtained from macroscopic symmetry arguments).

\section{ Biaxial order parameters \label{bop}}
 \subsection{\label{scalarops}Scalar biaxial order parameters}
As a common first step in the derivation of PFC models, we need to find an approximate expression for the density based on the symmetries of the system. The systematic method for doing this is to take an appropriate biaxial expansion of the density and truncate it after the second order. Higher orders of the expansion are not taken into account. The expansion coefficients will be related to a set of biaxial order parameters defined later on, however, the exact relationship depends, of course, on the expansion we choose. This relationship arises from the fact, that the order parameters are mathematically simply the projection of the density on some angular function, which might be part of a complete orthogonal basis. 
Common expansions are the expansion in Wigner D matrices as defined in \eqref{DefinitionWignerD}, which is the biaxial generalization of spherical harmonics and the Cartesian expansion whose second order is an expansion in the elements $R_{ij}$ of the rotation matrix for three Euler angles (as defined in \eqref{rotationmatrix}) 
and its coefficients are the elements of the Saup\'e matrix (at least almost, see \cite{turzi2011cartesian}).

However, this raises a question: how do we know which expansion coefficients of a general expansion can be relevant for biaxial phases or particles? The short answer is that our expansion relies on certain symmetry arguments, as a consequence of which some order parameters are automatically zero. This is not unlike the uniaxial nematic phase, where a transformation $\boldsymbol{l}_3 \to -\boldsymbol{l}_3$ of the director $\boldsymbol{l}_3$ needs to leave the system unchanged and thus does not allow polar order parameters of the first order of an angular expansion. The details on the required conditions for the order parameters $S,U,P,F$ employed here to be the only relevant ones are discussed below, however a derivation is found in \cite{rosso2007orientational}.

We first consider the case of constant directors, in which we can restrict ourselves to scalar order parameters. In addition to the orientation-averaged density ${\psi_1}$ and the uniaxial nematic order parameter $S$ familiar from previous work \cite{WittkowskiLB2010}, we employ the additional order parameters $U$, $P$ and $F$ that we have already used in Ref.\ \cite{el2024biaxial} to study the equilibrium case. These order parameters are defined as 
\begin{align}\label{DefinitionY}
    Y&=\frac{5}{8\pi^2 \rho_{0}}\INT{}{}{\boldsymbol{O}} \rho(\boldsymbol{O}) f_{Y}(\boldsymbol{O}),\\
    {\psi_1}&=\frac{1}{8\pi^2\rho_{0}}\INT{}{}{\boldsymbol{O}} \rho(\boldsymbol{O})
\end{align} 
with $Y$ $\in {S,U,P,F}$ and the functions
\begin{align}\label{3}
    f_{S}(\boldsymbol{O})&=\frac{3}{2}\cos^2(\theta)-\frac{1}{2},\cr
    f_{U}(\boldsymbol{O})&=\frac{\sqrt{3}}{2}\sin^2(\theta)\cos(2\chi),\cr
    f_{P}(\boldsymbol{O})&=\frac{\sqrt{3}}{2}\sin^2(\theta)\cos(2\phi),\cr
    f_{F}(\boldsymbol{O})&=\frac{1} {2}\left(1+\cos^2(\theta)\right)\cos(2\phi)\cos(2\chi)\cr
    &-\cos(\theta)\sin(2\phi)\sin(2\chi).
\end{align} 
Here, $\rho_0$ denotes the constant bulk density (which could also be interpreted as a spatial and orientational average over $\rho(\boldsymbol{R},\boldsymbol{O})$).
A more detailed discussion of the physical meaning of the order parameters will be provided in Sec. \ref{meaning} below. In the literature, there are different conventions for defining the order parameters. We here use a similar convention to the one employed in Ref.\ \cite{WittkowskiLB2011b}, other works  \cite{rosso2007orientational,el2024biaxial} omit the prefactor 5 in \cref{DefinitionY}. (For the difference regarding Ref.\cite{WittkowskiLB2011b}, see the discussion under \eqref{reference}.) Of course, the definitions are equivalent, if used consistently.
Note that we implicitly assume that there are no polar phases, as we would otherwise need to incorporate polar order parameters.

\subsection{Tensorial biaxial order parameters}
A further important aspect to take into account is the spatial dependence of the nematic directors, which would be relevant for instance when describing a nematic twist-bend phase. We therefore will now introduce general order parameters for a system with spatially varying nematic directors.\linebreak
 We begin by introducing the orthonormal tripods  ${\boldsymbol{m}_{1},\boldsymbol{m}_{2},\boldsymbol{m}_{3}}$, constituting the basis of the molecular frame, ${\boldsymbol{e}_{1},\boldsymbol{e}_{2},\boldsymbol{e}_{3}}$, constituting the basis of the fixed laboratory frame, and ${\boldsymbol{l}_{1},\boldsymbol{l}_{2},\boldsymbol{l}_{3}}$, consisting of the three nematic directors, 
 which we here explicitly allow to change over time and in space. Note that in uniaxial phases the nematic directors $\boldsymbol{l}_2$ and $\boldsymbol{l}_3$  are 
 irrelevant, as there is  
 ordering around only one axis, in which case the other nematic directors can be chosen arbitrarily. Finally, the molecular frame is particle-fixed in the sense that the orientation of a particle is constant in this frame.
 It can be written through the projection on the lab frame:
 \begin{align}
     \boldsymbol{m}_{1} = (\boldsymbol{m}_{1}\cdot \boldsymbol{e}_{1})\boldsymbol{e}_{1} + (\boldsymbol{m}_{1}\cdot \boldsymbol{e}_{2})\boldsymbol{e}_{2} + (\boldsymbol{m}_{1}\cdot \boldsymbol{e}_{3})\boldsymbol{e}_{3},\cr
     \boldsymbol{m}_{2} = (\boldsymbol{m}_{2}\cdot \boldsymbol{e}_{1})\boldsymbol{e}_{1} + (\boldsymbol{m}_{2}\cdot \boldsymbol{e}_{2})\boldsymbol{e}_{2} + (\boldsymbol{m}_{2}\cdot \boldsymbol{e}_{3})\boldsymbol{e}_{3},\cr
     \boldsymbol{m}_{3} = (\boldsymbol{m}_{3}\cdot \boldsymbol{e}_{1})\boldsymbol{e}_{1} + (\boldsymbol{m}_{3}\cdot \boldsymbol{e}_{2})\boldsymbol{e}_{2} + (\boldsymbol{m}_{3}\cdot \boldsymbol{e}_{3})\boldsymbol{e}_{3}.
\end{align} 
 The projection of $\boldsymbol{m}_{i}$ on $\boldsymbol{e}_{j}$ 
 is related to a rotation 
 by
 \begin{equation}
     \boldsymbol{m}_{i}\cdot \boldsymbol{e}_{j} = R_{ji},
 \end{equation}
 while the elements $R_{ji}$ of the corresponding rotation matrix 
 are given by
\begin{align}\label{rotationmatrix}
    R_{11} &= \cos(\phi)\cos(\theta)\cos(\chi)-\sin(\phi)\sin(\chi),\cr
    R_{12} &= -\cos(\chi)\sin(\phi)-\cos(\phi)\cos(\theta)\sin(\chi),\cr
    R_{13} &= \cos(\phi)\sin(\theta),\cr
    R_{21} &=\cos(\phi)\sin(\chi)+
    \cos(\theta)\cos(\chi)\sin(\phi),\cr
    R_{22} &= \cos(\phi)\cos(\chi) - \cos(\theta)\sin(\phi)\sin(\chi),\cr
    R_{23} &= \sin(\phi)\sin(\theta),\cr
    R_{31} &= -\cos(\chi)\sin(\theta),\cr
    R_{32} &= \sin(\theta)\sin(\chi),\cr
    R_{33} &= \cos(\theta).
\end{align} 
 Now we will move on by constructing the tensors 
 \begin{align}\label{tensorintroduction}
     \boldsymbol{M}_{0} &= \sqrt{\frac{3}{2}}\left(\boldsymbol{m}_{3}\otimes \boldsymbol{m}_{3}-\frac{1}{3}\boldsymbol{I}\right),\cr
     \boldsymbol{M}_{1} &=\sqrt{\frac{1}{2}}\left(\boldsymbol{m}_{1} \otimes \boldsymbol{m}_{1} - \boldsymbol{m}_{2} \otimes \boldsymbol{m}_{2}\right)
\end{align} 
 and
\begin{align}
     \boldsymbol{L}_{0} &= \sqrt{\frac{3}{2}}\left(\boldsymbol{l}_{3}\otimes \boldsymbol{l}_{3}-\frac{1}{3}\boldsymbol{I}\right),\cr
     \boldsymbol{L}_{1} &= \sqrt{\frac{1}{2}}\left(\boldsymbol{l}_{1} \otimes \boldsymbol{l}_{1} - \boldsymbol{l}_{2} \otimes \boldsymbol{l}_{2}\right).
\end{align}
Here, $\boldsymbol{I}$ denotes the identity matrix in three dimensions and $\otimes$ stands for the Kronecker product (also known as the tensor product).
The molecular tensors $\boldsymbol{M}_0 $ and $\boldsymbol{M}_1 $
are later averaged to become our tensor order parameters $\langle \boldsymbol{M}_{0} \rangle $ and $\langle \boldsymbol{M}_{1} \rangle $ and the tensors $\boldsymbol{L}_0$ and $\boldsymbol{L}_{1}$ will serve as a basis for an expansion of the tensor order parameters.
Note that the relations presented here are not valid for any kind of biaxial particle (or biaxial phase). The assumption made here is that the particle (or more accurately, its pair interaction potential) needs to possess a $D_{2h}$ symmetry. Informally, this corresponds to a "vertical flip-over" symmetry. It can be argued, as done in \cite{el2024biaxial}, that the results extend to certain other particles with different symmetry groups as well, since certain order parameters might be theoretically allowed, but not necessarily physically relevant. However, generally speaking, other order parameters could arise via other symmetry groups. One example would be a particle shape like a prism or a pyramid that would destroy the $\boldsymbol{l}_3 \to -\boldsymbol{l}_3$ symmetry. In this case, (first-order) polar order parameters would arise and potentially be physically relevant.
As another example, it might be of interest to study chiral particles or phases, which lack mirror symmetry. With this weaker symmetry assumption, there would be other second-order order parameters relevant. More discussion on this is given in \cite{rosso2007orientational}.

As we will note soon, the density can be fully expressed using only the biaxial tensors presented above. First, we obtain the following expression for the averaged molecular tensors $\langle \boldsymbol{M}_0 \rangle$ and $\langle \boldsymbol{M}_1\rangle$ by projecting them onto $\boldsymbol{L}_0$ and $\boldsymbol{L}_1$: 
 \begin{align}\label{directordef}
     \langle \boldsymbol{M}_{0} \rangle = S\boldsymbol{L}_{0} + P\boldsymbol{L}_{1},\cr
     \langle \boldsymbol{M}_{1} \rangle = U\boldsymbol{L}_{0} + F\boldsymbol{L}_{1}.
\end{align} 
 where we defined the $S,U,P,F$ scalar order parameters as
 \begin{align}
     S &= \frac{5}{8\pi^2}\langle \boldsymbol{M}_{0} : \boldsymbol{L}_{0} \rangle,\cr
     U &= \frac{5}{8\pi^2}\langle \boldsymbol{M}_{1} : \boldsymbol{L}_{0} \rangle,\cr
     P &= \frac{5}{8\pi^2}\langle \boldsymbol{M}_{0} : \boldsymbol{L}_{1} \rangle,\cr
     F &= \frac{5}{8\pi^2}\langle \boldsymbol{M}_{1} : \boldsymbol{L}_{1} \rangle,
\end{align} 
and $\langle... \rangle = \INT{}{}{\boldsymbol{O}}(\rho(\boldsymbol{R},\boldsymbol{O})-\rho_0)/\rho_0\cdot (...)$.
A discussion of their physical significance is given in Sec. \ref{meaning}. 
The inner product between two tensors $:$ denotes the operation
 \begin{equation}\label{innerproduct}
     \boldsymbol{A} : \boldsymbol{B} = \mathbf{Tr}(\boldsymbol{A}\boldsymbol{B}^{T})
 \end{equation}
 and the averaged molecular tensors $\langle \boldsymbol{M}_0 \rangle$ and $\langle \boldsymbol{M}_1 \rangle$ are
 \begin{align}
     \langle \boldsymbol{M}_{0} \rangle &= \frac{5}{8\pi^2\rho_0} \INT{}{}{\boldsymbol{O}}\rho(\boldsymbol{R},\boldsymbol{O})\sqrt{\frac{3}{2}}\left(\boldsymbol{m}_{3}\otimes \boldsymbol{m}_{3} -\frac{1}{3}\boldsymbol{I}\right),\cr
     \langle \boldsymbol{M}_{1}\rangle &= \frac{5}{8\pi^2\rho_0} \INT{}{}{\boldsymbol{O}}\rho(\boldsymbol{R},\boldsymbol{O})\sqrt{\frac{1}{2}}\left(\boldsymbol{m}_{2}\otimes \boldsymbol{m}_{2} - \boldsymbol{m}_{1} \otimes \boldsymbol{m}_{1}\right) .
\end{align} 
(Note that this differs from our usual orientational average $\langle ... \rangle = \INT{}{}{\boldsymbol{O}}\rho(\boldsymbol{O})/\rho_0 \cdot ...$ by a prefactor of $\frac{5}{8\pi^2}$. This is the only instance where it differs due to reasons of keeping a compact notation.)
Apart from the microscopic definitions of $\langle \boldsymbol{M}_{0} \rangle$ and $\langle \boldsymbol{M}_{1} \rangle $ presented above, we note that, as commonly done in literature, the directors $\boldsymbol{l}_i$ and the order parameters can be gained from the molecular tensors as their eigenvectors and eigenvalues respectively.
It is now easy to see
that for a constant nematic director $\boldsymbol{l}_{3} = (0,0,1)^\mathrm{T}$ this more general approach coincides with the previous definition in \eqref{3}.
 Now, generalizing some of the ideas outlined in \cite{rosso2007orientational} and \cite{Longa2005}, we express the biaxial density through the order parameters presented here by using orientational expansions. Although similar expressions appear in the earlier literature \cite{Longa2005}, previous work usually considered the case of constant directors (i.e. $\boldsymbol{l}_i$ coincides with $\boldsymbol{e}_i$), while we will later introduce a model for which we drop this requirement to find novel expressions for the description of biaxial phases.

 \section{Density expansions}

\subsection{Fixed directors \label{fdr}}
Now, having outlined the general theory for biaxial order, we want to apply this to our specific goal, namely the microscopic derivation of a field theory for biaxial liquid crystals. In order to do so, we first need to find an expansion of the one-particle-density in terms of the order parameters introduced earlier. 
This will later, as outlined in Sec. \ref{general}, allow us to gain closed-form expressions for the free energy in dependence of these order parameters. 

One possibility for an orientational expansion of the one-particle-density is to perform an expansion in elements of the rotation matrix  $R_{ij}$, also known as the Cartesian expansion. The $R_{ij}$ are given in \cref{rotationmatrix}.
We give the explicit form for a generic function $f$ up to second order here, more details can be found in \cite{turzi2011cartesian,te2020relations}:
\begin{align}
    f(\boldsymbol{R},\boldsymbol{O}) = \psi(\boldsymbol{R}) + \SUM{i,j=1}{3}P_{ij}(\boldsymbol{R})R_{ij} + \SUM{i,j,k,l=1}{3}Q_{ijkl}R_{ij}R_{kl}.
\end{align} 
with the orientation-averaged density $\psi$, the polarization $P_{ij}$ and the generalized nematic tensor $Q_{ijkl}$, which here (unlike in the uniaxial case) are a second- and a fourth-rank tensor, respectively. They are given by 
\begin{align}
    \psi &= \frac{1}{8\pi^2}\INT{}{}{\boldsymbol{O}}f(\boldsymbol{R},\boldsymbol{O}),\cr
    P_{ij} &= \frac{3}{8\pi^2}\INT{}{}{\boldsymbol{O}}f(\boldsymbol{R},\boldsymbol{O})R_{ij},\cr
    Q_{ijkl} &= \frac{5}{16\pi^2}\INT{}{}{\boldsymbol{O}}f(\boldsymbol{R}, \boldsymbol{O})\Big(R_{ij}R_{kl}+R_{il}R_{kj}-\frac{2}{3}\delta_{ik}\delta_{jl}\Big).\cr
\end{align} 
This expansion is closely related to the Saup\'e matrix \cite{turzi2011cartesian}, which is widely used in experiments  \cite{saupe1}.

Another widely used approach for a biaxial expansion of the density is the expansion in Wigner D matrices, also known as the angular multipole expansion, since it is the biaxial generalization of an expansion in spherical harmonics \cite{te2020relations}. 
The Wigner D matrices  $D_{mn}^l$ are defined as \cite{te2020orientational} 
\begin{equation}\label{DefinitionWignerD}
    D_{mn}^{l}=e^{-im\phi}e^{-in\chi}d_{mn}^l(\theta)
\end{equation}
where $d_{mn}^l$ is defined as
\begin{align}\label{DefinitionWignerD2}
    & d_{mn}^l(\theta) = \sqrt{(l+n)!(l-n)!(l+m)!(l-m)!} \cr
    &\times \sum_{k \in I_{mn}^l}\frac{(-1)^{k}\sin^{2l+m-n-2k}(\frac{\theta}{2})\cos^{2k-m+n}(\frac{\theta}{2}))}{(l+m-k)!(l-n-k)!k!(k-m+n)!}.
\end{align} 
Here, $I_{mn}^l$ is the set of all integers for which the arguments of the factorials appearing inside the sum in \eqref{DefinitionWignerD2} are greater or equal to zero. 
This special case of an expansion in spherical harmonics has been used multiple times in the literature before, such as in Refs.~\cite{WittkowskiLB2010,bickmann2020collective}.

Expressing the scalar order parameters $S,U,P,F$, discussed in Sec.~\ref{bop},  
in terms of Wigner D matrices results in 
\begin{align}
    S &= \frac{5}{8\pi^2}\langle D^{2}_{00} \rangle,\cr
    U &= \frac{5}{8\pi^2}\langle \frac{D^{2}_{02}+D^{2}_{0-2}}{\sqrt{2}} \rangle,\cr
    P &= \frac{5}{8\pi^2}\langle \frac{D^{2}_{20}+D^{2}_{-20}}{\sqrt{2}} \rangle,\cr
    F &= \frac{5}{8\pi^2}\langle \frac{D^{2}_{22}+D^{2}_{-22}+D^{2}_{2-2}+D^{2}_{-2-2}}{2} \rangle,
\end{align}
where $\langle ...\rangle$ denotes the orientation average $\INT{}{}{\boldsymbol{O}} \rho/ \rho_{0}$. These expressions can be derived from the definitions of $f_S, f_F, f_P, f_F$ provided in Ref.~\cite{mulder1989isotropic}. These relations will prove to be helpful in the expansion of the direct correlation function later on.

For the density we pursue another approach, as the relation between the order parameters and the resulting expansion coefficients is still quite complicated for Wigner D matrices (which is similarly true for an expansion in rotation matrix elements).
We proceed by making the ansatz
\begin{align} \label{Density}
    \rho(\boldsymbol{R},\boldsymbol{O}) &= \rho_{0}\Big({\psi_1}+S f_{S}(\boldsymbol{O})+U f_{U}(\boldsymbol{O})+ P f_{P}(\boldsymbol{O})\cr
    &+F f_{F}(\boldsymbol{O})\Big)
\end{align} 
for the density as a function of the scalar order parameters.
We now give the explicit justification of this approximation for the density by doing an expansion to the second order in a set of matrices introduced by \citet{mulder1989isotropic}, which we will refer to as \ZT{Mulder matrices}. The expansion is not done directly in $\rho$, but rather in $\rho/\rho_0$.

The Mulder matrices are an orthogonal function system and thus an orthogonal basis for a function $f$ (which in our case is given by $\rho/\rho_{0}$), defined by the relation
\begin{equation}\label{Mulder}
    \Delta_{mn}^l=\left(\frac{1}{\sqrt{2}}\right)^{2+\delta_{m0}+\delta_{n0}} \SUM{
   }{}D_{\sigma m, \sigma_{2}n}^l 
\end{equation}
where the sum is such that $\sigma,\sigma_{2}={-1,1}$, further $0\geq m, n \geq l$ and $l,n,m$ are even \cite{mulder1989isotropic}. They obey the orthogonality relations
\begin{equation}\label{MulderCoefficients}
     \INT{}{SO(3)}{\boldsymbol{O}}\Delta^{{l}_{1}}_{{m}_{1}n_{1}} \Delta^{{l}_{2}}_{{m}_{2}n_{2}}  = \frac{8\pi^2}{2l+1}\delta_{{l}_{1}{l}_{2}}\delta_{{m}_{1}{m}_{2}}\delta_{{l}_{1}{l}_{2}}.
\end{equation}
Here, the $D_{mn}^l$ are the Wigner D matrices defined above.

The resulting expansion of an arbitrary function  $f$ into Mulder matrices is then given by 
\begin{align}\label{MulderExpansion}
    f(\boldsymbol{R}_{1},\boldsymbol{R}_{2},\boldsymbol{O}_{1},&\boldsymbol{O}_{2})=\sum_{j=1, 2}\sum_{l_{j},m,_{l}n_{l}}f_{{l}_{1}{m}_{1}n_{1}{l}_{2}{m}_{2}n_{2}}\cr
    &\times \frac{(2{l}_{1}+1)(2{l}_{2}+1)}{64\pi^4}\Delta^{{l}_{1}}_{{m}_{1}{n_{1}}}
    \Delta^{{l}_{2}}_{{m}_{2}{n_{2}}},
\end{align} 
with 
\begin{align}\label{MulderCoefficients2}
    f_{{l}_{1}{m}_{1}n_{1}{l}_{2}{m}_{2}n_{2}}&=\INT{}{}{\boldsymbol{O}_{1}}\INT{}{}{{O}_{2}} \Delta_{{m}_{1}n_{1}}^{{l}_{1}}\Delta_{{m}_{2}n_{2}}^{{l}_{2}}\cr
    & \times f(\boldsymbol{O}_{1},\boldsymbol{O}_{2},\boldsymbol{R}_{1},\boldsymbol{R}_{2}).
\end{align} 
(The orientational expansion of the density depends only on a single orientation $\boldsymbol{O}$, yet other expansions later on will use a modified version of the general form in \eqref{MulderExpansion}.) 

After truncating the expansion in \eqref{MulderExpansion} at order $l$=2, we need to define our order parameters as averages of the Mulder matrices. Since we know that only the averages of the $f_{Y}$ and the orientational average of 1 (leading to the ${\psi_1}$ parameter) are relevant for the biaxial nematic phase, we need to find a way to express these averages through Mulder matrices. This can be achieved using the relations
\begin{align}\label{Eq:MulderSUPF}
{\psi_1} &=\frac{1}{8\pi^2}\langle 1 \rangle = \frac{1}{8\pi^2}\langle \Delta_{00}^0 \rangle,\cr
S &= \frac{5}{8\pi^2}\langle\frac{3}{2}\cos^2(\theta)-\frac{1}{2}\rangle=\frac{5}{8\pi^2}\langle \Delta_{00}^2 \rangle,\cr
U &= \frac{5}{8\pi^2}\langle \frac{\sqrt{3}}{2}\sin^2(\theta)\cos(2\chi)\rangle
= \frac{5}{8\pi^2}\langle \Delta_{02}^2 \rangle,\cr
P &= \frac{5}{8\pi^2}\langle \frac{\sqrt{3}}{2}\sin^2(\theta)\cos(2\phi)\rangle
= \frac{5}{8\pi^2}\langle \Delta_{20}^2 \rangle,\cr
F &= \frac{5}{8\pi^2}\langle \frac{1}{2}(1+\cos^2(\theta))\cos(2\phi)\cos(2\chi)\cr
&-\cos(\theta)\sin(2\phi)\sin(2\chi)\rangle
= \frac{5}{8\pi^2}\langle \Delta_{22}^2 \rangle,
\end{align} 
which, paired with \eqref{MulderCoefficients}, implies
\begin{align}\label{DensityMulder}
    \rho &\approx \rho_{0}(1+\frac{\Delta_{00}^{0}}{8\pi^2}\langle \Delta_{00}^{0}\rangle+\frac{5\langle \Delta_{00}^2 \rangle}{8\pi^2} \Delta_{00}^2 +\frac{5\langle \Delta_{02}^2 \rangle}{8\pi^2} \Delta_{02}^2\cr
    &+ \frac{5\langle \Delta_{20}^2 \rangle}{8\pi^2} \Delta_{20}^2
    +\frac{5\langle \Delta_{22}^2 \rangle}{8\pi^2} \Delta_{22}^2).
\end{align} 
Alternatively, we can explicitly convert back to the $S,U,P,F,{\psi_1}$ parameters and arrive back at \eqref{Density}. (Note that \eqref{Eq:MulderSUPF} clarifies how we can formally define $f_{\psi_1}:=1$. 
The absende of a factor of 5 comes from the fact that the order parameters $S, U, P, F$ appear in the second order of the orientational expansion, while  $\psi_1$ comes from the zeroth order of the orientational expansion.) 

\subsection{Spatially varying directors}\label{spatvardirectors}
We now generalize the expansion \cref{Density} for spatially varying directors. In order to do so, we expand in the elements of the two tensors $\boldsymbol{M}_0$ and $\boldsymbol{M}_1$, as their averages fully characterize the system. The case for constant directors is presented in \cite{Longa2005}, and thus it is easy to see how the more general method presented here 
reduces to the previous known case. Explicitly, the density can be written as
 \begin{align}\label{DensityDirector}
     \rho &= \rho_{0}\left(1+{\psi_1} +\langle \boldsymbol{M}_{0}\rangle : \boldsymbol{M}_{0} + \langle \boldsymbol{M}_{1}\rangle : \boldsymbol{M}_{1} \right) \cr
     &= \rho_{0} (1+\psi_{1}+\left(\langle \boldsymbol{M}_{0}: \boldsymbol{L}_{0} \rangle \boldsymbol{L}_{0} + \langle \boldsymbol{M}_{0} : \boldsymbol{L}_{1}\rangle \boldsymbol{L}_{1}\right): \boldsymbol{M}_{0}\cr
     & +\left(\langle \boldsymbol{M}_{1}: \boldsymbol{L}_{0} \rangle \boldsymbol{L}_{0} + \langle \boldsymbol{M}_{1} : \boldsymbol{L}_{1}\rangle \boldsymbol{L}_{1}\right): \boldsymbol{M}_{1})\cr
      &=\rho_{0}(1+\psi_{1}+  \langle \boldsymbol{M}_{0}: \boldsymbol{L}_{0}\rangle \boldsymbol{M}_{0}: \boldsymbol{L}_{0} + \langle \boldsymbol{M}_{0} : \boldsymbol{L}_{1}\rangle \boldsymbol{M}_{0}: \boldsymbol{L_{1}}\cr
     & + \langle \boldsymbol{M}_{1} : \boldsymbol{L}_{0}\rangle \boldsymbol{M}_{1}: \boldsymbol{L}_{0} + \langle \boldsymbol{M}_{1} : \boldsymbol{L}_{1}\rangle  \boldsymbol{M}_{1} : \boldsymbol{L}_{1})\cr
     &=\rho_{0}(1 + \psi_{1} + S \boldsymbol{M}_{0}: \boldsymbol{L}_{0} + U \boldsymbol{M}_{1}: \boldsymbol{L}_{0} + P \boldsymbol{M}_{0}: \boldsymbol{L}_{1} \cr
     & + F \boldsymbol{M}_{1} : \boldsymbol{L}_{1}),
\end{align} 
 which generalizes Eq.~\eqref{DensityMulder}. Here, $\langle \boldsymbol{M}_0 \rangle$ is (up to a conventional prefactor $\sqrt{3/2}$) the well-known nematic tensor $\boldsymbol{Q}$ 
also appearing, for instance, in the phenomenological Landau-de Gennes expansion of the free energy \cite{qtensormottram}. $\langle \boldsymbol{M}_1 \rangle$ is a similar tensor relevant for particles that are biaxial. (Notice that there can also be biaxial phases in a system of uniaxial particles, these correspond to a nonzero value of the order parameter $P$.)
 \subsection{Relation to expansions for uniaxial particles}
Next, we show that our biaxial expansion reduces to the uniaxial one employed in previous work \cite{WittkowskiLB2010}. The first step is to neglect the order parameters $U$, $P$, and $F$. Then, assuming $P=U=F=0$, the density from \cref{DensityDirector} has the form:
 \begin{equation} \label{DensityS}
     \rho = \rho_{0}\left( \psi_{1} + S \boldsymbol{M}_{0}: \boldsymbol{L}_{0} \right). 
 \end{equation}
The expression appearing in the last term reads
\begin{align}\label{sderiv}
    \boldsymbol{M}_{0} : \boldsymbol{L}_{0} &= \frac{3}{2}\Big((\boldsymbol{m}_{3}\cdot \boldsymbol{e}_{j})(\boldsymbol{m}_{3}\cdot \boldsymbol{e}_{i})(\boldsymbol{l}_{3}\cdot \boldsymbol{e}_{j})(\boldsymbol{l}_{3}\cdot \boldsymbol{e}_{i})-\frac{3}{9}\cr
    &-\frac{(\boldsymbol{m}_{3}\cdot \boldsymbol{e}_{j})(\boldsymbol{m}_{3}\cdot \boldsymbol{e}_{i})(\boldsymbol{e}_{i}\cdot \boldsymbol{e}_{j})+(\boldsymbol{l}_{3}\cdot \boldsymbol{e}_{j})(\boldsymbol{l}_{3}\cdot \boldsymbol{e}_{i})(\boldsymbol{e}_{i}\cdot \boldsymbol{e}_{j})}{3}\Big)\cr
    &= \frac{3}{2}\left((\boldsymbol{m}_{3}\cdot \boldsymbol{l}_{3})^2-\frac{2}{3}+\frac{1}{9}(\boldsymbol{e}_{i}\cdot \boldsymbol{e}_{i})(\boldsymbol{e}_{j}\cdot \boldsymbol{e}_{j})\right)\cr
    &=\frac{3}{2}\left((\boldsymbol{m}_{3}\cdot \boldsymbol{l}_{3})^2-\frac{1}{3}\right)\cr
    &= L_{2}(\boldsymbol{m}_{3}\cdot \boldsymbol{l}_{3}),
\end{align} 
where $L_{2}$ is the second Legendre polynomial. This agrees with the result by \citet{WittkowskiLB2010} (their $\hat{u}_{0}$ corresponds to our $\boldsymbol{l}_{3}$,our $\boldsymbol{m}_{3}$ corresponds to their $\hat{u}$ and finally our $\psi_1$ corresponds to their $1+\psi_1$):
\begin{equation}\label{reference}
    \rho = \rho_{0} \left( 1+\psi_1+S L_{2}(\boldsymbol{m}_{3} \cdot \boldsymbol{l}_3)\right).
\end{equation}
(Somewhat more subtle, our density still has a $\chi$-dependence, as seen by the remaining normalization factor proportional to $1/(8\pi^2)$, since we still integrate over all three Euler angles. However, the density distribution for $\chi$ is uniform, as we assumed $U=P=F=0$, 
and thus absorbed in the different prefactors for the order parameters as compared to \cite{WittkowskiLB2010} and \cite{WittkowskiLB2011b}, where the $\chi$-dependence was neglected entirely.)
Our method thus generalizes the results of Ref.\ \cite{WittkowskiLB2010} to both biaxial particles and biaxial phases. 

\subsection{Meaning of the scalar order parameters}
\label{meaning}
Analogously to \eqref{sderiv}, we recover the definitions in terms of $\boldsymbol{m}_i$ and $\boldsymbol{l}_i$ for $S$, $P$, $U$ and $F$ as given in \cite{rosso2007orientational}: 
\begin{align}
    S &= \frac{3}{2}\langle (\boldsymbol{m}_3 \cdot \boldsymbol{l}_3 - \frac{1}{3})\rangle,\cr
    U &= \frac{\sqrt{3}}{2}\langle(\boldsymbol{m}_{1}\cdot \boldsymbol{l}_{3})^2-(\boldsymbol{m}_{2}\cdot \boldsymbol{l}_{3})^2\rangle,\cr
    P &= \frac{\sqrt{3}}{2}\langle(\boldsymbol{m}_{3}\cdot \boldsymbol{l}_{1})^2-(\boldsymbol{m}_{3}\cdot \boldsymbol{l}_{2})^2\rangle,\cr
    F &= \frac{1}{2}\langle(\boldsymbol{m}_{1}\cdot \boldsymbol{l}_{1})^2-(\boldsymbol{m}_{1}\cdot \boldsymbol{l}_{2})^2-(\boldsymbol{m}_{2}\cdot \boldsymbol{l}_{1})^2\\
    &+(\boldsymbol{m}_{2}\cdot \boldsymbol{l}_{2})^2\rangle
\end{align} 
where the directors are now, in contrast to \cite{rosso2007orientational}, explicitly space- and time-dependent. This notation also helps us to develop some intuition for the physical meaning of the order parameters introduced here:
\begin{itemize}
    \item $S$ - order parameter for uniaxiality: this order parameter measures the alignment of the molecular axis $\boldsymbol{m}_{3}$ with the the director $\boldsymbol{l}_3$ (often referred to as \textit{the} nematic director). It is 0 in an isotropic phase and 1 in the perfectly nematic phase.  
    \item $U$ - order parameter for \textit{molecular} biaxiality: this order parameter is nonzero only if the particle shape is biaxial. However, $U$ can be nonzero even in a phase which is not biaxial. This can be understood by recognizing that the uniaxial (molecular) symmetry transformation $\boldsymbol{m}_{1} \to -\boldsymbol{m}_{1}, \boldsymbol{m}_{2} \to -\boldsymbol{m}_{2}$ does not leave the order parameter invariant, but the uniaxial (phase) symmetry transformation $\boldsymbol{l}_{1} \to -\boldsymbol{l}_{1}, \boldsymbol{l}_{2} \to -\boldsymbol{l}_{2}$ does leave the order parameter invariant, as expected.
    \item $P$ - order parameter for \textit{phase} biaxiality: here, the uniaxial phase symmetry transformation $\boldsymbol{l}_{1} \to -\boldsymbol{l}_{1}, \boldsymbol{l}_{2} \to -\boldsymbol{l}_{2}$ does not leave $P$ invariant, however, assuming uniaxial symmetry in the shape through $\boldsymbol{m}_{1} \to -\boldsymbol{m}_{1}, \boldsymbol{m}_{2} \to -\boldsymbol{m}_{2}$ leaves the order parameter invariant. As a result, this order parameter accounts for biaxiality in the phase. This means, this order parameter can be nonzero for both uniaxial and biaxial particle shapes, but it can only be nonzero in a biaxial phase. 
    \item $F$ - order parameter for \textit{full} biaxiality: this order parameter is only nonzero if both the particle shape and the phase are biaxial. Uniaxial symmetry transformations of the molecular axes or the director axes both do not leave the order parameter invariant. Thus, it can only be nonzero for full biaxiality.
\end{itemize}
Note that there is, as noted earlier already, in general a difference between \textit{biaxial phases} and \textit{biaxial particles}. A biaxial phase is a phase in which, in addition to $S$, $P$ is also nonzero -- which can at least mathematically also happen if the underlying particles are uniaxial. In contrast, $U$ and $F$ can only be nonzero if the particles are biaxial (an example for this would be hard cuboids). 

Finally, we note that we can recover the special case for constant directors {discussed in Sec.~\ref{fdr}} by simply setting $\boldsymbol{l_{i}}=\boldsymbol{e_{i}}$. Note that in this case the names \ZT{director frame} and \ZT{laboratory frame} can be and are used interchangeably \cite{rosso2007orientational,Longa2005}. This no longer holds in the general time-dependent case that we consider here. 

\section{Free energies  }
\subsection{Macroscopic model}\label{macro}
There have been multiple studies investigating the formulation of a macroscopic model incorporating full biaxiality by generalizing the early uniaxial macroscopic theory of de Gennes \cite{DeGennesP1993}, such as done in, e.g., Refs.\ \cite{Matteis2009,matteis,Xu2020,longa1987,longa1989}. Here, we give a short overview over these results. 
General methods for the derivation can be found in \cite{longa1989} and \cite{longa1987}.

Very often authors assume that $\boldsymbol{Q} = \sqrt{3/2}\langle \boldsymbol{M}_0 \rangle$ and $\boldsymbol{K} = \langle \boldsymbol{M}_1 \rangle/\sqrt{2}$
for their macroscopic models. This does not actually change the physics as the resulting model describing the coupling of (in principle) arbitrary tensors, that can especially have arbitrary prefactors. $\boldsymbol{Q}$ is often used for denoting the well-known nematic/uniaxial order tensor whose largest eigenvalue is equal to $S$, with the corresponding eigenvector being $\boldsymbol{l_{3}}$ \cite{phdrene}.
In a theory accounting for the coupling of the two tensors $\boldsymbol{Q}$ and $\boldsymbol{K}$,
the bulk energy is given by the following linear combination of tensor invariants up to fourth order where we combined expressions from \cite{matteis,Matteis2009} and \cite{EmmerichEtAl2012} to arrive at: 
\begin{align}
\mathcal{F} &= {b_1}\Tr(\boldsymbol{Q}^2)+{b_2}\Tr(\boldsymbol{Q}^3)+{b_3}\Tr(\boldsymbol{Q}^2)^2+{b_4}\Tr(\boldsymbol{K}^2)\cr
&+{b_5}\Tr(\boldsymbol{K}^2)^2+{b_6}\Tr(\boldsymbol{Q}\boldsymbol{K})^2+{b_7}\Tr(\boldsymbol{Q}^2\boldsymbol{K}^2)\cr
&+{b_8}\Tr(\boldsymbol{Q}\boldsymbol{K})^2+{b_9}\Tr(\boldsymbol{Q}\boldsymbol{K})+{b_{10}}\Tr(\boldsymbol{K}^3)+{b_{11}}\Tr(\boldsymbol{Q}^2\boldsymbol{K})\cr
&+{b_{12}}\Tr(\boldsymbol{Q}^3\boldsymbol{K})+{b_{13}}\Tr(\boldsymbol{Q}\boldsymbol{K}^3)+{b_{14}}\Tr(\boldsymbol{K}^2)\Tr(\boldsymbol{Q}^2)\cr
&+b_{15}\psi_1^2 +b_{16}\psi_1^3 +b_{17} \psi_1^4 + b_{18}\psi_1 \Tr(\boldsymbol{Q}\boldsymbol{K})+b_{19}\psi_1 \Tr(\boldsymbol{Q}^2)\cr
&+b_{20}\psi_1\Tr(\boldsymbol{K}^2)+b_{21}\psi_1\Tr(\boldsymbol{Q}^3)+b_{22}\psi_1 \Tr(\boldsymbol{K}^3)\cr
&+b_{23}\psi_1\Tr(\boldsymbol{K}^2\boldsymbol{Q})
+b_{24}\psi_1\Tr(\boldsymbol{Q}^2\boldsymbol{K})\cr
&+b_{25}\psi_1^2\Tr(\boldsymbol{Q}^2)+b_{26}\psi_1^2\Tr(\boldsymbol{K}^2) +b_{27}\psi_1^2\Tr(\boldsymbol{Q}\boldsymbol{K}).
\label{eq_Fmacro}
\end{align} 
The coefficients $b_1 - b_{27}$ are macroscopic parameters chosen in accordance with simulations or experimental data and $\Tr$ denotes the trace operator.
Clearly, the expansion \eqref{eq_Fmacro} is much more complicated than the uniaxial theory proposed by de Gennes, which is only based on $\boldsymbol{Q}$ \cite{DeGennesP1993} -- in \cref{eq_Fmacro}, we have a total of 27 independent terms! Now, following the notation introduced in \cite{Xu2020}, we include the elastic energy. For a tensor $\boldsymbol{A}$, we define gradient and divergence as 
\begin{align}
    (\nabla \boldsymbol{A})_{i,j_1...j_n} &= \partial_i \boldsymbol{A}_{j_1...j_n},\cr
    (\nabla \cdot \boldsymbol{A})_{j_1...j_{n-1}} &= \sum_i \partial_i \boldsymbol{A}_{j_1...j_{n-1}, i}
\end{align}
and for two tensors $\boldsymbol{A}$ and $\boldsymbol{B}$ use the inner product $:$ as defined in \eqref{innerproduct}.
In this notation, the form obtained by Xu and Chen \cite{Xu2020} for the elastic free energy density of a coupling of two distinct symmetric traceless tensors $\boldsymbol{Q}$ and $\boldsymbol{K}$ with each other and with a scalar field $\psi$ is 
\begin{align}\label{elasticenergy}
    f &= c_1 (\nabla \psi_1): (\nabla \psi_1) \cr
    &+c_2(\nabla \boldsymbol{Q}) : (\nabla \boldsymbol{Q})+ c_3 (\nabla \boldsymbol{Q}) : (\nabla \boldsymbol{K})\cr
    &+c_4 (\nabla \boldsymbol{K}) : (\nabla \boldsymbol{K}) + c_5 (\nabla \cdot \boldsymbol{Q}): (\nabla \cdot \boldsymbol{Q})\cr
    &+c_6 (\nabla \cdot \boldsymbol{Q}) : (\nabla \cdot \boldsymbol{K})+c_7 (\nabla \cdot \boldsymbol{K}): (\nabla \cdot \boldsymbol{K})\cr
    &+c_8 (\nabla \psi_1): (\nabla \cdot \boldsymbol{K})+ c_{9} (\nabla \psi_1): (\nabla \cdot \boldsymbol{Q}).
\end{align}
Once again, the coefficients $c_1-c_{9}$ are just macroscopic coefficients chosen in accordance with simulations or experimental data.
Note that this expression holds only for particles with symmetry group $D_{2h}$. 
Ref.\ \cite{Xu2020} only gives the bulk energy up to second order. Another theory is given in \cite{Matteis2009}, where both the full elastic energy and the bulk energy up to fourth order is presented.
However, unlike Ref.\ \cite{Xu2020}, Ref.\ \cite{Matteis2009} does not include the coupling of the space-dependent density with the nematic tensors.

\subsection{General route to microscopic models}\label{general}
We now present the general approach for systematic microscopic derivations of phase field crystal models, which naturally account for, in principle, all couplings up to any order, following previous work on the subject \cite{ElderPBSG2007,vanTeeffelenBVL2009,ArcherRRS2019,te2022derivation}. We then use the presented ideas to derive two models for biaxial liquid crystals.

First, we consider the \textbf{ideal gas free energy}:
\begin{enumerate}
    \item We insert the parametrization \eqref{DensityDirector} of the density in terms of the order parameters into the ideal gas free energy \eqref{idealgasfreeenergy}.
\item We Taylor expand the logarithm in \cref{idealgasfreeenergy} up to third order (such that the resulting expression for $\mathcal{F}_{\mathrm{id}}$ is a fourth-order polynomial). An expansion up to third order is common as it allows to model crystal formation \cite{EmmerichEtAl2012}.
\item We evaluate the angular integrals.
\end{enumerate}
For the \textbf{excess free energy}, the derivation is somewhat more complicated:
\begin{enumerate}
    \item We choose a suitable approximation for the excess free energy, the explicit form of which is not known. We will here work with the common Ramakrishnan-Yussouf approximation \cite{RamakrishnanY1979}, where the excess free energy is written in terms of the second-order direct correlation function. Other approaches, such as a higher order functional Taylor expansion, are also possible and have been investigated before \cite{WittkowskiL2011,WittkowskiLB2011b,te2022derivation}.
\item  We insert the parametrization \eqref{DensityDirector} of the density in terms of the order parameters into the excess free energy.
\item  Since the orientational dependence of the direct correlation function is generally unknown, we need to perform an orientational expansion respecting the internal symmetries of the system. Details are given later.
\item  We evaluate the angular integrals.
\item The resulting expression involves a convolution integral. In order to remove this nonlocality, we perform a gradient expansion up to second order (fourth order for terms only involving $\psi_1$). This is a common choice in PFC modeling \cite{EmmerichEtAl2012,te2022derivation}.
\end{enumerate}
This is the basic procedure for deriving phase field crystal models for liquid crystals. Of course, details may vary.
In what follows, we show explicitly how to use this method for a microscopic derivation of the free energy of a model system.
Specifically, we introduce a restricted model for full biaxiality (by fixing the directors) and a full model for phase biaxiality (by dropping order parameters related to particle biaxiality).
\subsection{Overview over the considered models}
In the following, we will obtain microscopic expressions for the coefficients of two different models:

\begin{itemize}
    \item \textbf{Model 1} incorporates all five scalar order parameters $\psi_1, S, U, P, F$, yet holds the directors constant, where we use the density expansion \eqref{Density} instead of \eqref{DensityDirector}. This means it describes both biaxial phases and biaxial particles, but does not incorporate elastic interactions, which are, for instance, responsible for the formation of twist-bend nematics.
    \item \textbf{Model 2} incorporates the scalar order parameters $\psi_1, S, P$ and does not hold the directors $\boldsymbol{l}_{i}$ constant, where we drop the other terms in the density expansion \eqref{DensityDirector}. As such, it fully incorporates the well-known tensor order parameter $\boldsymbol{Q}$ and its coupling with the orientation averaged density $\psi_1$. However, it is limited by not incorporating the second tensor order parameter $\boldsymbol{K}$ associated with the two scalar order parameters $U, F$ and thus is not able to describe the effects induced by molecular biaxiality that would be captured by $U$ and $F$. 
\end{itemize}
It is important to note that both of these models are special cases of the more general macroscopic theory presented in \ref{macro}. The elastic energy in \cref{elasticenergy} provides a good example: if we restrict ourselves to the term $(\nabla \boldsymbol{Q})\cdot (\nabla \boldsymbol{K})$ and assume the directors to be constant, as done in Model 1, we arrive at $(\nabla S)\cdot (\nabla U)+(\nabla P)\cdot (\nabla F)$.
In the case of neglecting shape biaxiality though, we can easily recognize that this simply means $\boldsymbol{K} = 0$ and the term will thus vanish completely. In summary, each of our simplified models capture one aspect of the more general macroscopic model.
However, while the macroscopic theory is the most general one of the models presented here, the microscopic derivation has the advantage of providing microscopic expressions for the coefficients and would, if carried out in full generality, reproduce the macroscopic model.

An explicit microscopic derivation of the full biaxial model would be practically difficult due to the high number of free parameters (see Sec. \ref{macro}) and the complex form of the density with a total of ten independent tensor elements.
However, after deriving the free energy for Model 1 in Sec.~\ref{model1section}, we develop a method to infer the full biaxial model from it in Sec.~\ref{modelfullsection}, where we combine the microscopic result with the macroscopic model from Sec.~\ref{macro}.
Then, in Sec.~\ref{model2section}, we introduce and discuss an important special case of the full biaxial model presented in Sec.~\ref{modelfullsection} 
(namely the case of uniaxial particle shapes). 
Later, in Sec.~\ref{sec_dynamics}, we only derive dynamic equations for the restricted Models 1 and 2, to avoid calculating a large number of integral necessary for dynamics.

\subsection{Model 1}\label{model1section}
We first consider the case where, as in previous work \cite{el2024biaxial}, directors are held constant, implying that we have to deal with a density $\rho(\boldsymbol{O}, \psi_{1}, S,U,P,F)$. Explicitly, we assume the form presented in \eqref{Density}.
Note that this considers both tensors $\langle \boldsymbol{M}_0 \rangle$ and $\langle \boldsymbol{M}_1 \rangle$ with constant directors $\boldsymbol{l}_i$. As an important fact, keep in mind that the exact form of $\boldsymbol{l}_i$ matters. Here, we choose $\boldsymbol{l}_3 = (0,0,1), \boldsymbol{l}_2 = (0,1,0)$. Other (constant) choices are possible, but will alter both the density and the resulting free energy. 

\subsubsection{Ideal gas free energy}

We begin with calculating the ideal gas free energy.
As discussed above, we substitute $x=\frac{\rho-\rho_{0}}{\rho_{0}}$ using \cref{DensityDirector} and Taylor expand the logarithm to third order, resulting in 
\begin{align}\label{idealexpansion}
    \beta \mathcal{F}_{\mathbf{id}}&=\INT{}{}{\boldsymbol{R}}\INT{}{}{\boldsymbol{O}}\rho (\ln(\rho \lambda^3)-1)\cr
    &=\rho_{0}\INT{}{}{\boldsymbol{R}} \INT{}{}{\boldsymbol{O}} (1+x)(\ln(\lambda^3 \rho_{0}(1+x))-1))\cr
    &\approx \mathcal{F}_{0}+\rho_{0}\INT{}{}{\boldsymbol{R}} \INT{}{}{\boldsymbol{O}} \left(\frac{x^2}{2}-\frac{x^3}{6}+\frac{x^4}{12}\right), 
\end{align}
where $\mathcal{F}_{0}$ consists of irrelevant constant terms the functional derivative of which vanishes. Now, we need to calculate the angular integrals by first inserting the density defined in \cref{Density} into the expression \cref{idealexpansion} and then calculating integrals of the form $\INT{}{}{\boldsymbol{O}} f_{Y}(\boldsymbol{O})^a f_{X}(\boldsymbol{O})^b$ with $a$ and $b$ being integers obeying $a+b \leq 4$. For $a$ or $b$ equal to 1 or 0, we can use the orthogonality relations \eqref{MulderCoefficients},  which, for example, imply that terms of the form $\INT{}{}{\boldsymbol{O}} f_{S}(\boldsymbol{O})f_{F}(\boldsymbol{O})$ vanish. Otherwise, the integrals are calculated numerically. The result is
\begin{align}\label{idealmicro}
    \beta \mathcal{F}_{\mathbf{id}} &= \mathcal{F}_{0}+8\pi^2\rho_{0}\INT{}{}{\boldsymbol{R}} \Bigg(\frac{1}{2}\left(\psi_{1}^2+\sum_{X}\frac{ X^2}{5}\right) \nonumber\\
    &-\frac{1}{6}\Big(\psi_{1}^3+\frac{3\psi_{1}}{5}\sum_{X} X^2 \nonumber\\
    &-\frac{2}{35}(3P^2S+3U^2S-3F^2S-6UPF-S^3)\Big) \nonumber \\
    &+\frac{1}{12}\Big(\psi_{1}^4+\sum_{X}\frac{3X^4}{35}+\psi_{1}^2\sum_{X}\frac{6X^2}{5} \nonumber \\
    &-\frac{2}{35}\psi_{1}(12P^2S+12U^2S-12F^2S-24UPF-4S^3) \nonumber \\
    &+\frac{1}{35}\Big(12F^2 S^2 +6F^2 U^2 +6F^2 P^2+ 6S^2 P^2\nonumber \\
    &+ 6S^2 U^2 + 12U^2 P^2 -12SUPF\Big)\Big)\Bigg),
\end{align} 
where $\mathcal{F}_0$ is an irrelevant constant.

\subsubsection{Excess free energy}\label{excess}
Now, we turn our attention towards $\mathcal{F}_{\mathrm{exc}}$. We employ the widely used Ramakrishnan-Yussouff approximation \cite{RamakrishnanY1979}
\begin{align}\label{ramakrsihnanyousseff}
    \beta \mathcal{F}_{\mathrm{exc}} &= \frac{1}{2}\INT{}{}{\boldsymbol{R}_{1}}\INT{}{}{\boldsymbol{R}_{2}}\INT{}{}{\boldsymbol{O}_{1}}\INT{}{}{\boldsymbol{O}_{2}} c^{(2)}(\boldsymbol{R}_{1},\boldsymbol{R}_{2},\boldsymbol{O}_{1},\boldsymbol{O}_{2})\cr
    &\times(\rho(\boldsymbol{R}_{1},\boldsymbol{O}_{1})-\rho_{0})(\rho(\boldsymbol{R}_{2},\boldsymbol{O}_{2})-\rho_{0}) 
\end{align} 
with the direct correlation function $c^{(2)}$ of the reference state $\rho_0$
that is defined as
\begin{equation}\label{directcorrelationfunction}
    \beta c^{(2)}(\boldsymbol{R}_{1},\boldsymbol{R}_{2}, \boldsymbol{O}_{1}, \boldsymbol{O}_{2}) = \frac{-\delta^2 \mathcal{F}_{\mathbf{exc}}}{\delta \rho_{1}(\boldsymbol{R}_{1},\boldsymbol{O}_{1})\delta \rho_{2}(\boldsymbol{R}_{2},\boldsymbol{O}_{2})}\bigg|_{\rho=
    \rho_{0}}.
\end{equation}
The definition \eqref{directcorrelationfunction} makes \cref{ramakrsihnanyousseff} a functional Taylor expansion of $\mathcal{F}_{\mathrm{exc}}$, where we expand around a constant bulk density $\rho_0$.

We can simplify $c^{(2)}$ by exploiting the symmetries that the system has in the state $\rho= \rho_0$ where it is evaluated. First, we employ spatial homogeneity, which implies that 
\begin{equation}\label{correlationhomogeniety}
     c^{(2)}(\boldsymbol{R}_{1},\boldsymbol{R}_{2},\boldsymbol{O}_{1},\boldsymbol{O}_{2})=c^{(2)}(\boldsymbol{R}_{1}+\boldsymbol{a},\boldsymbol{R}_{2}+\boldsymbol{a},\boldsymbol{O}_{1},\boldsymbol{O}_{2})
\end{equation}
with an arbitrary vector $\boldsymbol{a}$. Setting $\boldsymbol{a}=-\boldsymbol{R}_{2}$, it follows that
\begin{equation}\label{correlatioinhomogeniety}
    c^{(2)}(\boldsymbol{R}_{1},\boldsymbol{R}_{2},\boldsymbol{O}_{1},\boldsymbol{O}_{2})=c^{(2)}(\boldsymbol{R}_{1}-\boldsymbol{R}_{2},\boldsymbol{O}_{1},\boldsymbol{O}_{2}).
\end{equation}
Next, we exploit the fact that the system is spatially isotropic. Mathematically, this implies that it is invariant under a rotation $\mathbf{D}$, such that 
\begin{equation}\label{correlationistropy}
    c^{(2)}(\boldsymbol{R}_{1}-\boldsymbol{R}_{2},\boldsymbol{O}_{1},\boldsymbol{O}_{2})=c^{(2)}(\mathbf{D}\cdot (\boldsymbol{R}_{1}- \boldsymbol{R}_{2}), \mathbf{D} \boldsymbol{O}_{1},\mathbf{D} \boldsymbol{O}_{2}).
\end{equation}
Note that $\mathbf{D}$ is an abstract rotation operator here, which does not have to represented by the standard rotation matrix (as the orientation $\boldsymbol{O}$ also is, since three Euler angles and thus three molecular axis are needed, not represented by a single vector). We choose $\mathbf{D}$ in such a way that $\boldsymbol{R}_{1}-\boldsymbol{R}_{2}$ lies on the $z$-axis: 
\begin{align}\label{correlationisotropy2}
    c^{(2)}(\mathbf{D}\cdot (\boldsymbol{R}_{1}-\boldsymbol{R}_{2}), \mathbf{D}\boldsymbol{O}_{1},\mathbf{D}\boldsymbol{O}_{2})=c^{(2)}(&\abs{\boldsymbol{R}_{1}-\boldsymbol{R}_{2}}\boldsymbol{e}_{z},\cr
    &\mathbf{D}\boldsymbol{O}_{1},
    \mathbf{D}\boldsymbol{O}_{2}),
\end{align} 
where $\boldsymbol{e}_{z}$ is the unit vector in $z$-direction.
Now, we need to find a way to expand the direct correlation function in a way that allows us to evaluate the resulting angular integrals. We cannot choose a simple expansion in spherical harmonics as done in Ref.\ \cite{WittkowskiLB2010} since $c^{(2)}$ depends on three angles, such that an expansion in spherical harmonics is no longer sufficient.
Further, we need to choose an expansion that respects the internal symmetries of the system, such as the one presented in Appendix A of Ref.\ \cite{GrayG1984} (where however the biaxial expansion coefficients are not given) and applied (for uniaxial systems only) in Refs.\ \cite{Graf1999} and \cite{WittkowskiLB2010}. The chosen expansion for this purpose has (using the abbreviation $\boldsymbol{R}:=\boldsymbol{R}_1-\boldsymbol{R}_2$) the form 
\begin{align}\label{symmetryexpansion}
    &c^{(2)}(\abs{\boldsymbol{R}}\boldsymbol{e}_z,\boldsymbol{O}_{12},\boldsymbol{O}_1,\boldsymbol{O}_2) = \sum_{\lambda}\omega_{\lambda}(\abs{\boldsymbol{R}}\boldsymbol{e}_z)\phi_{\lambda}(\boldsymbol{O}_{12},\boldsymbol{O}_1,\boldsymbol{O}_2),
\end{align} 
with the orientational part
\begin{align}
    \phi_{\lambda}(\boldsymbol{O}_{12},\boldsymbol{O}_1,\boldsymbol{O}_2) = \sum_{{m}_{1},{m}_{2}, m}C({l}_{1},{l}_{2},l,{m}_{1},{m}_{2},m)\cr
  \times(D^{{l}_{1}}_{{m}_{1}n_1})^{*}(\boldsymbol{O}_1)(D^{{l}_{2}}_{{m}_{2}n_2})^{*}(\boldsymbol{O}_2) Y_{lm}^{*}(\boldsymbol{O}_{12}).
\end{align}
We have used the multiindex $\lambda = ({l}_{1},n_1,{l}_{2},n_{2},l)$ and the coefficients 
\begin{align}
&\omega_{\lambda} = \frac{(2l_1+1)(2l_2+1)}{64\pi^4}\sqrt{\frac{4\pi}{2l+1}}\INT{}{}{\boldsymbol{O}_{1}}\INT{}{}{\boldsymbol{O}_{2}}\cr
&\times \sum_{m=-\min({l}_{1},{l}_{2})}^{m = \min({l}_{1},{l}_{2})}C({l}_{1},m,{l}_{2},-m,l,0)\cr
&\times D^{{l}_{1}}_{{m}_{1}n_1}(\boldsymbol{O}_1)D^{{l}_{2}}_{{m}_{2}n_2}(\boldsymbol{O}_2)c^{(2)}(\abs{\boldsymbol{R}}\boldsymbol{e}_z,\boldsymbol{O}_1,\boldsymbol{O}_2).
\end{align} 
The essential idea of this expansion is that both $\phi_{\lambda}$ and $c^{(2)}$ are invariant under a rotation of all angles. The $C({l}_{1},m_1,{l}_{2},m_2,l,m)$ are the Clebsch-Gordan coefficients. This is sufficient for our next steps.

To obtain the excess free energy, we start by inserting \cref{symmetryexpansion} and \cref{DensityMulder} into \cref{ramakrsihnanyousseff} and integrating out the orientational degrees of freedom. This gives 
\begin{align} 
    \mathcal{F}_{\mathbf{exc}} &= \frac{1}{2}\INT{}{}{\boldsymbol{R}_{1}} \INT{}{}{\boldsymbol{R}_{2}} \sum_{X,Z,l}\frac{64\pi^4\rho_0^2}{(2{l}_{X}+1)(2{l}_{Z}+1)}X(\boldsymbol{R}_{1})Z(\boldsymbol{R}_{2})\cr
    &\times \omega_{\mathcal{S}(X)_1,\mathcal{S}(X)_3,\mathcal{S}(Z)_1,\mathcal{S}(Z)_3,l}(\boldsymbol{R}_1-\boldsymbol{R}_2)\cr
    &\times C({\mathcal{S}(X)_1,\mathcal{S}(X)_2,\mathcal{S}(Z)_1,\mathcal{S}(Z)_2,l, {m}_{X}+{m}_{Z}})
\end{align} 
with $X,Z \in \{\psi_{1},S,U/\sqrt{2},P/\sqrt{2},F/2\}$ and $\mathcal{S}(X) = (l_X,m_X,n_X)$ 
\begin{align}\label{Smap}
    \mathcal{S}(\psi_{1}) &= (0,0,0), \cr 
    \mathcal{S}(S) &= (2,0,0),\cr
    \mathcal{S}(U) &= (2,0,2) \phantom{t}\textbf{or}\phantom{t} (2,0,-2),\cr
    \mathcal{S}(P) &= (2,2,0) \phantom{t}\textbf{or}\phantom{t} (2,-2,0),\cr
    \mathcal{S}(F) &= (2,2,2) \phantom{t}\textbf{or}\phantom{t} (2,-2,-2)\cr
    &\phantom{t}\textbf{or}\phantom{t} (2,-2,2) \phantom{t}
    \textbf{or}\phantom{t} (2,2,-2).
\end{align} 
The notation $\mathcal{S}(X)_i$ is to be understood such that it denotes the $i$-th element of $\mathcal{S}(X)$ as given in \eqref{Smap}. As an example, if we are interested in the terms coupling $S$ and $P$, then the required Clebsch-Gordan coefficient would be either $C(2,0,2,2,2,2)$ or $C(2,0,2,-2,2,-2)$ (since $P$ is expressed through a sum of two Wigner D matrices), and the factor $\omega_{\lambda}$ would be $\omega_{20202}$ (in both cases). Explicitly,  $\mathcal{S}(X)_2$ would be either 2 or -2. Both combinations need to be realized such that a single order parameter might result in multiple terms.  
Note that $m = {m}_{1} +{m}_{2}$, as otherwise the Clebsch-Gordan coefficient would be 0. 
Due to the gradient expansion we will perform later, $l \in {0,2}$, since the integral $\INT{}{}{\boldsymbol{O}} Y_{lm} \boldsymbol{m}_3\otimes \boldsymbol{m}_3$ is non-vanishing only for $l \in {0,2}$. Thus, effectively, only the spherical harmonics $Y_{00}, Y_{20}, Y_{22}$ and $Y_{2-2}$ are allowed.
Afterwards, we perform a change of variables $\boldsymbol{R}=\boldsymbol{R}_{1}-\boldsymbol{R}_{2}$ and Taylor expand the order parameter $X(\boldsymbol{R}_{2}+\boldsymbol{R})$ with respect to $\boldsymbol{R}$.
(This last step is usually referred to as a 'gradient expansion'.) For a generic function $f$, we can make the following expansion (which we will truncate at order $l=2$ and order $l=4$ for terms only involving $\psi_1$)
\begin{equation}
    f(\boldsymbol{R}+\boldsymbol{\Bar{R}})=\sum_{l} \frac{\Bar{R}^l}{l!}(\Bar{u}\cdot \nabla)^l f(\boldsymbol{R})
\end{equation}
with $\boldsymbol{R}$, $\boldsymbol{\Bar{R}}= \Bar{R} \boldsymbol{\Bar{u}}$ being arbitrary vectors. Here, $\boldsymbol{\Bar{u}}$ is the unit vector with the same orientation as $\boldsymbol{\Bar{R}}$, while $\Bar{R}$ denotes the absolute value of $\boldsymbol{\Bar{R}}$. 
Finally, we evaluate the integral over $\boldsymbol{R}$ and arrive at
\begin{align}\label{constant}
    \mathcal{F}_{\mathbf{exc}} &= \frac{1}{2} \INT{}{}{\boldsymbol{R}_{1}} \sum_{X,Z,l}\frac{64\pi^4\rho_0^2 X(\boldsymbol{R}_{1})}{(2{l}_{X}+1)(2{l}_{Z}+1)}\cr
    &\times C({\mathcal{S}(X)_1,\mathcal{S}(X)_2,\mathcal{S}(Z)_1,\mathcal{S}(Z)_2,l, m})\cr
    &\times\Big(
    Z(\boldsymbol{R}_{1})
    \Big(\INT{}{}{\abs{\boldsymbol{R}}} \abs{\boldsymbol{R}}^2\omega_{\mathcal{S}(X)_1,\mathcal{S}(X)_3,\mathcal{S}(Z)_1,\mathcal{S}(Z)_3,l}(\abs{\boldsymbol{R}}\boldsymbol{e}_z)\cr
    &\times \INT{}{}{\boldsymbol{O}} Y_{lm} \Big)+\sum_{i,j}\partial_{i}\partial_{j}Z(\boldsymbol{R}_{1}) \cr
    &\Big(\INT{}{}{\abs{\boldsymbol{R}}} \frac{\abs{\boldsymbol{R}}^4}{2}\omega_{\mathcal{S}(X)_1,\mathcal{S}(X)_3,\mathcal{S}(Z)_1,\mathcal{S}(Z)_3,l}(\abs{\boldsymbol{R}}\boldsymbol{e}_z)\cr
    &\times \INT{}{}{\boldsymbol{O}} Y_{lm} {m}_{3,i} {m}_{3,j} \Big)
    \Big)
\end{align} 
with the constraints on $l,m$ and $\mathcal{S}$ as explained above. The values of $l_X, m_X, n_X$ and $l_Z, m_Z, n_Z$ are connected to the sum over $X, Z$ as given in \eqref{Smap}.  The integrals $\INT{\boldsymbol{S}_2}{}{\boldsymbol{O}} Y_{lm} $ and $\INT{\boldsymbol{S}_2}{}{\boldsymbol{O}} Y_{lm} m_{3,i} m_{3,j}$ are given in detail in the appendix \ref{taylor}.

\subsection{Inferring the full biaxial model}\label{modelfullsection}

While an explicit derivation would be very tedious and practically very difficult, the coefficients of the (macroscopic) full biaxial model from Sec.~\ref{macro}.
can be microscopically determined by a comparison to the special (and much less complex) case of constant directors (Model 1),
for which we provided an explicit microscopic derivation in Sec.~\ref{model1section}. This (quite surprising) result is obtained as follows: the key to establish a link between the corresponding free energies is the relation \eqref{directordef} of the averaged molecular tensors $\langle \boldsymbol{M}_0 \rangle$ and $\langle \boldsymbol{M}_1 \rangle$ to the scalar order parameters $S,U,P,F$. We proceed by inserting $\boldsymbol{l}_i=\boldsymbol{e}_i$ into \eqref{directordef}:
\begin{align}\label{insert}
\langle \boldsymbol{M}_0 \rangle &= \left(\begin{array}{ccc}
     -S/\sqrt{6}+P/\sqrt{2}&0  &0 \\
     0& -S/\sqrt{6}-P/\sqrt{2}  &0 \\
     0&0  & 2S/\sqrt{6}
\end{array}\right),\cr
\langle \boldsymbol{M}_1 \rangle &= \left(\begin{array}{ccc}
     -U/\sqrt{6}+F/\sqrt{2}&0  &0 \\
     0& -U/\sqrt{6}-F/\sqrt{2}  &0 \\
     0&0  & 2F/\sqrt{6}
\end{array}\right) .
\end{align}
Next we exploit the fact that we know the structure of \eqref{directormodel} and \eqref{fullmodel} in advance from the macroscopic considerations in Sec. \ref{macro}. 
We calculate all tensor invariants for constant directors by inserting \label{insert} into \eqref{directormodel} and \eqref{fullmodel}.
Since none of the terms of the macroscopic model go to zero (this was explicitly checked), we remain with all coefficients.
Thus, we could collect the appropriate terms, since we already know that both approaches need to be at least consistent, that is, if we assume constant directors in the macroscopic model, we need to arrive at \eqref{constant}. We can then finally collect all terms and compare coefficients and arrive at the full biaxial energy by using our new method combining a microscopic and macroscopic approach.

This method might help to greatly simplify calculations for even more complex symmetry groups, such as for particles exhibiting molecular chirality. (At least if the primary focus is deriving microscopic expressions for coefficients and not a first-principles derivation of the structure of the free energy. The structure of the free energy, i.e. the relevant tensor invariants, is however essentially a solved problem for all relevant symmetry groups, see \cite{Xu2020}.)
Another question to be addressed is whether the coefficients are dependent on any of the order parameters, which would render this method useless. Trivially, there cannot be an explicit dependence as only the tensor invariants depend on the order parameters. However, there might be an implicit dependence as the coefficients might change depending on what density we use in the derivation, similar to how certain approximations implicitly affect the resulting set of coefficients. We can, however, rule out that possibility as it is possible to use \eqref{DensityDirector} instead of \eqref{DensityMulder} (just with some restrictive assumptions on the tensor elements to ensure that the condition of constant directors is fulfilled) and once we evaluate the orientational integrals the coefficients still cannot depend on any tensor elements (or scalar order parameters). Since we know that both approaches have to give the same result for the restricted case, this has to be true for the general case as well. 

For the dynamic equations however, we do not use the general model but restrict ourselves to constant directors again, as this would result in a huge number of additional coefficients. 

\subsubsection{Ideal gas free energy}
The free energy of the ideal gas  \eqref{idealmicro} does not contain any gradient terms.
A direct comparison to the form of \eqref{eq_Fmacro} thus gives the general form
\begin{align}\label{idealmicroM}
    \beta \mathcal{F}_{id} &= 8\pi^2 \rho_0 \INT{}{}{\boldsymbol{R}}
    \Bigg(\frac{1}{2}\Big(\psi_1^2 +\frac{\Tr (\langle \boldsymbol{M}_0\rangle^2) +\Tr(\langle \boldsymbol{M}_1 \rangle^2)}{5}\Big)\cr
    &-\frac{1}{6}\Big(3\psi_1\frac{(\Tr (\langle \boldsymbol{M}_0 \rangle^2) +\Tr(\langle \boldsymbol{M}_1 \rangle^2))}{5} +\psi_1^3\cr
    &-\frac{2}{35}(-\sqrt{6}\Tr( \langle \boldsymbol{M}_0 \rangle^3)+3\sqrt{6}\Tr(\langle \boldsymbol{M}_0 \rangle \langle \boldsymbol{M}_1 \rangle^2)) \Big)\cr
    &+\frac{1}{12}(\psi_1^4 +\frac{3}{35}\psi_1^2(\Tr(\langle \boldsymbol{M}_0 \rangle^2) + \Tr(\langle \boldsymbol{M}_1 \rangle^2))\cr
    &-\frac{8\psi_1}{35}(-\sqrt{6}\Tr( \langle \boldsymbol{M}_0 \rangle^3)+3\sqrt{6}\Tr(\langle \boldsymbol{M}_0 \rangle \langle \boldsymbol{M}_1 \rangle^2))\cr
    &+\frac{1}{35}\Big(3\Tr(\langle \boldsymbol{M}_0 \rangle^2)^2+3\Tr(\langle \boldsymbol{M}_1 \rangle^2)^2\cr &-6\Tr(\langle \boldsymbol{M}_0\rangle \langle \boldsymbol{M}_1\rangle)^2
    +12\Tr(\langle \boldsymbol{M}_0\rangle^2)\Tr( \langle \boldsymbol{M}_1 \rangle)^2\Big) 
    \Bigg),
\end{align}
which is valid even for spatially varying directors, as discussed above.

\subsubsection{Excess free energy}

Next we calculate all tensor invariants presented in the bulk free energy \eqref{eq_Fmacro} (up to second order only) and the elastic part given in \eqref{elasticenergy} for constant directors to get the model for the excess free energy
\begin{align}\label{directormodel}
    \mathcal{F}_{exc} &= \frac{1}{2}\INT{}{}{\boldsymbol{R}} \Big(
    a_1 \psi_1^2 + a_2 (\nabla\psi_1)^2 +a_3 (\Delta \psi_1)^2\cr
    &a (S^2+P^2)+ b (U^2+F^2) + c (SU+PF)\cr
    &+ d \Big(\sqrt{\frac{3}{2}}((\boldsymbol{l}_3\cdot \nabla \psi_1)(\boldsymbol{l}_3 \cdot \nabla S)-\frac{\nabla \psi_1 \cdot \nabla S}{3})\cr
    &+\sqrt{\frac{1}{2}}((\boldsymbol{l}_1 \cdot \nabla \psi_1)(\boldsymbol{l}_1 \cdot \nabla P)-(\boldsymbol{l}_2 \cdot \nabla \psi_1)(\boldsymbol{l}_2 \cdot \nabla P))\Big)\cr
    &+ e \Big(\sqrt{\frac{3}{2}}((\boldsymbol{l}_3\cdot \nabla \psi_1)(\boldsymbol{l}_3 \cdot \nabla U)-\frac{\nabla \psi_1 \cdot \nabla U}{3})\cr
    &+\sqrt{\frac{1}{2}}((\boldsymbol{l}_1 \cdot \nabla \psi_1)(\boldsymbol{l}_1 \cdot \nabla F)-(\boldsymbol{l}_2 \cdot \nabla \psi_1)(\boldsymbol{l}_2 \cdot \nabla F))\Big)\cr
    &+f ((\nabla S)^2+(\nabla P)^2)\cr
    &+g \Big(\frac{(\nabla S)^2+(\boldsymbol{l}_3 \cdot \nabla S)^2}{2}+\frac{(\boldsymbol{l}_1 \cdot \nabla P)^2+ (\boldsymbol{l}_2 \cdot \nabla P)^2}{2}\Big)\cr
    &+h ((\nabla U)^2+(\nabla F)^2)\cr
    &+i \Big(\frac{(\nabla U)^2+(\boldsymbol{l}_3 \cdot \nabla U)^2}{2}+\frac{(\boldsymbol{l}_1 \cdot \nabla F)^2+ (\boldsymbol{l}_2 \cdot \nabla F)^2}{2}\Big)\cr
    &+j  ((\nabla S)(\nabla U)+(\nabla P)(\nabla F))\cr
    &+k \Big(\frac{(\nabla U)(\nabla S)+(\boldsymbol{l}_3 \cdot \nabla U)(\boldsymbol{l}_3 \cdot \nabla S)}{2}\cr
    &+\frac{(\boldsymbol{l}_1 \cdot \nabla F)(\boldsymbol{l}_1 \cdot \nabla P)+ (\boldsymbol{l}_2 \cdot \nabla F)(\boldsymbol{l}_2 \cdot \nabla P)}{2}\Big)
    \Big).
\end{align}
in terms of scalar order parameters and directors. 
Now we can compare the undetermined coefficients in \eqref{directormodel} to the predictions from \eqref{constant}
and finally arrive at: 
\begin{align}\label{fullmodel}
    \mathcal{F}_{exc} &= \INT{}{}{\boldsymbol{R}} \Big(a_1 \psi_1^2 + a_2 (\nabla \psi_1)^2 + a_3 (\Delta \psi_1)^2 \cr
     &+a \Tr(\langle \boldsymbol{M}_0 \rangle^2) + b\Tr( \langle \boldsymbol{M}_1 \rangle^2)\cr
    &+ c \Tr(\langle \boldsymbol{M}_0 \rangle \langle \boldsymbol{M}_1 \rangle)
    +d (\nabla\psi_1) : (\nabla \cdot \langle \boldsymbol{M}_0 \rangle) \cr
    &+ e (\nabla\psi_1) : (\nabla \cdot \langle \boldsymbol{M}_1 \rangle)+f (\nabla \langle \boldsymbol{M}_{0}\rangle):(\nabla \langle \boldsymbol{M}_{0}\rangle)\cr
    &+g (\nabla \cdot \langle \boldsymbol{M}_{0}\rangle): (\nabla\cdot \langle \boldsymbol{M}_{0}\rangle)+h (\nabla \langle \boldsymbol{M}_{1}\rangle ):(\nabla \langle \boldsymbol{M}_{1}\rangle)\cr
    &+i  (\nabla \cdot \langle \boldsymbol{M}_{1}\rangle): (\nabla\cdot \langle \boldsymbol{M}_{1}\rangle)+j  (\nabla \langle \boldsymbol{M}_{0}\rangle):(\nabla \langle \boldsymbol{M}_{1}\rangle)\cr
    &+k  (\nabla \cdot \langle \boldsymbol{M}_{0}\rangle):(\nabla \langle \boldsymbol{M}_{1}\rangle)
    \Big)
\end{align}
with the coefficients 
\begin{align}\label{coefficients}
    a_1 &= 64\pi^4 \rho_0^2\sqrt{4\pi}\INT{}{}{\abs{\boldsymbol{R}}}\abs{\boldsymbol{R}}^2\omega_{00000}(\abs{\boldsymbol{R}} \boldsymbol{e}_z),\cr
    a_2 &= -64\pi^4 \rho_0^2 \sqrt{4\pi}\INT{}{}{\abs{\boldsymbol{R}}}\frac{\abs{\boldsymbol{R}}^4}{2}\omega_{00000}(\abs{\boldsymbol{R}} \boldsymbol{e}_z),\cr
    a_3 &= 64\pi^4 \rho_0^2\sqrt{4\pi}\INT{}{}{\abs{\boldsymbol{R}}}\frac{\abs{\boldsymbol{R}}^6}{4!}\omega_{00000}(\abs{\boldsymbol{R}} \boldsymbol{e}_z),\cr
    a &= \frac{64\pi^4 \rho_0^2}{25\sqrt{5}}\sqrt{4\pi}\INT{}{}{\abs{\boldsymbol{R}}}\abs{\boldsymbol{R}}^2\omega_{20200}(\abs{\boldsymbol{R}} \boldsymbol{e}_z),\cr
    b &= \frac{64\pi^4 \rho_0^2}{50\sqrt{5}}\sqrt{4\pi}\INT{}{}{\abs{\boldsymbol{R}}}\abs{\boldsymbol{R}}^2\Big(2\omega_{222-20}(\abs{\boldsymbol{R}} \boldsymbol{e}_z)\cr
    &+\omega_{22220}(\abs{\boldsymbol{R}} \boldsymbol{e}_z)+\omega_{2-22-20}(\abs{\boldsymbol{R}} \boldsymbol{e}_z)\Big) ,\cr
    c &= \frac{64\pi^4 \rho_0^2}{25\sqrt{2}\sqrt{5}}\sqrt{4\pi}\INT{}{}{\abs{\boldsymbol{R}}}\abs{\boldsymbol{R}}^2(\omega_{20220}(\abs{\boldsymbol{R}} \boldsymbol{e}_z)\cr
    &+\omega_{202-20}(\abs{\boldsymbol{R}} \boldsymbol{e}_z)),\cr
    d &= -\frac{64\pi^4 \rho_0^2}{5}\sqrt{\frac{36\pi}{45}}\INT{}{}{\abs{\boldsymbol{R}}}\omega_{00202}\abs{\boldsymbol{R}}^2(\abs{\boldsymbol{R}} \boldsymbol{e}_z),\cr
    e &=- \frac{64\pi^4 \rho_0^2}{5}\sqrt{\frac{36\pi}{45}}\INT{}{}{\abs{\boldsymbol{R}}}\abs{\boldsymbol{R}}^2\Big(\omega_{00222}(\abs{\boldsymbol{R}} \boldsymbol{e}_z)\cr
    &+\omega_{002-22}(\abs{\boldsymbol{R}} \boldsymbol{e}_z)\Big),\cr
    f &= - \frac{64\pi^4 \rho_0^2}{25} \INT{}{}{\abs{\boldsymbol{R}}}\frac{\abs{\boldsymbol{R}}^4}{2}(\frac{4\sqrt{5\pi}}{15}\sqrt{2/7}\omega_{20202}(\abs{\boldsymbol{R}} \boldsymbol{e}_z)\cr
    &+\frac{4\pi}{3\sqrt{4\pi}}\sqrt{1/5}\omega_{20200}(\abs{\boldsymbol{R}} \boldsymbol{e}_z)),\cr
    g &=-2\Big( \frac{64\pi^4 \rho_0^2}{25} \INT{}{}{\abs{\boldsymbol{R}}}\frac{\abs{\boldsymbol{R}}^4}{2}(\frac{-2\sqrt{5\pi}}{15}\sqrt{2/7}\omega_{20202}(\abs{\boldsymbol{R}} \boldsymbol{e}_z)\cr
    &+\frac{4\pi}{3\sqrt{4\pi}}\sqrt{1/5}\omega_{20200}(\abs{\boldsymbol{R}} \boldsymbol{e}_z))-f\Big),\cr
    h &=- \frac{64\pi^4 \rho_0^2}{50} \INT{}{}{\abs{\boldsymbol{R}}}\frac{\abs{\boldsymbol{R}}^4}{2}(\frac{4\sqrt{5\pi}}{15}\sqrt{2/7}(\omega_{22222}(\abs{\boldsymbol{R}} \boldsymbol{e}_z)\cr
    &+\omega_{2-22-22}(\abs{\boldsymbol{R}} \boldsymbol{e}_z)+2\omega_{2-2222}(\abs{\boldsymbol{R}} \boldsymbol{e}_z))\cr
    &+\frac{4\pi}{3\sqrt{4\pi}}\sqrt{1/5}(\omega_{22220}(\abs{\boldsymbol{R}} \boldsymbol{e}_z)+\omega_{2-22-20}(\abs{\boldsymbol{R}} \boldsymbol{e}_z)\cr
    &+2\omega_{2-2220}(\abs{\boldsymbol{R}} \boldsymbol{e}_z))),\cr
    i &= -2\Big(\frac{64\pi^4 \rho_0^2}{50} \INT{}{}{\abs{\boldsymbol{R}}}\frac{\abs{\boldsymbol{R}}^4}{2}(\frac{-2\sqrt{5\pi}}{15}\sqrt{2/7}(\omega_{22222}(\abs{\boldsymbol{R}} \boldsymbol{e}_z)\cr
    &+\omega_{2-22-22}(\abs{\boldsymbol{R}} \boldsymbol{e}_z)+2\omega_{2-2222}(\abs{\boldsymbol{R}} \boldsymbol{e}_z))\cr
    &+\frac{4\pi}{3\sqrt{4\pi}}\sqrt{1/5}(\omega_{22220}(\abs{\boldsymbol{R}} \boldsymbol{e}_z)+\omega_{2-22-20}(\abs{\boldsymbol{R}} \boldsymbol{e}_z)\cr
    &+2\omega_{2-2220}(\abs{\boldsymbol{R}} \boldsymbol{e}_z)))-h\Big),\cr
    j &=- 2\frac{64\pi^4 \rho_0^2}{25\sqrt{2}} \sqrt{1/5}\INT{}{}{\abs{\boldsymbol{R}}}\frac{\abs{\boldsymbol{
    R}}^4}{2}(\frac{4\sqrt{5\pi}}{15}\sqrt{2/7}(\omega_{20222}(\abs{\boldsymbol{R}} \boldsymbol{e}_z)\cr
    &+\omega_{202-22}(\abs{\boldsymbol{R}} \boldsymbol{e}_z))+ \frac{4\pi}{3\sqrt{4\pi}}\sqrt{1/5}(\omega_{20222}+\omega_{202-22})),\cr
    k &= -4\Big(\frac{64\pi^4 \rho_0^2}{25\sqrt{2}} \INT{}{}{\abs{\boldsymbol{R}}}\frac{\abs{\boldsymbol{R}}^4}{2}(\frac{-2\sqrt{5\pi}}{15}\sqrt{2/7}(\omega_{20222}(\abs{\boldsymbol{R}} \boldsymbol{e}_z)\cr
    &+\omega_{202-22}(\abs{\boldsymbol{R}} \boldsymbol{e}_z))+ \frac{4\pi}{3\sqrt{4\pi}}\sqrt{1/5}(\omega_{20222}(\abs{\boldsymbol{R}} \boldsymbol{e}_z)\cr
    &+\omega_{202-22}(\abs{\boldsymbol{R}} \boldsymbol{e}_z)))-j\Big)
\end{align}
for \eqref{fullmodel}.
Combined with the ideal gas free energy \eqref{idealmicro}, we found the most general full biaxial model in a quasi-microscopic way
 by explicitly calculating all tensor invariants from the macroscopic model presented in Sec.~\ref{macro} for the case of constant directors and comparing terms.

\subsection{Model 2}\label{model2section}
The second microscopic model we will consider for the explicit 
derivation of dynamic equations is a full theory for $\psi_1$, $\langle\boldsymbol{M}_0\rangle_{ij}$ and their coupling. This means that we go beyond earlier PFC models only describing uniaxial phases (such as \cite{WittkowskiLB2010,EmmerichEtAl2012}) and introduce phase biaxiality by using the order parameter $P$. We however neglect the second tensor order parameter $\langle\boldsymbol{M}_1\rangle$ and work only with $\langle\boldsymbol{M}_0 \rangle$. Thus, we can arrive at this model by taking the results from Sec.~\ref{modelfullsection} and setting $\langle\boldsymbol{M}_1\rangle$ to $\boldsymbol{0}$.
This special case describes the 
important situation of uniaxial particles in the most general way, not fully presented in the PFC model literature so far. (As an alternate derivation for this model, on could also argue that the coefficients from \cite{WittkowskiLB2010,EmmerichEtAl2012} are, if formulated through the nematic tensor, still valid, as its symmetry does not change with the inclusion of the additional order parameter $P$. This is a bit similar to the discussion given below \eqref{coefficients}. In that case however, we would need to adapt our definition of the order parameters as they vary slightly in the chosen convention.)

We begin by writing 
\begin{align}
    \rho &= \rho_0 \left(\psi_{1}+\langle \boldsymbol{M}_0 \rangle: \boldsymbol{M}_0 \right)\cr
    &= \rho_{0} \left(\psi_{1}+\sum_{i,j} \boldsymbol{Q}_{ij}m_{3,i}m_{3,j}\right)
\end{align}
and consider the full three-dimensional nematic tensor for uniaxial particles 
\begin{equation}\label{tensor}
    \langle \boldsymbol{M}_0 \rangle = \sqrt{\frac{3}{2}}S\left(\boldsymbol{l}_3 \otimes \boldsymbol{l}_3 -\frac{1}{3}\boldsymbol{I}\right)+\frac{P}{\sqrt{2}} \left(\boldsymbol{l}_{1} \otimes \boldsymbol{l}_{1} -\boldsymbol{l}_{2} \otimes \boldsymbol{l}_{2}
\right) ,
\end{equation}
where by including the order parameter $P$ and the director $\boldsymbol{l}_2$ we also allow for the description of biaxial phases (see Sec. \ref{scalarops} for a discussion of the difference between biaxial phases and biaxial particles).
\subsubsection{Ideal gas free energy}
The ideal gas free energy is given by
\begin{align}\label{firstideal}
    \beta \mathcal{F}_{\mathbf{id}} &= 8\pi^2\rho_{0}\INT{}{}{\boldsymbol{R}} \Bigg(\frac{1}{2}\left(\psi_{1}^2+\frac{ S^2+P^2}{5}\right) \nonumber\\
    &-\frac{1}{6}\Big(\psi_{1}^3+\frac{3\psi_{1}}{5}(S^2+P^2) -\frac{2}{35}(3P^2S-S^3)\Big) \nonumber \cr
    &+\frac{1}{12}\Big(\psi_{1}^4+\frac{3(S^4+P^4)}{35}+\psi_{1}^2\frac{6(S^2+P^2)}{5}  \cr
    &-\frac{2}{35}\psi_{1}(12P^2S-4S^3)
    +\frac{6S^2 P^2}{35}\Big)\Bigg),
\end{align} 

 which equals \eqref{idealmicroM} when dropping the terms involving $\langle \boldsymbol{M}_1 \rangle$. This can be seen by inserting \eqref{tensor}. (The tensor invariants not involving any derivatives do not have any director dependence.)
 Alternatively, one can also use \eqref{idealmicro}and set $U=F=0$.
 This can be cross-checked by performing a Taylor expansion of the logarithm and integrating out orientational degrees of freedom  of \eqref{idealexpansion}, as discussed in Sec.\ref{general}.

\subsubsection{Excess free energy}
The excess free energy is then 
\begin{align}
    \mathcal{F}_{\mathbf{exc}} &=  \INT{}{}{\boldsymbol{R}} \Big(a_{1}\psi_{1}^2+a_{2}(\nabla \psi_{1})^2+a_{3} (\Laplace \psi_{1})^2\cr
    &+ a\Tr(\langle \boldsymbol{M}_0\rangle^2)+d(\nabla \psi_1):( \nabla \cdot \langle \boldsymbol{M}_0 \rangle) \cr
    &+ f(\nabla \cdot \langle \boldsymbol{M}_0 \rangle):(\nabla \cdot \langle \boldsymbol{M}_0 \rangle)\cr
    &+g (\nabla \langle \boldsymbol{M}_0 \rangle):( \nabla \langle \boldsymbol{M}_0 \rangle) \Big),
\end{align} 
where the parameters are given in \eqref{coefficients}. 
This could also be written explicitly in terms of $S$, $P$, and the three directors, as in \eqref{directormodel}.

This free energy does not only include the Frank elastic energy
\begin{align}
      \mathcal{F}_{\text{Frank}, \mathbf{l_3}} &= \frac{1}{2}\INT{}{}{\boldsymbol{R}} ( K_{1,S} (\nabla \cdot \boldsymbol{l}_{3})^2+K_{2,S} (\boldsymbol{l}_{3} \cdot (\nabla \times \boldsymbol{l}_{3}))^2\cr
    &+K_{3,S} (\boldsymbol{l}_{3} \times (\nabla \times \boldsymbol{l}_{3}))^2),
\end{align}
coming from contributions of the uniaxial director $\boldsymbol{l}_3$, but also gives expressions for the elastic energy for the directors $\boldsymbol{l}_{2}$ and $\boldsymbol{l}_{1}$. Taking, for instance, the director $\boldsymbol{l}_2$, the elastic energy would include, for example, the term 
\begin{align}
    \mathcal{F}_{\text{Frank}, \mathbf{l_2}} &= \frac{1}{2}\INT{}{}{\boldsymbol{R}} ( K_{1,P} (\nabla \cdot \boldsymbol{l}_{2})^2+K_{2,P} (\boldsymbol{l}_{2} \cdot (\nabla \times \boldsymbol{l}_{2}))^2\cr
    &+K_{3,P} (\boldsymbol{l}_{2} \times (\nabla \times \boldsymbol{l}_{2}))^2).
\end{align}  
The Frank coefficients are then given by
\begin{align}
    K_{1,S} &= \sqrt{\frac{3}{2}}S^2 \cdot \Tilde{K}_{1,S}  ,\cr
    K_{2,S} &= \sqrt{\frac{3}{2}}S^2 \cdot \Tilde{K}_{2,S},\cr
    K_{3,S} &= K_{1,S}  ,\cr
    K_{1,P} &= \sqrt{\frac{1}{2}}P^2 \cdot \Tilde{K}_{1,P},\cr
    K_{2,P} &= \sqrt{\frac{1}{2}}P^2 \cdot \Tilde{K}_{2,P},\cr
    K_{3,P} &= K_{1,P},
\end{align} 
where the unique relation between $K_i$ and the elastic coefficients $f$ and $g$  (also commonly denoted as $L_1$ and $L_2$ in the literature) in \eqref{fullmodel} is given by
\begin{align} 
    \Tilde{K}_{1,S} &= 2{f} + {g},\cr
    \Tilde{K}_{2,S} &= 2{f},\cr
    \Tilde{K}_{1,P} &= 2{h} + {i},\cr
    \Tilde{K}_{2,P} &= 2{h}
\end{align} 
The degeneracy of $K_{1,Y}=K_{3,Y}$ is an artifact of our expansion, which should be resolved in higher orders, as in the uniaxial case~\cite{phdrene}.
We see that there generally appear different kinds of Frank constants  in biaxial systems.
In the most general theory, involving both tensor order parameters, there will be  complex coupling between all four $S,U,P,F$ parameters and their directors as seen in Sec. \ref{macro}.

\section{Dynamical Field Theories \label{sec_dynamics}}

\subsection{General ingredients}
We also derive full dynamical equations for this model, as with Model 1 involving constant directors and varying order parameters $S,U,P,F$ the dynamics either of which has been fully derived in the literature yet.

Having expressed the free energy in terms of the orientational order parameters, we now derive the corresponding dynamic equations for Model 1 and Model 2.  
In both cases, our starting point is the DDFT for biaxial particles given by \cref{generalddft}. If we want to derive dynamical equations for the orientational order parameters, we will have to take an orientational average of \cref{generalddft}. For doing this, we need to know the form of the diffusion tensor $\boldsymbol{D}(\boldsymbol{O})$ appearing in this equation.

Since calculating the diffusion tensor is in general not analytically possible, we use as an approximation the form known for hard rods. This is reasonable as long as the particles' diffusion behavior is sufficiently similar to that of rods or spheres (the diffusion tensor for rods contains that of spheres as a limiting case), which should be the case if, for instance, the particles are long and thin cuboids (similar to rods) or if they are cubes (similar to spheres). If the particle shape is different, the qualitative form of the obtained field theory will still be correct as long as translation-diffusion-coupling can be neglected, but the values of the mobility coefficients may be less accurate. An explicit assumption about the form of $\boldsymbol{D}$ is required primarily to be able to analytically derive the dynamical equations for our order parameters in closed form.
The diffusion tensor of hard rods is given by \cite{Dhont1996}
\begin{align} \label{diffusionrods}
    \boldsymbol{D}_{\textbf{TR}} &= \boldsymbol{D}_{\textbf{RT}}=\boldsymbol{0},\cr
    \boldsymbol{D}_{\textbf{TT}} &= D_\parallel\boldsymbol{m}_3\otimes\boldsymbol{m}_3 +D_\perp(\boldsymbol{I} - \boldsymbol{m}_3\otimes\boldsymbol{m}_3),\cr 
    \boldsymbol{D}_{\textbf{RR}} &= D_\text{R} \boldsymbol{I}.
\end{align} 
Here, $\boldsymbol{0}$ is a tensor where all entries are zero, and $\boldsymbol{m}_3$  and could be represented by
\begin{equation}\label{nematicdirector}
    \boldsymbol{m}_3={R}\cdot (0,0,1)^T,
\end{equation}
which, using \cref{rotationmatrix}, is given by
\begin{equation}\label{nematicdirector2}
    \boldsymbol{m}_3=( \cos(\phi)\sin(\theta), \sin(\phi)\sin(\theta), \cos(\theta) ).
\end{equation}

The explicit formulas for $D_\parallel$ (parallel translational diffusion coefficient), $D_\perp$ (perpendicular translational diffusion coefficient) and $D_\text{R}$ (rotational diffusion coefficient) are (see Ref.\ \cite{Dhont1996})
\begin{align}\label{diffusionrods2}
    D_\parallel&=\frac{k_{\mathrm{B}}T}{\gamma_\parallel} ,\cr
    D_\perp&=\frac{k_{\mathrm{B}}T}{\gamma_\perp} ,\cr
    D_\text{R}&=\frac{k_{\mathrm{B}}T}{\gamma_\text{R}}  .
\end{align} 
Further, 
\begin{align}\label{diffusionrods3}
    \gamma_\text{R}&=\frac{\pi \eta L^3}{3}\ln\left(\frac{L}{D}\right) ,\cr
    \gamma_\parallel&=\frac{2\pi\eta L}{\ln\left(\frac{L}{D}\right)} ,\cr
    \gamma_\perp&=2\gamma_\parallel .
\end{align} 
In this case, $L$ and $D$ are the length and diameter of the rod, respectively, and $\eta$ denotes the viscosity of the surrounding solvent.

\subsection{Model 1: PFC for scalar order parameters}\label{dynamicderivation}

\subsubsection{Full formulation}

Having clarified the necessary prerequisites for the usage of DDFT, we can derive dynamical equations for  Model 1 in terms of our scalar order parameters.
We start by taking the first time derivative of the definition \eqref{DefinitionY} of a generic scalar order parameter $Y$ 
and insert \eqref{generalddft}: 
\begin{equation}\label{timeevolution}
    \partial_{t} Y =\frac{2l_Y+1}{8\pi^2 \rho_{0}} \INT{}{}{\boldsymbol{O}} f_{Y}(\boldsymbol{O}) \beta \nabla_{{\boldsymbol{\mathcal{R}}}} \cdot \left(D(\boldsymbol{O})\rho \nabla_{{\boldsymbol{\mathcal{R}}}} \frac{\delta \mathcal{F}}{\delta \rho}\right).
\end{equation}
Here, the free energy has the form given in \eqref{constant} added to \eqref{firstideal}. We assume $l=2$ for $Y \in S,U,P,F$ and $l=0$ for $Y=\psi_1$.
Next, we use the relation
\begin{equation}\label{chainrule}
    \frac{\delta \mathcal{F}}{\delta \rho}=\sum_{Y}\frac{2l_Y+1}{\rho_0 8\pi^2}f_{Y}\frac{\delta \mathcal{F}} {\delta Y}
\end{equation}
and insert
\begin{equation}\label{timeevoltiondensity}
    \rho=\rho_{0}\sum_{Y}f_{Y} Y
\end{equation}
with Y $\in {\psi_{1}, S,U,P,F}$
into \cref{timeevolution}. Moreover, we define the coefficients
\begin{align}\label{transrot}
    \Tilde{M}_{XYZ}^{(D_\textbf{TT})} &= \frac{(2l_Y+1)(2l_Z+1)}{64\pi^4 \rho_0}\INT{\mathrm{SO(3)}}{}{\boldsymbol{O}}D_{\textbf{TT}}(\boldsymbol{O}) f_{Y}(\boldsymbol{O})f_X(\boldsymbol{O})\cr
    & \times f_Z(\boldsymbol{O}),\cr
    \Tilde{M}_{XYZ}^{(D_{\textbf{TT}})} &= \frac{(2l_Y+1)(2l_Z+1)}{64\pi^4 \rho_0}\cr&\INT{\mathrm{SO(3)}}{}{\boldsymbol{O}}D_{\textbf{RR}}f_Y(\boldsymbol{O})\nabla_{\boldsymbol{O}}\cdot(f_X(\boldsymbol{O}) \nabla_{\boldsymbol{O}}f_Z(\boldsymbol{O})).
\end{align} 
with $X, Y, Z \in $ $ \{\psi_1,S,U,P,F\}$ and (once again) the convention  $f_{\psi_1} = 1$. The parameter $l_X$ is 0 for $X=\psi_1$ and 2 else. After evaluating the orientational integrals in \cref{timeevolution}, we arrive at the final result
\begin{equation}\label{FINALEQUATION}
    \partial_{t}Y=\beta \nabla \cdot \left(\sum_{X,Z}X\Tilde{M}^{(D_{\textbf{TT}})}_{YXZ}\nabla \frac{\delta \mathcal{F}}{\delta Z}\right) + \beta \sum_{X, Z}\Tilde{M}^{(D_{\textbf{RR}})}_{YXZ}X \frac{\delta \mathcal{F}}{\delta Z}.
\end{equation}
The coefficients can be easily calculated with the integrals presented in Appendix \ref{coefficientsmodel1}.
\subsubsection{Constant mobility approximation (CMA)}\label{CMA}
One important approximation used in many PFC models is the so-called "constant mobility approximation, where we assume the density-dependent mobility of \eqref{generalddft} to be constant, that is, we set $\beta D \rho \approx \beta D \rho_0$ to arrive at
\begin{equation}\label{constantmobilityDDFT}
    \partial_{t}\rho = \rho_{0}\beta \nabla_{\boldsymbol{\mathcal{R}}} \cdot  D(\boldsymbol{O})\cdot \nabla_{\boldsymbol{\mathcal{R}}} \frac{\delta \mathcal{F}}{\delta \rho}
\end{equation}
which we will now use to derive "simplified" PFC models using \eqref{constantmobilityDDFT} by inserting it into the first time derivative of \eqref{DefinitionY} to proceed with the derivation of our new simplified models. 

Using the constant mobility approximation, we arrive at 
\begin{equation}\label{FINALEQUATION2}
    \partial_{t}Y=\beta  \nabla \cdot \sum_{Z}M^{(D_\textbf{{TT}})}_{Y\psi_1 Z}\nabla \frac{\delta F}{\delta Z} + \beta \sum_{Z}M^{(D_\textbf{{RR}})}_{Y1Z} \frac{\delta F}{\delta Z}
\end{equation}
This model is drastically simpler than the previously presented full DDFT model given by \cref{FINALEQUATION}, since the summation over $X$ vanishes completely. This leads to a much smaller number of terms in \eqref{FINALEQUATION} and thus less coefficients to be calculated.

\subsection{Model 2: PFC for tensorial order parameters} 
Here, we derive equations for the  
dynamics of the uniaxial nematic tensor $\langle {M}_0\rangle_{ij}$ in Model 2.
As the number of required coefficients is quite large, we only consider the constant mobility approximation explained in Sec.~\ref{dynamicderivation}. Doing so, 
we arrive at the following dynamic equations, extending results obtained for the special case $P=0$ in \cite{yabunak}:
\begin{align} 
    \partial_{t}\psi_{1} &= \frac{1}{8\pi^2 \rho_0}\INT{}{}{\boldsymbol{\textbf{O}}} \Big(\beta  \nabla^2 \Big(\rho_0 D_{\textbf{TT}}(\boldsymbol{\textbf{O}})\frac{\delta \mathcal{F}}{\delta \rho}\Big)\cr
    &+D_{\textbf{RR}}\nabla_{\textbf{O}}^2 \rho_0\frac{\delta \mathcal{F}}{\delta \rho}\Big),\cr
    \partial_{t} \langle M_0\rangle_{ij} &= \frac{5}{ 8\pi^2 \rho_0}\INT{}{}{\boldsymbol{\textbf{O}}}\sqrt{\frac{3}{2}}\left({m}_{3,i}{m}_{3,j}-\frac{1}{3}\right)\cr
    &\times \Big(\beta \nabla^2 \rho_0 \left(D_\textbf{TT}(\boldsymbol{\textbf{O}})\frac{\delta \mathcal{F}}{\delta \rho}\right) \cr
    &+D_{\textbf{RR}}\nabla_{\boldsymbol{\textbf{O}}}^2 \rho_0 \frac{\delta \mathcal{F}}{\delta \rho}\Big).
\end{align} 
By using the relation
\begin{equation}
    \frac{\delta \mathcal{F}}{\delta \rho} = \frac{1}{8\pi^2 \rho_0}\frac{\delta \mathcal{F}}{\delta \psi_{1}}+\frac{5}{8\pi^2 \rho_0}\sqrt{\frac{3}{2}}\left(m_{3,i}{m}_{3,j}-\frac{1}{3}\right)\frac{\delta \mathcal{F}}{\delta \langle {M}_0 \rangle_{ij}}
\end{equation}
we can use the microscopic definition in \eqref{DefinitionY} of the order parameters to arrive at
\begin{align} \label{psiMdyn}
    \partial_{t}\psi_{1} &= \beta  \nabla \cdot \sum_{Z}\Tilde{M}^{(D_\textbf{TT})}_{\psi_1,Z}\nabla \frac{\delta F}{\delta Z},\cr
    \partial_{t}\langle {M}_0\rangle_{ij} &= \beta  \nabla \cdot \sum_{Z}\Tilde{M}^{(D_\textbf{TT})}_{{\langle M_0\rangle_{ij}}, Z}\nabla \frac{\delta F}{\delta Z} + \beta \sum_{Z}\Tilde{M}^{(D_\textbf{RR})}_{\langle\boldsymbol{M}_0\rangle_{ij}, Z} \frac{\delta F}{\delta Z},\cr
    Z &\in \{\psi_{1}, \langle{M}_0\rangle_{ij}\},
\end{align} 
where the coefficients are defined as
\begin{align}\label{transrot2}
    \Tilde{M}_{X, Z}^{(D_\textbf{TT})} &= \frac{(2l_{X}+1)(2l_{Z}+1)}{64\pi^4\rho_{0}}\INT{\mathrm{SO(3)}}{}{\boldsymbol{O}} f_{X}(\boldsymbol{O})f_{Z}(\boldsymbol{O})D_{\textbf{TT}}(\boldsymbol{O}),\cr
    \Tilde{M}_{X, Z}^{(D_\textbf{RR})} &= \frac{(2l_{X}+1)(2l_{Z}+1)}{64\pi^4\rho_{0}}\INT{\mathrm{SO(3)}}{}{\boldsymbol{O}} f_{X}(\boldsymbol{O})\nabla^2_{\boldsymbol{O}}f_{Z}(\boldsymbol{O}),\cr
    f_{\langle\boldsymbol{M}_0\rangle_{ij}} &= \sqrt{\frac{3}{2}}(m_{3,i}{m}_{3,j}-\frac{1}{3}),\cr
    f_{\psi_1} &= 1.
\end{align} 
The integrals necessary for the coefficients are given in detail in Appendix \ref{coefficientsmodel2}. (One might wonder why we chose to integrate over SO(3), instead of $\boldsymbol{S}_2$, as only two angles are relevant. This is done simply to ensure a unified notation -- since the distributions do not depend on $\chi$ here, the additional integral over $\chi$ simply contributes an extra factor $2\pi$, which is canceled by our normalization, as explained in the comment below \eqref{reference}.)

Note that the form of \eqref{psiMdyn} implies
that only $\psi_{1}$ is conserved (due to mass conservation), but not $\langle{M}_0 \rangle_{ij}$. 
This is because the coefficients $\Tilde{M}_{X,Z}^{(D_\textbf{RR})}$ 
vanish only for $X=\psi_{1}$. In this case,  coefficients of this type do not appear, as they are proportional to the angular derivative of a constant function, which is obviously 0.
This is, however, not the case for all other order parameters, whose microscopic definition is weighted with some angular function. For the scalar order parameters $S,U,P,F$, this would be the corresponding Mulder matrix, for a tensor order parameter element $\langle {M}_0\rangle_{ij}$, this would be $\sqrt{\frac{3}{2}}({m}_{3,i}{m}_{3,j}-\frac{1}{3})$.

\section{Conclusion}
Starting from the general form of a DDFT for biaxial particles \cite{WittkowskiL2011}, we have derived PFC models for the coupled dynamics of the orientational order parameters $S$, $U$, $P$, and $F$ relevant for biaxial particles \cite{rosso2007orientational,el2024biaxial} (model 1) and for the dynamics of the full nematic tensor of uniaxial particles that can form biaxial phases (model 2). We moreover demonstrated that the derivation gives the correct coefficients in the free energy also for general biaxial liquid crystals. The models derived here provide a general description of spatially inhomogeneous orientational ordering dynamics in non-polar systems and allow to simulate it more efficiently than previously existing theories, such as DDFT. 
It is also more general than previous PFC models for uniaxial liquid crystals~\cite{EmmerichEtAl2012}.
Despite this, it still remains a theory based on first principles, unlike macroscopic theories that have to get numerical values for the coefficients elsewhere, such as in experiments or simulations~\cite{Xu2020}.

The next step would be a numerical implementation and investigation of the models derived here. Moreover, one could extend the theory towards the active case to obtain a biaxial active PFC model. Finally, one could investigate particles with lower symmetry class. The method proposed in this work allows to do this with a significantly reduced derivation effort, allowing to perform previously unattempted derivations.

\appendix

\section{Coefficients for Model 1.}\label{coefficientsmodel1}
In this section, we list and compute all of the integrals needed for the coefficients $\Tilde{M}^{(D_\textbf{TT})}_{XYZ}$ and $\Tilde{M}^{(D_\textbf{RR})}_{XYZ}$ defined in \eqref{transrot} and the spherical harmonics integrals mentioned earlier.
\subsection{Taylor expansion}\label{taylor}
In Sec.\ref{excess} we derived the excess free energy for a system of biaxial liquid crystals with constant directors. In order to approximately calculate a convolution integral appearing in this  derivation, we need to perform a gradient expansion, which involves calculating several angular integrals. We now give all integrals necessary for the gradient expansion. 
For the second order of the gradient expansion we need the following integrals:
\begin{align}
\INT{\mathrm{\boldsymbol{S}_2}}{}{\boldsymbol{O}} \boldsymbol{m}_3\otimes \boldsymbol{m}_3 Y_{22}(\boldsymbol{O}) &= \left(\begin{array}{ccc} \sqrt{2\pi/15} &\ii\sqrt{2\pi/15}&0\cr
\ii\sqrt{2\pi/15}&-\sqrt{2\pi/15}&0\\
0&0&0
\end{array} \right),\cr
\INT{\mathrm{\boldsymbol{S}_2}}{}{\boldsymbol{O}} \boldsymbol{m}_3 \otimes \boldsymbol{m}_3 Y_{20}(\boldsymbol{O}) &= \boldsymbol{\mathbf{diag}}\left(-2\frac{\sqrt{5\pi}}{15}, -2\frac{\sqrt{5\pi}}{15},4\frac{\sqrt{5\pi}}{15}\right),\cr
\INT{\mathrm{\boldsymbol{S}_2}}{}{\boldsymbol{O}} \boldsymbol{m}_3 \otimes \boldsymbol{m}_3 Y_{2-2}(\boldsymbol{O}) &=  \left(\begin{array}{ccc}\sqrt{2\pi/15}&-\ii\sqrt{2\pi/15}&0\cr
-\ii\sqrt{2\pi/15}&-\sqrt{2\pi/15}&0\cr
0&0&0 \end{array}\right),\cr
\INT{\mathrm{\boldsymbol{S}_2}}{}{\boldsymbol{O}} \boldsymbol{m}_3 \otimes \boldsymbol{m}_3 Y_{00}(\boldsymbol{O}) &= \boldsymbol{\mathbf{diag}}\left(\frac{4\pi}{3\sqrt{4\pi}},\frac{4\pi}{3\sqrt{4\pi}},\frac{4\pi}{3\sqrt{4\pi}}\right).
\end{align} 
For zeroth order of the gradient expansion we need the following integral:
\begin{equation}
    \INT{\mathrm{\boldsymbol{S}_2}}{}{\boldsymbol{O}} Y_{lm}(\boldsymbol{O})= \delta_{l0}\delta_{m0}\sqrt{4\pi}.
\end{equation}
Here, $\ii$ denotes the imaginary unit. 
\subsection{Translational diffusion}
When deriving dynamic equations for the order parameters, the coefficients of the resulting terms can be calculated analytically, which involves a number of angular integrals. This involves two distinct parts: translational and rotational diffusion, which are calculated in two different manners, as explained in \eqref{transrot}.
Now, we give all integrals necessary for the translational diffusion coefficients:
\begin{widetext}
\begin{align*}\label{integralcoefficients1}
    &\INT{\mathrm{SO(3)}}{}{\boldsymbol{O}}\boldsymbol{m}_3\otimes\boldsymbol{m}_3f_S(\boldsymbol{O})f_S(\boldsymbol{O})f_S(\boldsymbol{O}) = \frac{8\pi^2}{105}\boldsymbol{\textbf{diag}}\left(-1, -1, 8\right),\\
    &\INT{\mathrm{SO(3)}}{}{\boldsymbol{O}}\boldsymbol{m}_3\otimes\boldsymbol{m}_3f_S(\boldsymbol{O}) f_U(\boldsymbol{O})f_U(\boldsymbol{O}) = \boldsymbol{\textbf{diag}}\left(\frac{-8\pi^2}{35},\frac{-8\pi^2}{35},0\right),\\
    &\INT{\mathrm{SO(3)}}{}{\boldsymbol{O}} \boldsymbol{m}_3 \otimes \boldsymbol{m}_3 \cdot f_S(\boldsymbol{O}) f_U(\boldsymbol{O}) f_P(\boldsymbol{O}) = \boldsymbol{0},\\
    &\INT{\mathrm{SO(3)}}{}{\boldsymbol{O}} \boldsymbol{m}_3 \otimes \boldsymbol{m}_3 \cdot f_S(\boldsymbol{O}) f_U(\boldsymbol{O}) f_F(\boldsymbol{O}) = \frac{4\sqrt{3}\pi^2}{105}\boldsymbol{\textbf{diag}}\left(-1,1,0\right),\\
    &\INT{\mathrm{SO(3)}}{}{\boldsymbol{O}} \boldsymbol{m}_3 \otimes \boldsymbol{m}_3\cdot f_S(\boldsymbol{O}) f_S(\boldsymbol{O}) f_U(\boldsymbol{O}) = \boldsymbol{0},\\
    &\INT{\mathrm{SO(3)}}{}{\boldsymbol{O}} \boldsymbol{m}_3 \otimes \boldsymbol{m}_3 f_S(\boldsymbol{O}) f_S(\boldsymbol{O}) f_P(\boldsymbol{O}) = \frac{8\sqrt{3}\pi^2}{105}\boldsymbol{\textbf{diag}}\left(1,-1,0\right),\\
    &\INT{\mathrm{SO(3)}}{}{\boldsymbol{O}} \boldsymbol{m}_3 \otimes \boldsymbol{m}_3\cdot f_S(\boldsymbol{O}) f_P(\boldsymbol{O})f_P(\boldsymbol{O}) = \boldsymbol{\textbf{diag}}\left(\frac{-8\pi^2}{35},\frac{-8\pi^2}{35},0\right),\\
    &\INT{\mathrm{SO(3)}}{}{\boldsymbol{O}} \boldsymbol{m}_3 \otimes \boldsymbol{m}_3 \cdot f_S(\boldsymbol{O}) f_P(\boldsymbol{O}) f_F(\boldsymbol{O}) = \boldsymbol{0},\\
    &\INT{\mathrm{SO(3)}}{}{\boldsymbol{O}} \boldsymbol{m}_3 \otimes \boldsymbol{m}_3\cdot f_S(\boldsymbol{O}) f_S(\boldsymbol{O}) f_F(\boldsymbol{O}) = \boldsymbol{0},\\
    &\INT{\mathrm{SO(3)}}{}{\boldsymbol{O}}\boldsymbol{m}_3 \otimes \boldsymbol{m}_3\cdot f_S(\boldsymbol{O}) f_F(\boldsymbol{O})f_F(\boldsymbol{O}) = \boldsymbol{\textbf{diag}}\left(0, 0,\frac{16\pi^2}{35}\right),\\
    &\INT{\mathrm{SO(3)}}{}{\boldsymbol{O}} \boldsymbol{m}_3 \otimes \boldsymbol{m}_3 \cdot f_S(\boldsymbol{O}) f_{\psi_1}(\boldsymbol{O}) f_U(\boldsymbol{O}) =\boldsymbol{0},\\
    &\INT{\mathrm{SO(3)}}{}{\boldsymbol{O}} \boldsymbol{m}_3 \otimes \boldsymbol{m}_3 \cdot f_S(\boldsymbol{O}) f_{\psi_1}(\boldsymbol{O}) f_F(\boldsymbol{O}) =\boldsymbol{0},\\
    &\INT{\mathrm{SO(3)}}{}{\boldsymbol{O}} \boldsymbol{m}_3 \otimes \boldsymbol{m}_3 \cdot f_S(\boldsymbol{O}) f_{\psi_1}(\boldsymbol{O}) f_S(\boldsymbol{O}) =\boldsymbol{\textbf{diag}}\left(\frac{8\pi^2}{21},\frac{8\pi^2}{21},\frac{88\pi^2}{105}\right),\\
    &\INT{\mathrm{SO(3)}}{}{\boldsymbol{O}} \boldsymbol{m}_3 \otimes \boldsymbol{m}_3\cdot f_S(\boldsymbol{O}) f_{\psi_1}(\boldsymbol{O}) f_P(\boldsymbol{O}) =\frac{16\sqrt{3}\pi^2}{105}\boldsymbol{\textbf{diag}}\left(-1, 1,0\right),\\
    &\INT{\mathrm{SO(3)}}{}{\boldsymbol{O}} \boldsymbol{m}_3 \otimes \boldsymbol{m}_3 \cdot f_{S}(\boldsymbol{O})f_{\psi_1}(\boldsymbol{O})f_{\psi_1}(\boldsymbol{O})=\frac{8\pi^2}{15}\boldsymbol{\textbf{diag}}\left(-1,-1,2\right),\\
    &\INT{\mathrm{SO(3)}}{}{\boldsymbol{O}} \boldsymbol{m}_3 \otimes \boldsymbol{m}_3 \cdot f_P(\boldsymbol{O})f_P(\boldsymbol{O})f_P(\boldsymbol{O}) = \frac{8\sqrt{3}\pi^2}{35} \boldsymbol{\textbf{diag}}\left(1, -1,0\right),\\
    &\INT{\mathrm{SO(3)}}{}{\boldsymbol{O}} \boldsymbol{m}_3 \otimes \boldsymbol{m}_3 \cdot f_P(\boldsymbol{O})f_P(\boldsymbol{O}) f_{\psi_1}(\boldsymbol{O}) = \boldsymbol{\textbf{diag}}\left(\frac{24\pi^2}{35},\frac{24\pi^2}{35},\frac{8\pi^2}{35}\right),\\
    &\INT{\mathrm{SO(3)}}{}{\boldsymbol{O}} \boldsymbol{m}_3 \otimes \boldsymbol{m}_3\cdot f_P(\boldsymbol{O}) f_{\psi_1}(\boldsymbol{O})f_{\psi_1}(\boldsymbol{O}) = \boldsymbol{\textbf{diag}}\left(\frac{8\sqrt{3}\pi^2}{15},-\frac{8\sqrt{3}\pi^2}{15},0\right),\\
    &\INT{\mathrm{SO(3)}}{}{\boldsymbol{O}} \boldsymbol{m}_3 \otimes \boldsymbol{m}_3 \cdot f_P(\boldsymbol{O}) f_{\psi_1}(\boldsymbol{O}) f_F(\boldsymbol{O}) = \boldsymbol{0},\\
    &\INT{\mathrm{SO(3)}}{}{\boldsymbol{O}} \boldsymbol{m}_3 \otimes \boldsymbol{m}_3  \cdot f_P(\boldsymbol{O}) f_{\psi_1}(\boldsymbol{O}) f_U(\boldsymbol{O}) = \boldsymbol{0},\\
    &\INT{\mathrm{SO(3)}}{}{\boldsymbol{O}} \boldsymbol{m}_3 \otimes \boldsymbol{m}_3 f_P(\boldsymbol{O}) f_F(\boldsymbol{O}) f_U(\boldsymbol{O}) = \boldsymbol{\textbf{diag}}\left(\frac{4\pi^2}{21},\frac{4\pi^2}{21},\frac{8\pi^2}{105}\right),\\
    &\INT{\mathrm{SO(3)}}{}{\boldsymbol{O}} \boldsymbol{m}_3 \otimes \boldsymbol{m}_3 \cdot f_P(\boldsymbol{O}) f_P(\boldsymbol{O}) f_U(\boldsymbol{O}) = \boldsymbol{0},\\
    &\INT{\mathrm{SO(3)}}{}{\boldsymbol{O}} \boldsymbol{m}_3 \otimes \boldsymbol{m}_3\cdot f_P(\boldsymbol{O}) f_P(\boldsymbol{O}) f_F(\boldsymbol{O}) = \boldsymbol{0}\\
    &\INT{\mathrm{SO(3)}}{}{\boldsymbol{O}} \boldsymbol{m}_3 \otimes \boldsymbol{m}_3 \cdot f_U(\boldsymbol{O})f_U(\boldsymbol{O})f_U(\boldsymbol{O}) = \boldsymbol{0},\\
    &\INT{\mathrm{SO(3)}}{}{\boldsymbol{O}} \boldsymbol{m}_3 \otimes \boldsymbol{m}_3 \cdot f_U(\boldsymbol{O})
    f_U(\boldsymbol{O}) f_{\psi_1}(\boldsymbol{O}) = \boldsymbol{\textbf{diag}}\left(\frac{24\pi^2}{35},\frac{24\pi^2}{35},\frac{8\pi^2}{35}\right),\\
    &\INT{\mathrm{SO(3)}}{}{\boldsymbol{O}} \boldsymbol{m}_3 \otimes \boldsymbol{m}_3 \cdot f_U(\boldsymbol{O}) f_{\psi_1}(\boldsymbol{O})f_{\psi_1}(\boldsymbol{O}) = \boldsymbol{0},\\
    &\INT{\mathrm{SO(3)}}{}{\boldsymbol{O}} \boldsymbol{m}_3 \otimes \boldsymbol{m}_3 \cdot f_U(\boldsymbol{O}) f_{\psi_1}(\boldsymbol{O}) f_F(\boldsymbol{O}) = \frac{16\sqrt{3}\pi^2}{105}\boldsymbol{\textbf{diag}}\left(1,-1,0\right),\\
    &\INT{\mathrm{SO(3)}}{}{\boldsymbol{O}} \boldsymbol{m}_3 \otimes \boldsymbol{m}_3 \cdot f_U(\boldsymbol{O}) f_U(\boldsymbol{O}) f_F(\boldsymbol{O})= \boldsymbol{0},\\
    &\INT{\mathrm{SO(3)}}{}{\boldsymbol{O}} \boldsymbol{m}_3 \otimes \boldsymbol{m}_3 \cdot f_U(\boldsymbol{O}) f_F(\boldsymbol{O}) f_F(\boldsymbol{O}) = \boldsymbol{0},\\
    &\INT{\mathrm{SO(3)}}{}{\boldsymbol{O}} \boldsymbol{m}_3 \otimes \boldsymbol{m}_3 \cdot f_F(\boldsymbol{O})f_F(\boldsymbol{O})f_F(\boldsymbol{O}) = \boldsymbol{0},\\
    &\INT{\mathrm{SO(3)}}{}{\boldsymbol{O}} \boldsymbol{m}_3 \otimes \boldsymbol{m}_3 \cdot f_F(\boldsymbol{O}) f_F(\boldsymbol{O}) f_{\psi_1}(\boldsymbol{O}) = \boldsymbol{\textbf{diag}}\left(\frac{8\pi^2}{21},\frac{8\pi^2}{21},\frac{88\pi^2}{105}\right),\\
    &\INT{\mathrm{SO(3)}}{}{\boldsymbol{O}} \boldsymbol{m}_3 \otimes \boldsymbol{m}_3 \cdot f_F(\boldsymbol{O}) f_{\psi_1}(\boldsymbol{O})f_{\psi_1}(\boldsymbol{O}) = \boldsymbol{0},\\
    &\INT{\mathrm{SO(3)}}{}{\boldsymbol{O}} \boldsymbol{m}_3 \otimes \boldsymbol{m}_3 \cdot f_{\psi_1}(\boldsymbol{O}) f_{\psi_1}(\boldsymbol{O}) f_{\psi_1}(\boldsymbol{O}) = \boldsymbol{\textbf{diag}}\left(\frac{4\pi^2}{3},\frac{4\pi^2}{3},\frac{4\pi^2}{3}\right).
\end{align*}
\end{widetext}
All other integrals for translational diffusion follow from symmetry, e.g.\ $\INT{}{}{\boldsymbol{O}} \boldsymbol{m}_3 \otimes \boldsymbol{m}_3\cdot f_{X}(O)f_{Y}(O)f_{Z}(O)$ is equal to  $\INT{}{}{\boldsymbol{O}} \boldsymbol{m}_3 \otimes \boldsymbol{m}_3\cdot f_{Y}(O)f_{X}(O)f_{Z}(O)$.

\subsection{Rotational diffusion}
The aforementioned symmetry does not apply to rotational diffusion due to the different structure of the integrals involving angular derivatives. We now give all integrals necessary for evaluating the coefficients in Eq.\ \eqref{transrot}.\\
\begin{align*}
    &\INT{\mathrm{SO(3)}}{}{\boldsymbol{O}}f_{\psi_1}(\boldsymbol{O})\nabla_{\boldsymbol{O}}\cdot\left(f_{\psi_1}(\boldsymbol{O})\nabla_{\boldsymbol{O}}f_{S}(\boldsymbol{O})\right)= 8\pi^2 ,\\
    &\INT{\mathrm{SO(3)}}{}{\boldsymbol{O}}f_{S}(\boldsymbol{O})\nabla_{\boldsymbol{O}}\cdot\left(f_{\psi_1}(\boldsymbol{O})\nabla_{\boldsymbol{O}}f_{S}(\boldsymbol{O})\right) = -\frac{3}{2} \pi^2 ,\\
    &\INT{\mathrm{SO(3)}}{}{\boldsymbol{O}}f_{S}(\boldsymbol{O})\nabla_{\boldsymbol{O}}\cdot\left(f_{\psi_1}(\boldsymbol{O})\nabla_{\boldsymbol{O}}f_{U}(\boldsymbol{O})\right) = 0,\\
    &\INT{\mathrm{SO(3)}}{}{\boldsymbol{O}}f_{S}(\boldsymbol{O})\nabla_{\boldsymbol{O}}\cdot\left(f_{\psi_1}(\boldsymbol{O})\nabla_{\boldsymbol{O}}f_{P}(\boldsymbol{O})\right) = 0,\\
    &\INT{\mathrm{SO(3)}}{}{\boldsymbol{O}}f_{S}(\boldsymbol{O})\nabla_{\boldsymbol{O}}\cdot\left(f_{\psi_1}(\boldsymbol{O})\nabla_{\boldsymbol{O}}f_{F}(\boldsymbol{O})\right) = 0,\\
    &\INT{\mathrm{SO(3)}}{}{\boldsymbol{O}}f_{S}(\boldsymbol{O})\nabla_{\boldsymbol{O}}\cdot\left(f_{S}(\boldsymbol{O})\nabla_{\boldsymbol{O}}f_{\psi_1}(\boldsymbol{O})\right) = \frac{16}{5}\pi^2,\\
    &\INT{\mathrm{SO(3)}}{}{\boldsymbol{O}}f_{S}(\boldsymbol{O})\nabla_{\boldsymbol{O}}\cdot\left(f_{S} (\boldsymbol{O})\nabla_{\boldsymbol{O}}f_{S}(\boldsymbol{O})\right) = \frac{64}{35}\pi^2,\\
    &\INT{\mathrm{SO(3)}}{}{\boldsymbol{O}}f_{S}(\boldsymbol{O})\nabla_{\boldsymbol{O}}\cdot\left(f_{S}(\boldsymbol{O})\nabla_{\boldsymbol{O}}f_{U}(\boldsymbol{O})\right) = 0,\\
    &\INT{\mathrm{SO(3)}}{}{\boldsymbol{O}}f_{S}(\boldsymbol{O})\nabla_{\boldsymbol{O}}\cdot\left(f_{S}(\boldsymbol{O})\nabla_{\boldsymbol{O}}f_{P}(\boldsymbol{O})\right) = 0,\\
    &\INT{\mathrm{SO(3)}}{}{\boldsymbol{O}}f_{S}(\boldsymbol{O})\nabla_{\boldsymbol{O}}\cdot\left(f_{S}(\boldsymbol{O})\nabla_{\boldsymbol{O}}f_{F}(\boldsymbol{O})\right) = 0,\\
    &\INT{\mathrm{SO(3)}}{}{\boldsymbol{O}}f_{S}(\boldsymbol{O})\nabla_{\boldsymbol{O}}\cdot\left(f_{U}(\boldsymbol{O})\nabla_{\boldsymbol{O}}f_{\psi_1}(\boldsymbol{O})\right) = 0,\\
    &\INT{\mathrm{SO(3)}}{}{\boldsymbol{O}}f_{S}(\boldsymbol{O})\nabla_{\boldsymbol{O}}\cdot\left(f_{U}(\boldsymbol{O})\nabla_{\boldsymbol{O}}f_{S}(\boldsymbol{O})\right) = 0,\\
    &\INT{\mathrm{SO(3)}}{}{\boldsymbol{O}}f_{S}(\boldsymbol{O})\nabla_{\boldsymbol{O}}\cdot\left(f_{U}(\boldsymbol{O})\nabla_{\boldsymbol{O}}f_{U}(\boldsymbol{O})\right) = \frac{48}{35}\pi^2,\\
    &\INT{\mathrm{SO(3)}}{}{\boldsymbol{O}}f_{S}(\boldsymbol{O})\nabla_{\boldsymbol{O}}\cdot\left(f_{U}(\boldsymbol{O})\nabla_{\boldsymbol{O}}f_{P}(\boldsymbol{O})\right)= 0,\\
    &\INT{\mathrm{SO(3)}}{}{\boldsymbol{O}}f_{S}(\boldsymbol{O})\nabla_{\boldsymbol{O}}\cdot\left(f_{U}(\boldsymbol{O})\nabla_{\boldsymbol{O}}f_{F}(\boldsymbol{O})\right) = 0,\\
    &\INT{\mathrm{SO(3)}}{}{\boldsymbol{O}}f_{S}(\boldsymbol{O})\nabla_{\boldsymbol{O}}\cdot\left(f_{P}(\boldsymbol{O})\nabla_{\boldsymbol{O}}f_{\psi_1}(\boldsymbol{O})\right) = 0,\\
    &\INT{\mathrm{SO(3)}}{}{\boldsymbol{O}}f_{S}(\boldsymbol{O})\nabla_{\boldsymbol{O}}\cdot\left(f_{P}(\boldsymbol{O})\nabla_{\boldsymbol{O}}f_{S}(\boldsymbol{O})\right) = 0,\\
    &\INT{\mathrm{SO(3)}}{}{\boldsymbol{O}}f_{S}(\boldsymbol{O})\nabla_{\boldsymbol{O}}\cdot\left(f_{P}(\boldsymbol{O})\nabla_{\boldsymbol{O}}f_{U}(\boldsymbol{O})\right) = 0,\\
    &\INT{\mathrm{SO(3)}}{}{\boldsymbol{O}}f_{S}(\boldsymbol{O})\nabla_{\boldsymbol{O}}\cdot\left(f_{P}(\boldsymbol{O})\nabla_{\boldsymbol{O}}f_{P}(\boldsymbol{O})\right) = \frac{48}{35}\pi^2,\\
    &\INT{\mathrm{SO(3)}}{}{\boldsymbol{O}}f_{S}(\boldsymbol{O})\nabla_{\boldsymbol{O}}\cdot\left(f_{P}(\boldsymbol{O})\nabla_{\boldsymbol{O}}f_{F}(\boldsymbol{O})\right) = 0,\\
    &\INT{\mathrm{SO(3)}}{}{\boldsymbol{O}}f_{S}(\boldsymbol{O})\nabla_{\boldsymbol{O}}\cdot\left(f_{F}(\boldsymbol{O})\nabla_{\boldsymbol{O}}f_{\psi_1}(\boldsymbol{O})\right) = 0,\\
    &\INT{\mathrm{SO(3)}}{}{\boldsymbol{O}}f_{S}(\boldsymbol{O})\nabla_{\boldsymbol{O}}\cdot\left(f_{F}(\boldsymbol{O})\nabla_{\boldsymbol{O}}f_{S}(\boldsymbol{O})\right) = 0,\\
    &\INT{\mathrm{SO(3)}}{}{\boldsymbol{O}}f_{S}(\boldsymbol{O})\nabla_{\boldsymbol{O}}\cdot\left(f_{F}(\boldsymbol{O})\nabla_{\boldsymbol{O}}f_{U}(\boldsymbol{O})\right)= 0,\\
    &\INT{\mathrm{SO(3)}}{}{\boldsymbol{O}}f_{S}(\boldsymbol{O})\nabla_{\boldsymbol{O}}\cdot\left(f_{F}(\boldsymbol{O})\nabla_{\boldsymbol{O}}f_{P}(\boldsymbol{O})\right) = 0,\\
    &\INT{\mathrm{SO(3)}}{}{\boldsymbol{O}}f_{S}(\boldsymbol{O})\nabla_{\boldsymbol{O}}\cdot\left(f_{F}(\boldsymbol{O})\nabla_{\boldsymbol{O}}f_{F}(\boldsymbol{O})\right) = \frac{64}{35}\pi^2,\\
    &\INT{\mathrm{SO(3)}}{}{\boldsymbol{O}}f_{U}(\boldsymbol{O})\nabla_{\boldsymbol{O}}\cdot\left(f_{\psi_1}(\boldsymbol{O})\nabla_{\boldsymbol{O}}f_{\psi_1}(\boldsymbol{O})\right) = 0,\\
    &\INT{\mathrm{SO(3)}}{}{\boldsymbol{O}}f_{U}(\boldsymbol{O})\nabla_{\boldsymbol{O}}\cdot\left(f_{\psi_1}(\boldsymbol{O})\nabla_{\boldsymbol{O}}f_{S}(\boldsymbol{O})\right) = 0,\\
    &\INT{\mathrm{SO(3)}}{}{\boldsymbol{O}}f_{U}(\boldsymbol{O})\nabla_{\boldsymbol{O}}\cdot\left(f_{\psi_1}(\boldsymbol{O})\nabla_{\boldsymbol{O}}f_{U}(\boldsymbol{O})\right) = -\frac{28}{5} \pi^2,\\
    &\INT{\mathrm{SO(3)}}{}{\boldsymbol{O}}f_{U}(\boldsymbol{O})\nabla_{\boldsymbol{O}}\cdot\left(f_{\psi_1}(\boldsymbol{O})\nabla_{\boldsymbol{O}}f_{P}(\boldsymbol{O})\right) = 0,\\
    &\INT{\mathrm{SO(3)}}{}{\boldsymbol{O}}f_{U}(\boldsymbol{O})\nabla_{\boldsymbol{O}}\cdot\left(f_{\psi_1}(\boldsymbol{O})\nabla_{\boldsymbol{O}}f_{F}(\boldsymbol{O})\right) = 0,\\
    &\INT{\mathrm{SO(3)}}{}{\boldsymbol{O}}f_{U}(\boldsymbol{O})\nabla_{\boldsymbol{O}}\cdot\left(f_{S}(\boldsymbol{O})\nabla_{\boldsymbol{O}}f_{\psi_1}(\boldsymbol{O})\right) = 0,\\
    &\INT{\mathrm{SO(3)}}{}{\boldsymbol{O}}f_{U}(\boldsymbol{O})\nabla_{\boldsymbol{O}}\cdot\left(f_{S}(\boldsymbol{O})\nabla_{\boldsymbol{O}}f_{S}(\boldsymbol{O})\right) = 0,\\
    &\INT{\mathrm{SO(3)}}{}{\boldsymbol{O}}f_{U}(\boldsymbol{O})\nabla_{\boldsymbol{O}}\cdot\left(f_{S}(\boldsymbol{O})\nabla_{\boldsymbol{O}}f_{U}(\boldsymbol{O})\right) = \frac{4}{7}\pi^2,\\
    &\INT{\mathrm{SO(3)}}{}{\boldsymbol{O}}f_{U}(\boldsymbol{O})\nabla_{\boldsymbol{O}}\cdot\left(f_{S}(\boldsymbol{O})\nabla_{\boldsymbol{O}}f_{P}(\boldsymbol{O})\right) = 0,\\
    &\INT{\mathrm{SO(3)}}{}{\boldsymbol{O}}f_{U}(\boldsymbol{O})\nabla_{\boldsymbol{O}}\cdot\left(f_{S}(\boldsymbol{O})\nabla_{\boldsymbol{O}}f_{F}(\boldsymbol{O})\right) = 0,\\
    &\INT{\mathrm{SO(3)}}{}{\boldsymbol{O}}f_{U}(\boldsymbol{O})\nabla_{\boldsymbol{O}}\cdot\left(f_{U}(\boldsymbol{O})\nabla_{\boldsymbol{O}}f_{\psi_1}(\boldsymbol{O})\right) = -\frac{4}{5}\pi^2,\\
    &\INT{\mathrm{SO(3)}}{}{\boldsymbol{O}}f_{U}(\boldsymbol{O})\nabla_{\boldsymbol{O}}\cdot\left(f_{U}(\boldsymbol{O})\nabla_{\boldsymbol{O}}f_{S}(\boldsymbol{O})\right) = \frac{4}{7}\pi^2,\\
    &\INT{\mathrm{SO(3)}}{}{\boldsymbol{O}}f_{U}(\boldsymbol{O})\nabla_{\boldsymbol{O}}\cdot\left(f_{U}(\boldsymbol{O})\nabla_{\boldsymbol{O}}f_{U}(\boldsymbol{O})\right) = 0,\\
    &\INT{\mathrm{SO(3)}}{}{\boldsymbol{O}}f_{U}(\boldsymbol{O})\nabla_{\boldsymbol{O}}\cdot\left(f_{U}(\boldsymbol{O})\nabla_{\boldsymbol{O}}f_{P}(\boldsymbol{O})\right) = 0,\\
    &\INT{\mathrm{SO(3)}}{}{\boldsymbol{O}}f_{U}(\boldsymbol{O})\nabla_{\boldsymbol{O}}\cdot\left(f_{U}(\boldsymbol{O})\nabla_{\boldsymbol{O}}f_{P}(\boldsymbol{O})\right) = 0,\\
    &\INT{\mathrm{SO(3)}}{}{\boldsymbol{O}}f_{U}(\boldsymbol{O})\nabla_{\boldsymbol{O}}\cdot\left(f_{P}(\boldsymbol{O})\nabla_{\boldsymbol{O}}f_{\psi_1}(\boldsymbol{O})\right) = 0,\\
    &\INT{\mathrm{SO(3)}}{}{\boldsymbol{O}}f_{U}(\boldsymbol{O})\nabla_{\boldsymbol{O}}\cdot\left(f_{P}(\boldsymbol{O})\nabla_{\boldsymbol{O}}f_{S}(\boldsymbol{O})\right) = 0,\\
    &\INT{\mathrm{SO(3)}}{}{\boldsymbol{O}}f_{U}(\boldsymbol{O})\nabla_{\boldsymbol{O}}\cdot\left(f_{P}(\boldsymbol{O})\nabla_{\boldsymbol{O}}f_{U}(\boldsymbol{O})\right) = 0,\\
    &\INT{\mathrm{SO(3)}}{}{\boldsymbol{O}}f_{U}(\boldsymbol{O})\nabla_{\boldsymbol{O}}\cdot\left(f_{P}(\boldsymbol{O})\nabla_{\boldsymbol{O}}f_{P}(\boldsymbol{O})\right) = 0,\\\
    &\INT{\mathrm{SO(3)}}{}{\boldsymbol{O}}f_{U}(\boldsymbol{O})\nabla_{\boldsymbol{O}}\cdot\left(f_{P}(\boldsymbol{O})\nabla_{\boldsymbol{O}}f_{F}(\boldsymbol{O})\right) = -\frac{34}{35}\pi^2,\\
    &\INT{\mathrm{SO(3)}}{}{\boldsymbol{O}}f_{U}(\boldsymbol{O})\nabla_{\boldsymbol{O}}\cdot\left(f_{F}(\boldsymbol{O})\nabla_{\boldsymbol{O}}f_{\psi_1}(\boldsymbol{O})\right) = 0,\\
    &\INT{\mathrm{SO(3)}}{}{\boldsymbol{O}}f_{U}(\boldsymbol{O})\nabla_{\boldsymbol{O}}\cdot\left(f_{F}(\boldsymbol{O})\nabla_{\boldsymbol{O}}f_{S}(\boldsymbol{O})\right) = 0,\\
    &\INT{\mathrm{SO(3)}}{}{\boldsymbol{O}}f_{U}(\boldsymbol{O})\nabla_{\boldsymbol{O}}\cdot\left(f_{F}(\boldsymbol{O})\nabla_{\boldsymbol{O}}f_{U}(\boldsymbol{O})\right) = 0,\\
    &\INT{\mathrm{SO(3)}}{}{\boldsymbol{O}}f_{U}(\boldsymbol{O})\nabla_{\boldsymbol{O}}\cdot\left(f_{F}(\boldsymbol{O})\nabla_{\boldsymbol{O}}f_{P}(\boldsymbol{O})\right) = -\frac{34}{35}\pi^2,\\
    &\INT{\mathrm{SO(3)}}{}{\boldsymbol{O}}f_{U}(\boldsymbol{O})\nabla_{\boldsymbol{O}}\cdot\left(f_{F}(\boldsymbol{O})\nabla_{\boldsymbol{O}}f_{F}(\boldsymbol{O})\right) = 0,\\
    &\INT{\mathrm{SO(3)}}{}{\boldsymbol{O}}f_{P}(\boldsymbol{O})\nabla_{\boldsymbol{O}}\cdot\left(f_{\psi_1}(\boldsymbol{O})\nabla_{\boldsymbol{O}}f_{\psi_1}(\boldsymbol{O})\right) = 0,\\
    &\INT{\mathrm{SO(3)}}{}{\boldsymbol{O}}f_{P}(\boldsymbol{O})\nabla_{\boldsymbol{O}}\cdot\left(f_{\psi_1}(\boldsymbol{O})\nabla_{\boldsymbol{O}}f_{S}(\boldsymbol{O})\right) = 0 ,\\
    &\INT{\mathrm{SO(3)}}{}{\boldsymbol{O}}f_{P}(\boldsymbol{O})\nabla_{\boldsymbol{O}}\cdot\left(f_{\psi_1}(\boldsymbol{O})\nabla_{\boldsymbol{O}}f_{U}
    \boldsymbol{O})\right) = 0, \\
    &\INT{\mathrm{SO(3)}}{}{\boldsymbol{O}}f_{P}(\boldsymbol{O})\nabla_{\boldsymbol{O}}\cdot\left(f_{\psi_1}(\boldsymbol{O})\nabla_{\boldsymbol{O}}f_{P}
    \boldsymbol{O})\right) = -\frac{28}{5}\pi^2,\\
    &\INT{\mathrm{SO(3)}}{}{\boldsymbol{O}}f_{P}(\boldsymbol{O})\nabla_{\boldsymbol{O}}\cdot\left(f_{\psi_1}(\boldsymbol{O})\nabla_{\boldsymbol{O}}f_{F}(\boldsymbol{O})\right) = 0,\\
    &\INT{\mathrm{SO(3)}}{}{\boldsymbol{O}}f_{P}(\boldsymbol{O})\nabla_{\boldsymbol{O}}\cdot\left(f_{S}(\boldsymbol{O})\nabla_{\boldsymbol{O}}f_{\psi_1}(\boldsymbol{O})\right)  = 0,\\
    &\INT{\mathrm{SO(3)}}{}{\boldsymbol{O}}f_{P}(\boldsymbol{O})\nabla_{\boldsymbol{O}}\cdot\left(f_{S}(\boldsymbol{O})\nabla_{\boldsymbol{O}}f_{S}(\boldsymbol{O})\right)  = 0,\\
    &\INT{\mathrm{SO(3)}}{}{\boldsymbol{O}}f_{P}(\boldsymbol{O})\nabla_{\boldsymbol{O}}\cdot\left(f_{S}(\boldsymbol{O})\nabla_{\boldsymbol{O}}f_{U}(\boldsymbol{O})\right)  = 0,\\
    &\INT{\mathrm{SO(3)}}{}{\boldsymbol{O}}f_{P}(\boldsymbol{O})\nabla_{\boldsymbol{O}}\cdot\left(f_{S}(\boldsymbol{O})\nabla_{\boldsymbol{O}}f_{P}(\boldsymbol{O})\right)  = \frac{20}{35}\pi^2,\\
    &\INT{\mathrm{SO(3)}}{}{\boldsymbol{O}}f_{P}(\boldsymbol{O})\nabla_{\boldsymbol{O}}\cdot\left(f_{S}(\boldsymbol{O})\nabla_{\boldsymbol{O}}f_{F}(\boldsymbol{O})\right)  = 0,\\
    &\INT{\mathrm{SO(3)}}{}{\boldsymbol{O}}f_{P}(\boldsymbol{O})\nabla_{\boldsymbol{O}}\cdot\left(f_{U}(\boldsymbol{O})\nabla_{\boldsymbol{O}}f_{\psi_1}(\boldsymbol{O})\right)  = 0,\\
    &\INT{\mathrm{SO(3)}}{}{\boldsymbol{O}}f_{P}(\boldsymbol{O})\nabla_{\boldsymbol{O}}\cdot\left(f_{U}(\boldsymbol{O})\nabla_{\boldsymbol{O}}f_{S}(\boldsymbol{O})\right)  = 0,\\
    &\INT{\mathrm{SO(3)}}{}{\boldsymbol{O}}f_{P}(\boldsymbol{O})\nabla_{\boldsymbol{O}}\cdot\left(f_{U}(\boldsymbol{O})\nabla_{\boldsymbol{O}}f_{U}(\boldsymbol{O})\right)  = 0,\\
    &\INT{\mathrm{SO(3)}}{}{\boldsymbol{O}}f_{P}(\boldsymbol{O})\nabla_{\boldsymbol{O}}\cdot\left(f_{U}(\boldsymbol{O})\nabla_{\boldsymbol{O}}f_{P}(\boldsymbol{O})\right)  = 0,\\
    &\INT{\mathrm{SO(3)}}{}{\boldsymbol{O}}f_{P}(\boldsymbol{O})\nabla_{\boldsymbol{O}}\cdot\left(f_{U}(\boldsymbol{O})\nabla_{\boldsymbol{O}}f_{F}(\boldsymbol{O})\right)  = -\frac{34}{35}\pi^2 ,\\
    &\INT{\mathrm{SO(3)}}{}{\boldsymbol{O}}f_{P}(\boldsymbol{O})\nabla_{\boldsymbol{O}}\cdot\left(f_{P}(\boldsymbol{O})\nabla_{\boldsymbol{O}}f_{\psi_1}(\boldsymbol{O})\right)  =-\frac{28}{35}\pi^2,\\
    &\INT{\mathrm{SO(3)}}{}{\boldsymbol{O}}f_{P}(\boldsymbol{O})\nabla_{\boldsymbol{O}}\cdot\left(f_{P}(\boldsymbol{O})\nabla_{\boldsymbol{O}}f_{S}(\boldsymbol{O})\right)  = \frac{20}{35}\pi^2,\\
    &\INT{\mathrm{SO(3)}}{}{\boldsymbol{O}}f_{P}(\boldsymbol{O})\nabla_{\boldsymbol{O}}\cdot\left(f_{P}(\boldsymbol{O})\nabla_{\boldsymbol{O}}f_{U}(\boldsymbol{O})\right)  = 0,\\
    &\INT{\mathrm{SO(3)}}{}{\boldsymbol{O}}f_{P}(\boldsymbol{O})\nabla_{\boldsymbol{O}}\cdot\left(f_{P}(\boldsymbol{O})\nabla_{\boldsymbol{O}}f_{P}(\boldsymbol{O})\right)  = 0,\\
    &\INT{\mathrm{SO(3)}}{}{\boldsymbol{O}}f_{P}(\boldsymbol{O})\nabla_{\boldsymbol{O}}\cdot\left(f_{P}(\boldsymbol{O})\nabla_{\boldsymbol{O}}f_{F}(\boldsymbol{O})\right)  = 0,\\
    &\INT{\mathrm{SO(3)}}{}{\boldsymbol{O}}f_{P}(\boldsymbol{O})\nabla_{\boldsymbol{O}}\cdot\left(f_{F}(\boldsymbol{O})\nabla_{\boldsymbol{O}}f_{\psi_1}(\boldsymbol{O})\right)  = 0,\\
    &\INT{\mathrm{SO(3)}}{}{\boldsymbol{O}}f_{P}(\boldsymbol{O})\nabla_{\boldsymbol{O}}\cdot\left(f_{F}(\boldsymbol{O})\nabla_{\boldsymbol{O}}f_{S}(\boldsymbol{O})\right)  = 0,\\
    &\INT{\mathrm{SO(3)}}{}{\boldsymbol{O}}f_{P}(\boldsymbol{O})\nabla_{\boldsymbol{O}}\cdot\left(f_{F}(\boldsymbol{O})\nabla_{\boldsymbol{O}}f_{U}(\boldsymbol{O})\right)  = -\frac{34}{35}\pi^2,\\
    &\INT{\mathrm{SO(3)}}{}{\boldsymbol{O}}f_{P}(\boldsymbol{O})\nabla_{\boldsymbol{O}}\cdot\left(f_{F}(\boldsymbol{O})\nabla_{\boldsymbol{O}}f_{P}(\boldsymbol{O})\right)  = 0,\\
    &\INT{\mathrm{SO(3)}}{}{\boldsymbol{O}}f_{P}(\boldsymbol{O})\nabla_{\boldsymbol{O}}\cdot\left(f_{F}(\boldsymbol{O})\nabla_{\boldsymbol{O}}f_{F}(\boldsymbol{O})\right)  = 0,\\
    &\INT{\mathrm{SO(3)}}{}{\boldsymbol{O}}f_{F}(\boldsymbol{O})\nabla_{\boldsymbol{O}}\cdot\left(f_{\psi_1}(\boldsymbol{O})\nabla_{\boldsymbol{O}}f_{\psi_1}(\boldsymbol{O})\right)  = 0,\\
    &\INT{\mathrm{SO(3)}}{}{\boldsymbol{O}}f_{F}(\boldsymbol{O})\nabla_{\boldsymbol{O}}\cdot\left(f_{\psi_1}(\boldsymbol{O})\nabla_{\boldsymbol{O}}f_{S}(\boldsymbol{O})\right) = 0,\\
    &\INT{\mathrm{SO(3)}}{}{\boldsymbol{O}}f_{F}(\boldsymbol{O})\nabla_{\boldsymbol{O}}\cdot\left(f_{\psi_1}(\boldsymbol{O})\nabla_{\boldsymbol{O}}f_{U}(\boldsymbol{O})\right) = 0,\\
    &\INT{\mathrm{SO(3)}}{}{\boldsymbol{O}}f_{F}(\boldsymbol{O})\nabla_{\boldsymbol{O}}\cdot\left(f_{\psi_1}(\boldsymbol{O})\nabla_{\boldsymbol{O}}f_{P}(\boldsymbol{O})\right) = 0,\\
    &\INT{\mathrm{SO(3)}}{}{\boldsymbol{O}}f_{F}(\boldsymbol{O})\nabla_{\boldsymbol{O}}\cdot\left(f_{\psi_1}(\boldsymbol{O})\nabla_{\boldsymbol{O}}f_{F}(\boldsymbol{O})\right) = -\frac{126}{35}\pi^2,\\
    &\INT{\mathrm{SO(3)}}{}{\boldsymbol{O}}f_{F}(\boldsymbol{O})\nabla_{\boldsymbol{O}}\cdot\left(f_{S}(\boldsymbol{O})\nabla_{\boldsymbol{O}}f_{\psi_1}(\boldsymbol{O})\right) = 0,\\
    &\INT{\mathrm{SO(3)}}{}{\boldsymbol{O}}f_{F}(\boldsymbol{O})\nabla_{\boldsymbol{O}}\cdot\left(f_{S}(\boldsymbol{O})\nabla_{\boldsymbol{O}}f_{S}(\boldsymbol{O})\right) = 0,\\
    &\INT{\mathrm{SO(3)}}{}{\boldsymbol{O}}f_{F}(\boldsymbol{O})\nabla_{\boldsymbol{O}}\cdot\left(f_{S}(\boldsymbol{O})\nabla_{\boldsymbol{O}}f_{U}(\boldsymbol{O})\right) = 0,\\
    &\INT{\mathrm{SO(3)}}{}{\boldsymbol{O}}f_{F}(\boldsymbol{O})\nabla_{\boldsymbol{O}}\cdot\left(f_{S}(\boldsymbol{O})\nabla_{\boldsymbol{O}}f_{P}(\boldsymbol{O})\right) = 0,\\
    &\INT{\mathrm{SO(3)}}{}{\boldsymbol{O}}f_{F}(\boldsymbol{O})\nabla_{\boldsymbol{O}}\cdot\left(f_{S}(\boldsymbol{O})\nabla_{\boldsymbol{O}}f_{F}(\boldsymbol{O})\right) = -\frac{6}{35}\pi^2,\\
    &\INT{\mathrm{SO(3)}}{}{\boldsymbol{O}}f_{F}(\boldsymbol{O})\nabla_{\boldsymbol{O}}\cdot\left(f_{U}(\boldsymbol{O})\nabla_{\boldsymbol{O}}f_{\psi_1}(\boldsymbol{O})\right) = 0,\\
    &\INT{\mathrm{SO(3)}}{}{\boldsymbol{O}}f_{F}(\boldsymbol{O})\nabla_{\boldsymbol{O}}\cdot\left(f_{U}(\boldsymbol{O})\nabla_{\boldsymbol{O}}f_{S}(\boldsymbol{O})\right)  = 0,\\
    &\INT{\mathrm{SO(3)}}{}{\boldsymbol{O}}f_{F}(\boldsymbol{O})\nabla_{\boldsymbol{O}}\cdot\left(f_{U}(\boldsymbol{O})\nabla_{\boldsymbol{O}}f_{U}(\boldsymbol{O})\right)  = 0,\\
    &\INT{\mathrm{SO(3)}}{}{\boldsymbol{O}}f_{F}(\boldsymbol{O})\nabla_{\boldsymbol{O}}\cdot\left(f_{U}(\boldsymbol{O})\nabla_{\boldsymbol{O}}f_{P}(\boldsymbol{O})\right)  = -\frac{20}{35}\pi^2,\\
    &\INT{\mathrm{SO(3)}}{}{\boldsymbol{O}}f_{F}(\boldsymbol{O})\nabla_{\boldsymbol{O}}\cdot\left(f_{U}(\boldsymbol{O})\nabla_{\boldsymbol{O}}f_{F}(\boldsymbol{O})\right)  = 0,\\
&\INT{\mathrm{SO(3)}}{}{\boldsymbol{O}}f_{F}(\boldsymbol{O})\nabla_{\boldsymbol{O}}\cdot\left(f_{P}(\boldsymbol{O})\nabla_{\boldsymbol{O}}f_{\psi_1}(\boldsymbol{O})\right)  = 0,\\
&\INT{\mathrm{SO(3)}}{}{\boldsymbol{O}}f_{F}(\boldsymbol{O})\nabla_{\boldsymbol{O}}\cdot\left(f_{P}(\boldsymbol{O})\nabla_{\boldsymbol{O}}f_{S}(\boldsymbol{O})\right) = 0,\\
&\INT{\mathrm{SO(3)}}{}{\boldsymbol{O}}f_{F}(\boldsymbol{O})\nabla_{\boldsymbol{O}}\cdot\left(f_{P}(\boldsymbol{O})\nabla_{\boldsymbol{O}}f_{U}(\boldsymbol{O})\right) = -\frac{4}{7}\pi^2,\\
&\INT{\mathrm{SO(3)}}{}{\boldsymbol{O}}f_{F}(\boldsymbol{O})\nabla_{\boldsymbol{O}}\cdot\left(f_{P}(\boldsymbol{O})\nabla_{\boldsymbol{O}}f_{P}(\boldsymbol{O})\right) = 0,\\
&\INT{\mathrm{SO(3)}}{}{\boldsymbol{O}}f_{F}(\boldsymbol{O})\nabla_{\boldsymbol{O}}\cdot\left(f_{P}(\boldsymbol{O})\nabla_{\boldsymbol{O}}f_{F}(\boldsymbol{O})\right) = 0,\\
&\INT{\mathrm{SO(3)}}{}{\boldsymbol{O}}f_{F}(\boldsymbol{O})\nabla_{\boldsymbol{O}}\cdot\left(f_{F}(\boldsymbol{O})\nabla_{\boldsymbol{O}}f_{\psi_1}(\boldsymbol{O})\right) = \frac{42}{35}\pi^2,\\
&\INT{\mathrm{SO(3)}}{}{\boldsymbol{O}}f_{F}(\boldsymbol{O})\nabla_{\boldsymbol{O}}\cdot\left(f_{F}(\boldsymbol{O})\nabla_{\boldsymbol{O}}f_{S}(\boldsymbol{O})\right) = -\frac{6}{35}\pi^2,\\
&\INT{\mathrm{SO(3)}}{}{\boldsymbol{O}}f_{F}(\boldsymbol{O})\nabla_{\boldsymbol{O}}\cdot\left(f_{F}(\boldsymbol{O})\nabla_{\boldsymbol{O}}f_{U}(\boldsymbol{O})\right) = 0,\\
&\INT{\mathrm{SO(3)}}{}{\boldsymbol{O}}f_{F}(\boldsymbol{O})\nabla_{\boldsymbol{O}}\cdot\left(f_{F}(\boldsymbol{O})\nabla_{\boldsymbol{O}}f_{P}(\boldsymbol{O})\right) = 0,\\
&\INT{\mathrm{SO(3)}}{}{\boldsymbol{O}}f_{F}(\boldsymbol{O})\nabla_{\boldsymbol{O}}\cdot\left(f_{F}(\boldsymbol{O})\nabla_{\boldsymbol{O}}f_{F}(\boldsymbol{O})\right) = 0,\\
&\INT{\mathrm{SO(3)}}{}{\boldsymbol{O}}f_{\psi_1}(\boldsymbol{O})\nabla_{\boldsymbol{O}}\cdot\left(f_{\psi_1}(\boldsymbol{O})\nabla_{\boldsymbol{O}}f_{\psi_1}(\boldsymbol{O})\right) = 0,\\
&\INT{\mathrm{SO(3)}}{}{\boldsymbol{O}}f_{\psi_1}(\boldsymbol{O})\nabla_{\boldsymbol{O}}\cdot\left(f_{\psi_1}(\boldsymbol{O})\nabla_{\boldsymbol{O}}f_{S}(\boldsymbol{O})\right)  = 0,\\
&\INT{\mathrm{SO(3)}}{}{\boldsymbol{O}}f_{\psi_1}(\boldsymbol{O})\nabla_{\boldsymbol{O}}\cdot\left(f_{\psi_1}(\boldsymbol{O})\nabla_{\boldsymbol{O}}f_{U}(\boldsymbol{O})\right)  = 0 ,\\
&\INT{\mathrm{SO(3)}}{}{\boldsymbol{O}}f_{\psi_1}(\boldsymbol{O})\nabla_{\boldsymbol{O}}\cdot\left(f_{\psi_1}(\boldsymbol{O})\nabla_{\boldsymbol{O}}f_{P}(\boldsymbol{O})\right)  = 0,\\
&\INT{\mathrm{SO(3)}}{}{\boldsymbol{O}}f_{\psi_1}(\boldsymbol{O})\nabla_{\boldsymbol{O}}\cdot\left(f_{\psi_1}(\boldsymbol{O})\nabla_{\boldsymbol{O}}f_{F}(\boldsymbol{O})\right)  = 0,\\
&\INT{\mathrm{SO(3)}}{}{\boldsymbol{O}}f_{\psi_1}(\boldsymbol{O})\nabla_{\boldsymbol{O}}\cdot\left(f_{S}(\boldsymbol{O})\nabla_{\boldsymbol{O}}f_{\psi_1}(\boldsymbol{O})\right)  = 0,\\
&\INT{\mathrm{SO(3)}}{}{\boldsymbol{O}}f_{\psi_1}(\boldsymbol{O})\nabla_{\boldsymbol{O}}\cdot\left(f_{S}(\boldsymbol{O})\nabla_{\boldsymbol{O}}f_{S}(\boldsymbol{O})\right)  = 0,\\
&\INT{\mathrm{SO(3)}}{}{\boldsymbol{O}}f_{\psi_1}(\boldsymbol{O})\nabla_{\boldsymbol{O}}\cdot\left(f_{S}(\boldsymbol{O})\nabla_{\boldsymbol{O}}f_{U}(\boldsymbol{O})\right)  = 0,\\
&\INT{\mathrm{SO(3)}}{}{\boldsymbol{O}}f_{\psi_1}(\boldsymbol{O})\nabla_{\boldsymbol{O}}\cdot\left(f_{S}(\boldsymbol{O})\nabla_{\boldsymbol{O}}f_{U}(\boldsymbol{O})\right)  = 0,\\
&\INT{\mathrm{SO(3)}}{}{\boldsymbol{O}}f_{\psi_1}(\boldsymbol{O})\nabla_{\boldsymbol{O}}\cdot\left(f_{S}(\boldsymbol{O})\nabla_{\boldsymbol{O}}f_{F}(\boldsymbol{O})\right)  = 0,\\
&\INT{\mathrm{SO(3)}}{}{\boldsymbol{O}}f_{\psi_1}(\boldsymbol{O})\nabla_{\boldsymbol{O}}\cdot\left(f_{U}(\boldsymbol{O})\nabla_{\boldsymbol{O}}f_{\psi_1}(\boldsymbol{O})\right)  = 0,\\
&\INT{\mathrm{SO(3)}}{}{\boldsymbol{O}}f_{\psi_1}(\boldsymbol{O})\nabla_{\boldsymbol{O}}\cdot\left(f_{U}(\boldsymbol{O})\nabla_{\boldsymbol{O}}f_{\psi_1}(\boldsymbol{O})\right)  = 0,\\
&\INT{\mathrm{SO(3)}}{}{\boldsymbol{O}}f_{\psi_1}(\boldsymbol{O})\nabla_{\boldsymbol{O}}\cdot\left(f_{U}(\boldsymbol{O})\nabla_{\boldsymbol{O}}f_{U}(\boldsymbol{O})\right)  = 0,\\
&\INT{\mathrm{SO(3)}}{}{\boldsymbol{O}}f_{\psi_1}(\boldsymbol{O})\nabla_{\boldsymbol{O}}\cdot\left(f_{U}(\boldsymbol{O})\nabla_{\boldsymbol{O}}f_{P}(\boldsymbol{O})\right)  = 0,\\
&\INT{\mathrm{SO(3)}}{}{\boldsymbol{O}}f_{\psi_1}(\boldsymbol{O})\nabla_{\boldsymbol{O}}\cdot\left(f_{U}(\boldsymbol{O})\nabla_{\boldsymbol{O}}f_{F}(\boldsymbol{O})\right)  = 0,\\
&\INT{\mathrm{SO(3)}}{}{\boldsymbol{O}}f_{\psi_1}(\boldsymbol{O})\nabla_{\boldsymbol{O}}\cdot\left(f_{P}(\boldsymbol{O})\nabla_{\boldsymbol{O}}f_{\psi_1}(\boldsymbol{O})\right)  = 0,\\
&\INT{\mathrm{SO(3)}}{}{\boldsymbol{O}}f_{\psi_1}(\boldsymbol{O})\nabla_{\boldsymbol{O}}\cdot\left(f_{P}(\boldsymbol{O})\nabla_{\boldsymbol{O}}f_{S}(\boldsymbol{O})\right)  = 0,\\
&\INT{\mathrm{SO(3)}}{}{\boldsymbol{O}}f_{\psi_1}(\boldsymbol{O})\nabla_{\boldsymbol{O}}\cdot\left(f_{P}(\boldsymbol{O})\nabla_{\boldsymbol{O}}f_{U}(\boldsymbol{O})\right)  = 0,\\
&\INT{\mathrm{SO(3)}}{}{\boldsymbol{O}}f_{\psi_1}(\boldsymbol{O})\nabla_{\boldsymbol{O}}\cdot\left(f_{P}(\boldsymbol{O})\nabla_{\boldsymbol{O}}f_{P}(\boldsymbol{O})\right)  = 0,\\
&\INT{\mathrm{SO(3)}}{}{\boldsymbol{O}}f_{\psi_1}(\boldsymbol{O})\nabla_{\boldsymbol{O}}\cdot\left(f_{P}(\boldsymbol{O})\nabla_{\boldsymbol{O}}f_{F}(\boldsymbol{O})\right)  = 0,\\
&\INT{\mathrm{SO(3)}}{}{\boldsymbol{O}}f_{\psi_1}(\boldsymbol{O})\nabla_{\boldsymbol{O}}\cdot\left(f_{F}(\boldsymbol{O})\nabla_{\boldsymbol{O}}f_{\psi_1}(\boldsymbol{O})\right)  = 0,\\
&\INT{\mathrm{SO(3)}}{}{\boldsymbol{O}}f_{\psi_1}(\boldsymbol{O})\nabla_{\boldsymbol{O}}\cdot\left(f_{F}(\boldsymbol{O})\nabla_{\boldsymbol{O}}f_{S}(\boldsymbol{O})\right) = 0,\\
&\INT{\mathrm{SO(3)}}{}{\boldsymbol{O}}f_{\psi_1}(\boldsymbol{O})\nabla_{\boldsymbol{O}}\cdot\left(f_{F}(\boldsymbol{O})\nabla_{\boldsymbol{O}}f_{U}(\boldsymbol{O})\right) = 0,\\
&\INT{\mathrm{SO(3)}}{}{\boldsymbol{O}}f_{\psi_1}(\boldsymbol{O})\nabla_{\boldsymbol{O}}\cdot\left(f_{F}(\boldsymbol{O})\nabla_{\boldsymbol{O}}\psi_{P}(\boldsymbol{O})\right) = 0,\\
&\INT{\mathrm{SO(3)}}{}{\boldsymbol{O}}f_{\psi_1}(\boldsymbol{O})\nabla_{\boldsymbol{O}}\cdot\left(f_{F}(\boldsymbol{O})\nabla_{\boldsymbol{O}}f_{F}(\boldsymbol{O})\right) = 0.
\end{align*}
\section{Coefficients for Model 2.}\label{coefficientsmodel2}
As with Model 1, here evaluate the integrals required for calculating the coefficients in Eq.\ \eqref{transrot2}.
\subsection{Translational diffusion}
We first calculate all integrals of the form $\INT{}{}{\boldsymbol{O}} \frac{3}{2}(m_{3,i} m_{3,j} -\frac{1}{3})m_{3,k} m_{3,l} \boldsymbol{m}_3 \otimes \boldsymbol{m}_3$. Those are necessary to calculate the coefficients for translational diffusion:
\begin{widetext}
\begin{align*}
\INT{\mathrm{\boldsymbol{S}_2}}{}{\boldsymbol{O}} m_{3,1} m_{3,1} \frac{3}{2}(m_{3,1} m_{3,1} -\frac{1}{3})\boldsymbol{m}_3 \otimes \boldsymbol{m}_3  &=\left(\begin{array}{ccc}
     16 \pi /35& 0& 0  \\
     0& 4 \pi /105& 0 \\
     0& 0& 4 \pi /105
\end{array}\right),\\
\INT{\mathrm{\boldsymbol{S}_2}}{}{\boldsymbol{O}} m_{3,1} m_{3,1} \frac{3}{2}(m_{3,1} m_{3,2} -\frac{1}{3})\boldsymbol{m}_3 \otimes \boldsymbol{m}_3    &= \left(\begin{array}{ccc}
     -2\pi /5& 6\pi /15& 0  \\
     6\pi /15& -2\pi / 15& 0 \\
     0& 0& -2\pi / 15
\end{array}\right),\\
\INT{\mathrm{\boldsymbol{S}_2}}{}{\boldsymbol{O}} m_{3,1} m_{3,1} \frac{3}{2}(m_{3,1} m_{3,3} -\frac{1}{3})\boldsymbol{m}_3 \otimes \boldsymbol{m}_3    &= \left(\begin{array}{ccc}
-2\pi / 5& 0&6 \pi /35\\
0& -2\pi / 15& 0\\
6\pi /15& 0& -2\pi/ 15
\end{array}\right),\\
\INT{\mathrm{\boldsymbol{S}_2}}{}{\boldsymbol{O}} m_{3,1} m_{3,1} \frac{3}{2}(m_{3,2} m_{3,3} -\frac{1}{3})\boldsymbol{m}_3 \otimes \boldsymbol{m}_3    &= 
\left(\begin{array}{ccc}
-2\pi /5& 0& 0\\
0& -2\pi / 15& 2\pi/ 35\\
0& 2\pi / 35& -2\pi/ 15
\end{array}\right),\\
\INT{\mathrm{\boldsymbol{S}_2}}{}{\boldsymbol{O}} m_{3,1} m_{3,1} \frac{3}{2}(m_{3,2} m_{3,2} -\frac{1}{3}) \boldsymbol{m}_3 \otimes \boldsymbol{m}_3   &= 
\left(\begin{array}{ccc}
-8\pi / 35& 0& 0\\
0& 4 \pi /105& 0\\
0& 0& -8\pi /105
\end{array}\right),\\
\INT{\mathrm{\boldsymbol{S}_2}}{}{\boldsymbol{O}} m_{3,1} m_{3,1} \frac{3}{2}(m_{3,3} m_{3,3} -\frac{1}{3})\boldsymbol{m}_3 \otimes \boldsymbol{m}_3    &=
\left(\begin{array}{ccc}
-8\pi / 35& 0& 0\\
0& -8\pi /105& 0\\
0& 0& 4 \pi /105
\end{array}\right),\\
\INT{\mathrm{\boldsymbol{S}_2}}{}{\boldsymbol{O}} m_{3,1} m_{3,2}2 \frac{3}{2}(m_{3,1} m_{3,1} -\frac{1}{3})\boldsymbol{m}_3 \otimes \boldsymbol{m}_3    &= 
\left(\begin{array}{ccc}
0& 4 \pi /105& 0\\
4 \pi /105& 0& 0\\
0& 0& 0
\end{array}\right),\\
\INT{\mathrm{\boldsymbol{S}_2}}{}{\boldsymbol{O}} m_{3,1} m_{3,2} \frac{3}{2}(m_{3,1} m_{3,2} -\frac{1}{3}) \boldsymbol{m}_3 \otimes \boldsymbol{m}_3   &= 
\left(\begin{array}{ccc}
6 \pi /35& -2\pi / 15& 0\\
-2\pi / 15&6\pi /15& 0\\
0& 0& 2\pi / 35
\end{array}\right),\\
\INT{\mathrm{\boldsymbol{S}_2}}{}{\boldsymbol{O}} m_{3,1} m_{3,2} \frac{3}{2}(m_{3,1} m_{3,3} -\frac{1}{3}) \boldsymbol{m}_3 \otimes \boldsymbol{m}_3   &= 
\left(\begin{array}{ccc}
0& -2\pi / 15& 0\\
-2\pi / 15& 0& 2\pi / 35\\
0& 2\pi / 35& 0
\end{array}\right),\\
\INT{\mathrm{\boldsymbol{S}_2}}{}{\boldsymbol{O}} m_{3,1} m_{3,2} \frac{3}{2}(m_{3,2} m_{3,3} -\frac{1}{3})\boldsymbol{m}_3 \otimes \boldsymbol{m}_3    &= 
\left(\begin{array}{ccc}
0& -2\pi / 15& 2\pi / 35\\
-2\pi / 15&0&0\\
2\pi/ 35& 0& 0
\end{array}\right),\\
\INT{\mathrm{\boldsymbol{S}_2}}{}{\boldsymbol{O}} m_{3,1} m_{3,2} \frac{3}{2}(m_{3,2} m_{3,2} -\frac{1}{3}) \boldsymbol{m}_3 \otimes \boldsymbol{m}_3   &= 
\left(\begin{array}{ccc}
0& 4\pi /105& 0\\
4 \pi /105& 0& 0\\
0& 0& 0
\end{array}\right),\\
\INT{\mathrm{\boldsymbol{S}_2}}{}{\boldsymbol{O}} m_{3,1} m_{3,2} \frac{3}{2}(m_{3,3} m_{3,3} -\frac{1}{3}) \boldsymbol{m}_3 \otimes \boldsymbol{m}_3   &= 
\left(\begin{array}{ccc}
0& -8\pi /105& 0\\
-8\pi /105& 0& 0\\
0& 0&0
\end{array}\right),\\
\INT{\mathrm{\boldsymbol{S}_2}}{}{\boldsymbol{O}} m_{3,1} \boldsymbol{m}_3, \frac{3}{2}(m_{3,1} m_{3,1} -\frac{1}{3}) \boldsymbol{m}_3 \otimes \boldsymbol{m}_3   &= 
\left(\begin{array}{ccc}
0& 0& 4\pi /105\\
0& 0& 0\\
4\pi /105& 0& 0
\end{array}\right),\\
\INT{\mathrm{\boldsymbol{S}_2}}{}{\boldsymbol{O}} m_{3,1} m_{3,3} \frac{3}{2}(m_{3,1} m_{3,2} -\frac{1}{3})\boldsymbol{m}_3 \otimes \boldsymbol{m}_3    &= 
\left(\begin{array}{ccc}
0& 0& -2\pi / 15\\
0& 0& 2\pi / 35\\
-2\pi / 15& 2\pi / 35& 0
\end{array}\right),\\
\INT{\mathrm{\boldsymbol{S}_2}}{}{\boldsymbol{O}} m_{3,1} m_{3,3} \frac{3}{2}(m_{3,1} m_{3,3} -\frac{1}{3})\boldsymbol{m}_3 \otimes \boldsymbol{m}_3    &= \left(\begin{array}{ccc}6 \pi /35& 0& -2\pi / 15\\
0& 2\pi / 35& 0\\
-2\pi / 15& 0&6\pi /15\end{array}\right),\\
\INT{\mathrm{\boldsymbol{S}_2}}{}{\boldsymbol{O}} m_{3,1} m_{3,3} \frac{3}{2}(m_{3,2} m_{3,3} -\frac{1}{3})\boldsymbol{m}_3 \otimes \boldsymbol{m}_3    &= \left(\begin{array}{ccc}0& 2\pi / 35& -2\pi / 15\\
2\pi / 35& 0& 0\\
-2\pi / 15& 0&0\end{array}\right),\\
\INT{\mathrm{\boldsymbol{S}_2}}{}{\boldsymbol{O}} m_{3,1} m_{3,3} \frac{3}{2}(m_{3,2} m_{3,2} -\frac{1}{3}) \boldsymbol{m}_3 \otimes \boldsymbol{m}_3   &= \left(\begin{array}{ccc}0& 0& -8 \pi /105\\
0& 0& 0\\
-8\pi /105&0& 0\end{array}\right),\\
\INT{\mathrm{\boldsymbol{S}_2}}{}{\boldsymbol{O}} m_{3,1} m_{3,3} \frac{3}{2}(m_{3,3} m_{3,3} -\frac{1}{3})\boldsymbol{m}_3 \otimes \boldsymbol{m}_3    &= \left(\begin{array}{ccc}0& 0& 4 \pi /105\\
0& 0& 0\\
4 \pi /105&0& 0\end{array}\right),\\
\INT{\mathrm{\boldsymbol{S}_2}}{}{\boldsymbol{O}} m_{3,2} m_{3,3} \frac{3}{2}(m_{3,1} m_{3,1} -\frac{1}{3})\boldsymbol{m}_3 \otimes \boldsymbol{m}_3    &= \left(\begin{array}{ccc}0& 0& 0\\
0& 0& -8 \pi/105\\
0& -8 \pi /105& 0\end{array}\right),\\
\INT{\mathrm{\boldsymbol{S}_2}}{}{\boldsymbol{O}} m_{3,2} m_{3,3} \frac{3}{2}(m_{3,1} m_{3,2} -\frac{1}{3})\boldsymbol{m}_3 \otimes \boldsymbol{m}_3    &= \left(\begin{array}{ccc}0& 0& 2\pi / 35\\
0& 0& -2\pi / 15\\
2\pi / 35& -2\pi / 15& 0\end{array}\right),\\
\INT{\mathrm{\boldsymbol{S}_2}}{}{\boldsymbol{O}} m_{3,2} m_{3,3} \frac{3}{2}(m_{3,1} m_{3,3} -\frac{1}{3})\boldsymbol{m}_3 \otimes \boldsymbol{m}_3    &= \left(\begin{array}{ccc}0& 2\pi / 35& 0\\
2\pi / 35& 0& -2\pi / 15\\
0& -2\pi / 15& 0\end{array}\right),\\
\INT{\mathrm{\boldsymbol{S}_2}}{}{\boldsymbol{O}} m_{3,2} m_{3,3} \frac{3}{2}(m_{3,2} m_{3,1} -\frac{1}{3}) \boldsymbol{m}_3 \otimes \boldsymbol{m}_3   &= \left(\begin{array}{ccc}2\pi / 35& 0& 0\\ 
0&6 \pi /35& -2\pi / 15\\
0& -2\pi / 15&6 \pi /35\end{array}\right),\\
\INT{\mathrm{\boldsymbol{S}_2}}{}{\boldsymbol{O}} m_{3,2} m_{3,3} \frac{3}{2}(m_{3,2} m_{3,2} -\frac{1}{3})\boldsymbol{m}_3 \otimes \boldsymbol{m}_3    &= \left(\begin{array}{ccc}0& 0& 0\\
0& 0& 4 \pi /105\\ 
0& 4 \pi /105& 0\end{array}\right),\\
\INT{\mathrm{\boldsymbol{S}_2}}{}{\boldsymbol{O}} m_{3,2} m_{3,3} \frac{3}{2}(m_{3,3} m_{3,3} -\frac{1}{3})\boldsymbol{m}_3 \otimes \boldsymbol{m}_3    &= \left(\begin{array}{ccc}0& 0& 0\\
0& 0& 4 \pi /105\\ 
0& 4 \pi /105& 0\end{array}\right),\\
\INT{\mathrm{\boldsymbol{S}_2}}{}{\boldsymbol{O}} m_{3,2} m_{3,2} \frac{3}{2}(m_{3,1} m_{3,1} -\frac{1}{3}) \boldsymbol{m}_3 \otimes \boldsymbol{m}_3   &= \left(\begin{array}{ccc}4 \pi /105& 0& 0\\ 
0& -8 \pi / 35& 0\\
0& 0& -8 \pi /105\end{array}\right),\\
\INT{\mathrm{\boldsymbol{S}_2}}{}{\boldsymbol{O}} m_{3,2} m_{3,2} \frac{3}{2}(m_{3,1} m_{3,2} -\frac{1}{3})\boldsymbol{m}_3 \otimes \boldsymbol{m}_3     &= \left(\begin{array}{ccc}-2\pi / 15&6 \pi /35& 0\\
6 \pi /35& -2\pi /5& 0\\
0& 0& -2\pi / 15\end{array}\right),\\
\INT{\mathrm{\boldsymbol{S}_2}}{}{\boldsymbol{O}} m_{3,2} m_{3,2} \frac{3}{2}(m_{3,1} m_{3,3} -\frac{1}{3})\boldsymbol{m}_3 \otimes \boldsymbol{m}_3     &= \left(\begin{array}{ccc}-2\pi/ 15& 0& 2\pi / 35\\
0& -2\pi / 5& 0\\
2\pi / 35& 0& -2\pi / 15\end{array}\right),\\
\INT{\mathrm{\boldsymbol{S}_2}}{}{\boldsymbol{O}} m_{3,2} m_{3,2} \frac{3}{2}(m_{3,2} m_{3,3} -\frac{1}{3})\boldsymbol{m}_3 \otimes \boldsymbol{m}_3     &= \left(\begin{array}{ccc}-2\pi / 15&0& 0\\
0& -2\pi /5&6 \pi /35\\
0&6 \pi /35& -2\pi / 15\end{array}\right),\\
\INT{\mathrm{\boldsymbol{S}_2}}{}{\boldsymbol{O}} m_{3,2} m_{3,2} \frac{3}{2}(m_{3,2} m_{3,2} -\frac{1}{3}) \boldsymbol{m}_3 \otimes \boldsymbol{m}_3    &= \left(\begin{array}{ccc}4 \pi /105& 0& 0\\
0& 16 \pi / 35& 0\\
0& 0& 4 \pi /105\end{array}\right),\\
\INT{\mathrm{\boldsymbol{S}_2}}{}{\boldsymbol{O}} m_{3,2} m_{3,2} \frac{3}{2}(m_{3,3} m_{3,3} -\frac{1}{3}) \boldsymbol{m}_3 \otimes \boldsymbol{m}_3    &= \left(\begin{array}{ccc}-8 \pi /105& 0& 0\\
0& -8 \pi / 35& 0\\ 
0&0&4 \pi /105\end{array}\right),\\
\INT{\mathrm{\boldsymbol{S}_2}}{}{\boldsymbol{O}} m_{3,3} m_{3,3} \frac{3}{2}(m_{3,1} m_{3,1} -\frac{1}{3}) \boldsymbol{m}_3 \otimes \boldsymbol{m}_3    &= \left(\begin{array}{ccc}4 \pi /105& 0& 0\\
0& -8 \pi /105& 0\\
0& 0& -8 \pi/ 35\end{array}\right),\\
\INT{\mathrm{\boldsymbol{S}_2}}{}{\boldsymbol{O}} m_{3,3} m_{3,3} \frac{3}{2}(m_{3,1} m_{3,2} -\frac{1}{3})\boldsymbol{m}_3 \otimes \boldsymbol{m}_3    &= \left(\begin{array}{ccc}-2\pi / 15& 2\pi / 35& 0\\
2\pi / 35& -2\pi / 15& 0\\
0& 0&-2\pi /5\end{array}\right),\\
\INT{\mathrm{\boldsymbol{S}_2}}{}{\boldsymbol{O}} m_{3,3} m_{3,3} \frac{3}{2}(m_{3,1} m_{3,1} -\frac{1}{3})\boldsymbol{m}_3 \otimes \boldsymbol{m}_3    &= \left(\begin{array}{ccc}-2\pi / 15& 0& 6\pi / 35\\
0& -2\pi / 15& 0\\
6 \pi /35& 0& -2\pi / 5\end{array}\right),\\
\INT{\mathrm{\boldsymbol{S}_2}}{}{\boldsymbol{O}} m_{3,3} m_{3,3} \frac{3}{2}(m_{3,2} m_{3,3} -\frac{1}{3}) \boldsymbol{m}_3 \otimes \boldsymbol{m}_3   &= \left(\begin{array}{ccc}-2\pi / 15& 0& 0\\ 
0& -2\pi / 15&6 \pi /35\\
0&6 \pi /35& -2\pi /5\end{array}\right),\\
\INT{\mathrm{\boldsymbol{S}_2}}{}{\boldsymbol{O}} m_{3,3} m_{3,3} \frac{3}{2}(m_{3,2} m_{3,2} -\frac{1}{3}) \boldsymbol{m}_3 \otimes \boldsymbol{m}_3   &= \left(\begin{array}{ccc}-8\pi /105&0&0\\
0& 4 \pi /105& 0\\
0& 0&-8\pi/35\end{array}\right),\\
\INT{\mathrm{\boldsymbol{S}_2}}{}{\boldsymbol{O}} m_{3,3} m_{3,3} \frac{3}{2}(m_{3,3} m_{3,3} -\frac{1}{3})\boldsymbol{m}_3 \otimes \boldsymbol{m}_3    &= \left(\begin{array}{ccc}4\pi /105& 0& 0\\
0& 4 \pi /105& 0\\
0& 0& 16 \pi/35\end{array}\right),\\
\INT{\mathrm{\boldsymbol{S}_2}}{}{\boldsymbol{O}}  \frac{3}{2}(m_{3,1} m_{3,1} -\frac{1}{3})\boldsymbol{m}_3 \otimes \boldsymbol{m}_3    &= 
\left(\begin{array}{ccc}8 \pi /15& 0& 0\\
0& 4 \pi/15&0\\ 
0& 0& 4 \pi /15\end{array}\right),\\
\INT{\mathrm{\boldsymbol{S}_2}}{}{\boldsymbol{O}}  \frac{3}{2}(m_{3,1} m_{3,2} -\frac{1}{3})\boldsymbol{m}_3 \otimes \boldsymbol{m}_3    &= \left(\begin{array}{ccc}
-2 \pi /3& 2 \pi /5& 0\\
2 \pi /5& -2 \pi /3& 0\\
0& 0& -2 \pi /3\end{array}\right),\\
\INT{\mathrm{\boldsymbol{S}_2}}{}{\boldsymbol{O}}  \frac{3}{2}(m_{3,1} m_{3,3} -\frac{1}{3})\boldsymbol{m}_3 \otimes \boldsymbol{m}_3    &= \left(\begin{array}{ccc}-2 \pi /3& 0& 2 \pi /5\\
0& -2 \pi /3& 0\\
2 \pi /5& 0& -2 \pi /3\end{array}\right),\\
\INT{\mathrm{\boldsymbol{S}_2}}{}{\boldsymbol{O}}  \frac{3}{2}(m_{3,1} m_{3,1} -\frac{1}{3})\boldsymbol{m}_3 \otimes \boldsymbol{m}_3    &= \left(\begin{array}{ccc}
-2 \pi /3& 0&0\\
0& -2 \pi /3& 2 \pi /5\\ 
0& 2 \pi /5& -2 \pi /3\end{array}\right),\\
\INT{\mathrm{\boldsymbol{S}_2}}{}{\boldsymbol{O}}  \frac{3}{2}(m_{3,2} m_{3,2} -\frac{1}{3})\boldsymbol{m}_3 \otimes \boldsymbol{m}_3    &= \left(\begin{array}{ccc}4 \pi/15& 0& 0\\
0& 8 \pi /15& 0\\
0& 0& 4 \pi /15\end{array}\right),\\
\INT{\mathrm{\boldsymbol{S}_2}}{}{\boldsymbol{O}}  \frac{3}{2}(m_{3,3} m_{3,3} -\frac{1}{3})\boldsymbol{m}_3 \otimes \boldsymbol{m}_3    &= \left(\begin{array}{ccc}-4\pi /15& 0&0\\
0& -4\pi /15& 0\\
0& 0& 8 \pi /15\end{array}\right).
\end{align*}
\end{widetext}
Finally, we calculate all integrals of the form $\INT{}{}{\boldsymbol{O}}m_{3,k} m_{3,l} \boldsymbol{m}_3 \otimes \boldsymbol{m}_3$ : \begin{align*}
\INT{\mathrm{\boldsymbol{S}_2}}{}{\boldsymbol{O}}   \boldsymbol{m}_3 \otimes \boldsymbol{m}_3 &= \left(\begin{array}{ccc}
4\pi /3& 0& 0\\
0& 4\pi /3& 0\\
0& 0& 4\pi /3
\end{array}\right),\\
\INT{\mathrm{\boldsymbol{S}_2}}{}{\boldsymbol{O}}  m_{3,1} m_{3,1}  \boldsymbol{m}_3 \otimes \boldsymbol{m}_3 &=
\left(\begin{array}{ccc}
4\pi /5& 0& 0\\
0& 4 \pi /15& 0\\
0& 0& 4 \pi /15
\end{array}\right),\\
\INT{\mathrm{\boldsymbol{S}_2}}{}{\boldsymbol{O}}  m_{3,2} m_{3,2}  \boldsymbol{m}_3 \otimes \boldsymbol{m}_3 &= 
\left(\begin{array}{ccc}
4 \pi /15&0& 0\\
0& 4 \pi /5& 0\\
0& 0& 4 \pi /15
\end{array}\right),\\
\INT{\mathrm{\boldsymbol{S}_2}}{}{\boldsymbol{O}}  m_{3,3} m_{3,3}  \boldsymbol{m}_3 \otimes \boldsymbol{m}_3 &= 
\left(\begin{array}{ccc}
4 \pi /15& 0& 0\\
0& 4 \pi /15& 0\\
0& 0& 4 \pi /5
\end{array}\right),\\
\INT{\mathrm{\boldsymbol{S}_2}}{}{\boldsymbol{O}} m_{3,1} m_{3,2}  \boldsymbol{m}_3 \otimes \boldsymbol{m}_3 &= 
\left(\begin{array}{ccc}
0& 4 \pi /15& 0\\
4 \pi /15& 0& 0\\
0&0& 0
\end{array}\right),\\
\INT{\mathrm{\boldsymbol{S}_2}}{}{\boldsymbol{O}} m_{3,1} m_{3,3}  \boldsymbol{m}_3 \otimes \boldsymbol{m}_3 &= 
\left(\begin{array}{ccc}
0& 0& 4 \pi /15\\
0& 0& 0\\
4 \pi /15& 0& 0
\end{array}\right),\\
\INT{\mathrm{\boldsymbol{S}_2}}{}{\boldsymbol{O}} m_{3,2} m_{3,3}  \boldsymbol{m}_3 \otimes \boldsymbol{m}_3 &= 
\left(\begin{array}{ccc}
0& 0& 0\\
0& 0&4 \pi /15\\
0& 4 \pi /15& 0
\end{array}\right).
\end{align*}
\subsection{Rotational diffusion}
Now, we need to calculate the coefficients for the rotational diffusion. 
We start with the integrals of the form $\INT{\boldsymbol{S}_2}{}{\boldsymbol{O}}\nabla_{\boldsymbol{O}}^2 u_i u_k$: 
\begin{align*}
\INT{\mathrm{\boldsymbol{S}_2}}{}{\boldsymbol{O}} \nabla_{\boldsymbol{O}}^2 m_{3,1} m_{3,1} &=-\frac{8\pi}{3},\\
\INT{\mathrm{\boldsymbol{S}_2}}{}{\boldsymbol{O}} \nabla_{\boldsymbol{O}}^2 m_{3,2} m_{3,2} &=-\frac{8\pi}{3},\\
\INT{\mathrm{\boldsymbol{S}_2}}{}{\boldsymbol{O}} \nabla_{\boldsymbol{O}}^2 m_{3,3} m_{3,3} &=\frac{16\pi}{3},\\
\INT{\mathrm{\boldsymbol{S}_2}}{}{\boldsymbol{O}}\nabla_{\boldsymbol{O}}^2 m_{3,1} m_{3,2} &=0,\\
\INT{\mathrm{\boldsymbol{S}_2}}{}{\boldsymbol{O}}\nabla_{\boldsymbol{O}}^2 m_{3,1} m_{3,3} &=0,\\
\INT{\mathrm{\boldsymbol{S}_2}}{}{\boldsymbol{O}}\nabla_{\boldsymbol{O}}^2 m_{3,2} m_{3,3} &=0.
\end{align*}
Now we move on to all integrals of the form $\INT{\boldsymbol{S}_2}{}{\boldsymbol{O}}\frac{3}{2}(m_{3,k} m_{3,l} - \frac{1}{3})\nabla_{\boldsymbol{O}}^2 u_i u_j$:
\begin{align*}
\INT{\mathrm{\boldsymbol{S}_2}}{}{\boldsymbol{O}} \frac{3}{2}(m_{3,1} m_{3,1} - \frac{1}{3})\nabla_{\boldsymbol{O}}^2 m_{3,1} m_{3,1}  &=-\frac{46\pi}{15},\\
\INT{\mathrm{\boldsymbol{S}_2}}{}{\boldsymbol{O}} \frac{3}{2}(m_{3,1} m_{3,1} - \frac{1}{3})\nabla_{\boldsymbol{O}}^2 m_{3,1} m_{3,2}  &= 0,\\
\INT{\mathrm{\boldsymbol{S}_2}}{}{\boldsymbol{O}} \frac{3}{2}(m_{3,1} m_{3,1} - \frac{1}{3})\nabla_{\boldsymbol{O}}^2 m_{3,1} m_{3,3}  &= 0,\\
\INT{\mathrm{\boldsymbol{S}_2}}{}{\boldsymbol{O}} \frac{3}{2}(m_{3,1} m_{3,1} - \frac{1}{3})\nabla_{\boldsymbol{O}}^2 m_{3,2} m_{3,3}  &=0,\\
\INT{\mathrm{\boldsymbol{S}_2}}{}{\boldsymbol{O}} \frac{3}{2}(m_{3,1} m_{3,1} - \frac{1}{3})\nabla_{\boldsymbol{O}}^2 m_{3,2} m_{3,2} &=\frac{38\pi}{15},\\
\INT{\mathrm{\boldsymbol{S}_2}}{}{\boldsymbol{O}} \frac{3}{2}(m_{3,1} m_{3,1} - \frac{1}{3})\nabla_{\boldsymbol{O}}^2 m_{3,3} m_{3,3} &=\frac{8\pi}{15},\\
\INT{\mathrm{\boldsymbol{S}_2}}{}{\boldsymbol{O}} \frac{3}{2}(m_{3,1} m_{3,2} - \frac{1}{3})\nabla_{\boldsymbol{O}}^2 m_{3,1} m_{3,1} &=\frac{4\pi}{3},\\
\INT{\mathrm{\boldsymbol{S}_2}}{}{\boldsymbol{O}} \frac{3}{2}(m_{3,1} m_{3,2} - \frac{1}{3})\nabla_{\boldsymbol{O}}^2 m_{3,1} m_{3,2} &=-\frac{14\pi}{5},\\
\INT{\mathrm{\boldsymbol{S}_2}}{}{\boldsymbol{O}} \frac{3}{2}(m_{3,1} m_{3,2} - \frac{1}{3})\nabla_{\boldsymbol{O}}^2 m_{3,1} m_{3,3} &= 0,\\
\INT{\mathrm{\boldsymbol{S}_2}}{}{\boldsymbol{O}} \frac{3}{2}(m_{3,1} m_{3,2} - \frac{1}{3})\nabla_{\boldsymbol{O}}^2 m_{3,2} m_{3,3} &=0,\\
\INT{\mathrm{\boldsymbol{S}_2}}{}{\boldsymbol{O}} \frac{3}{2}(m_{3,1} m_{3,2} - \frac{1}{3})\nabla_{\boldsymbol{O}}^2 m_{3,2} m_{3,2} &=\frac{4 \pi}{3},\\
\INT{\mathrm{\boldsymbol{S}_2}}{}{\boldsymbol{O}} \frac{3}{2}(m_{3,1} m_{3,2} - \frac{1}{3})\nabla_{\boldsymbol{O}}^2 m_{3,3} m_{3,3} &=-\frac{8\pi}{3},\\
\INT{\mathrm{\boldsymbol{S}_2}}{}{\boldsymbol{O}} \frac{3}{2}(m_{3,1} m_{3,3} - \frac{1}{3})\nabla_{\boldsymbol{O}}^2 m_{3,1} m_{3,1}  &=\frac{4\pi}{3},\\
\INT{\mathrm{\boldsymbol{S}_2}}{}{\boldsymbol{O}} \frac{3}{2}(m_{3,1} m_{3,3} - \frac{1}{3})\nabla_{\boldsymbol{O}}^2 m_{3,1} m_{3,2}  &= 0,\\
\INT{\mathrm{\boldsymbol{S}_2}}{}{\boldsymbol{O}} \frac{3}{2}(m_{3,1} m_{3,3} - \frac{1}{3})\nabla_{\boldsymbol{O}}^2 m_{3,1} m_{3,3}  &= -\frac{14\pi}{5},\\
\INT{\mathrm{\boldsymbol{S}_2}}{}{\boldsymbol{O}} \frac{3}{2}(m_{3,1} m_{3,3} - \frac{1}{3})\nabla_{\boldsymbol{O}}^2 m_{3,2} m_{3,3}  &= 0,\\
\INT{\mathrm{\boldsymbol{S}_2}}{}{\boldsymbol{O}} \frac{3}{2}(m_{3,1} m_{3,3} - \frac{1}{3})\nabla_{\boldsymbol{O}}^2 m_{3,2} m_{3,2}  &=  \frac{4\pi}{3},\\
\INT{\mathrm{\boldsymbol{S}_2}}{}{\boldsymbol{O}} \frac{3}{2}(m_{3,1} m_{3,3} - \frac{1}{3})\nabla_{\boldsymbol{O}}^2 m_{3,3} m_{3,3} &=-\frac{8\pi}{3},\\
\INT{\mathrm{\boldsymbol{S}_2}}{}{\boldsymbol{O}} \frac{3}{2}(m_{3,2} m_{3,3} - \frac{1}{3})\nabla_{\boldsymbol{O}}^2 m_{3,1} m_{3,1}  &=  \frac{4\pi}{3},\\
\INT{\mathrm{\boldsymbol{S}_2}}{}{\boldsymbol{O}} \frac{3}{2}(m_{3,2} m_{3,3} - \frac{1}{3})\nabla_{\boldsymbol{O}}^2 m_{3,1} m_{3,2}  &= 0,\\
\INT{\mathrm{\boldsymbol{S}_2}}{}{\boldsymbol{O}} \frac{3}{2}(m_{3,2} m_{3,3} - \frac{1}{3})\nabla_{\boldsymbol{O}}^2 m_{3,1} m_{3,3}  &= 0,\\
\INT{\mathrm{\boldsymbol{S}_2}}{}{\boldsymbol{O}} \frac{3}{2}(m_{3,2} m_{3,3} - \frac{1}{3})\nabla_{\boldsymbol{O}}^2 m_{3,2} m_{3,3}  &= -\frac{14\pi}{5},\\
\INT{\mathrm{\boldsymbol{S}_2}}{}{\boldsymbol{O}} \frac{3}{2}(m_{3,2} m_{3,3} - \frac{1}{3})\nabla_{\boldsymbol{O}}^2 m_{3,2} m_{3,2}  &=  \frac{4\pi}{3},\\
\INT{\mathrm{\boldsymbol{S}_2}}{}{\boldsymbol{O}} \frac{3}{2}(m_{3,2} m_{3,3} - \frac{1}{3})\nabla_{\boldsymbol{O}}^2 m_{3,3} m_{3,3}  &= -\frac{8\pi}{3} ,\\
\INT{\mathrm{\boldsymbol{S}_2}}{}{\boldsymbol{O}} \frac{3}{2}(m_{3,2} m_{3,2} - \frac{1}{3})\nabla_{\boldsymbol{O}}^2 m_{3,1} m_{3,1}  &= \frac{38\pi}{15},\\
\INT{\mathrm{\boldsymbol{S}_2}}{}{\boldsymbol{O}} \frac{3}{2}(m_{3,2} m_{3,2} - \frac{1}{3})\nabla_{\boldsymbol{O}}^2 m_{3,1} m_{3,2}  &= 0,\\
\INT{\mathrm{\boldsymbol{S}_2}}{}{\boldsymbol{O}} \frac{3}{2}(m_{3,2} m_{3,2} - \frac{1}{3})\nabla_{\boldsymbol{O}}^2 m_{3,1} m_{3,3}  &= 0,\\
\INT{\mathrm{\boldsymbol{S}_2}}{}{\boldsymbol{O}} \frac{3}{2}(m_{3,2} m_{3,2} - \frac{1}{3})\nabla_{\boldsymbol{O}}^2 m_{3,2} m_{3,3}  &= 0,\\
\INT{\mathrm{\boldsymbol{S}_2}}{}{\boldsymbol{O}} \frac{3}{2}(m_{3,2} m_{3,2} - \frac{1}{3})\nabla_{\boldsymbol{O}}^2 m_{3,2} m_{3,2}  &= -\frac{46\pi}{15},\\
\INT{\mathrm{\boldsymbol{S}_2}}{}{\boldsymbol{O}} \frac{3}{2}(m_{3,2} m_{3,2} - \frac{1}{3})\nabla_{\boldsymbol{O}}^2 m_{3,3} m_{3,3}  &= \frac{8\pi}{15},\\
\INT{\mathrm{\boldsymbol{S}_2}}{}{\boldsymbol{O}} \frac{3}{2}(m_{3,3} m_{3,3}- \frac{1}{3})\nabla_{\boldsymbol{O}}^2 m_{3,1} m_{3,1}  &= \frac{8\pi}{15},\\
\INT{\mathrm{\boldsymbol{S}_2}}{}{\boldsymbol{O}} \frac{3}{2}(m_{3,3} m_{3,3} - \frac{1}{3})\nabla_{\boldsymbol{O}}^2 m_{3,1} m_{3,2} &= 0,\\
\INT{\mathrm{\boldsymbol{S}_2}}{}{\boldsymbol{O}} \frac{3}{2}(m_{3,3} m_{3,3} - \frac{1}{3})\nabla_{\boldsymbol{O}}^2 m_{3,1} m_{3,3}  &= 0,\\
\INT{\mathrm{\boldsymbol{S}_2}}{}{\boldsymbol{O}} \frac{3}{2}(m_{3,3} m_{3,3} - \frac{1}{3})\nabla_{\boldsymbol{O}}^2 m_{3,2} m_{3,3}  &= 0,\\
\INT{\mathrm{\boldsymbol{S}_2}}{}{\boldsymbol{O}} \frac{3}{2}(m_{3,3} m_{3,3} - \frac{1}{3})\nabla_{\boldsymbol{O}}^2 m_{3,2} m_{3,2}  &= \frac{8\pi}{15},\\ 
\INT{\mathrm{\boldsymbol{S}_2}}{}{\boldsymbol{O}} \frac{3}{2}(m_{3,3} m_{3,3} - \frac{1}{3})\nabla_{\boldsymbol{O}}^2 m_{3,3} m_{3,3}  &= -\frac{16\pi}{15}. 
\end{align*}

\acknowledgments{We thank Raphael Wittkowski and Giovanni De Matteis for helpful discussions. A.E. would like to thank Amelie Keller for helpful discussions and her continuous support and the Studienstiftung des Deutschen Volkes as well as the Clarendon Trust for financial support.
Funding by the Deutsche Forschungsgemeinschaft (DFG) through the SPP 2265 under grant numbers WI 5527/1-2 (RW) and LO 418/25-2 (HL) and through the SFB 1552 under grant number  465145163 (MtV)  is gratefully acknowledged.}

\end{document}